\documentclass[aps,
prd,
superscriptaddress,
showpacs,
preprintnumbers,
nofootinbib,
notitlepage,
floatfix]{revtex4-1}

\usepackage{amsmath,amssymb,hyperref,subfigure,xspace}
\usepackage{mciteplus,color,mathtools,graphicx}
\usepackage{braket,bbm,slashed,bm}
\usepackage{color}
\usepackage{multirow}
\usepackage[english]{babel}
\usepackage{mathrsfs}
\usepackage{enumerate}
\usepackage{xspace}
\usepackage{slashed}
\usepackage{wrapfig}
\usepackage{float}
\usepackage[export]{adjustbox}
\usepackage[utf8]{inputenc}
\usepackage{comment}
\usepackage[normalem]{ulem}

\renewcommand{\k}{\bm{k}}
\newcommand{\p}{\bm{p}}
\newcommand{\q}{\bm{q}}

\newcommand{\one}{\mathbbm{1}}

\newcommand{\Cc}{\mathcal{C}}
\newcommand{\Dc}{\mathcal{D}}

\newcommand{\Kc}{\mathcal{K}}

\newcommand{\Mc}{\mathcal{M}}

\newcommand{\Oc}{\mathcal{O}}

\newcommand{\Rc}{\mathcal{R}}

\DeclareMathOperator{\re}{Re}
\DeclareMathOperator{\im}{Im}

\newcommand{\bcol}{\left[ \begin{array}{c}}
\newcommand{\ecol}{\end{array} \right]}
\newcommand{\beq}{\begin{eqnarray}}
\newcommand{\eeq}{\end{eqnarray}}

\usepackage{mfirstuc} 
\newcommand{\addReviewer}[2]{
  \expandafter\newcommand\csname #1\endcsname[1]{{\bf \color{#2} \capitalisewords{#1}:\,##1}}
  \expandafter\newcommand\csname #1cor\endcsname[2]{{\color{#2} \capitalisewords{#1}:\,\st{##1}{\bf ##2}}}
  \expandafter\newcommand\csname #1color\endcsname{#2}
}

\usepackage{soul,color}
\definecolor{cardinal}{rgb}{0.77, 0.12, 0.23}

\addReviewer{sebastian}{cardinal}
\addReviewer{raul}{blue}
\addReviewer{digonto}{red}

\hypersetup{ 
    pdfnewwindow=true,   
    colorlinks=true,     
    linkcolor=blue,      
    citecolor=blue,      
    filecolor=blue,      
    urlcolor=blue        
} 

\begin{document}


\newcommand{\UnivWash}{Physics Department, University of Washington, Seattle, WA 98195-1560, USA}
\newcommand{\jlab}{Thomas Jefferson National Accelerator Facility, 12000 Jefferson Avenue, Newport News, Virginia 23606, USA}
\newcommand{\odu}{Department of Physics, Old Dominion University, Norfolk, Virginia 23529, USA}
\newcommand{\UCB}{Department of Physics, University of California, Berkeley, CA 94720, USA}    
\newcommand{\LBNL}{Nuclear Science Division, Lawrence Berkeley National Laboratory, Berkeley, CA 94720, USA}  


\title{
Analytic continuation of the relativistic three-particle scattering amplitudes 
}



\author{Sebastian~M.~Dawid}
\email[email: ]{dawids@uw.edu}
\affiliation{\UnivWash}

\author{Md Habib E Islam}
\email[e-mail: ]{m2islam@odu.edu}
\affiliation{\odu}
\affiliation{\jlab}

\author{Ra\'ul A.~Brice\~no}
\email[e-mail: ]{rbriceno@berkeley.edu}
\affiliation{\UCB}
\affiliation{\LBNL} 


\begin{abstract}
We investigate the relativistic scattering of three identical scalar bosons interacting via pair-wise interactions. Extending techniques from the non-relativistic three-body scattering theory, we provide a detailed and general prescription for solving and analytically continuing integral equations describing the three-body reactions. We use these techniques to study a system with zero angular momenta described by a single scattering length leading to a bound state in a two-body sub-channel. We obtain bound-state--particle and three-particle amplitudes in the previously unexplored kinematical regime; in particular, for real energies below elastic thresholds and complex energies in the physical and unphysical Riemann sheets. We extract positions of three-particle bound-states that agree with previous finite-volume studies, providing further evidence for the consistency of the relativistic finite-volume three-body quantization conditions. We also determine previously unobserved virtual bound states in this theory. Finally, we find numerical evidence of the breakdown of the two-body finite-volume formalism in the vicinity of the left-hand cuts and argue for the generalization of the existing formalism.
\end{abstract}

\date{\today}
\maketitle

\section{Introduction}
\label{sec:introduction}

The need for a non-perturbative and relativistic framework to describe the dynamics of three-hadron systems is pressing and encompasses a broad class of hadronic and nuclear physics subfields, ranging from the lattice quantum chromodynamics (QCD) computations to experimental searches for the spectrum of strong interactions. The majority of QCD states are unstable resonances that reveal themselves in reactions with final products consisting of three and more particles \cite{Ketzer:2019wmd, LHCb:2021tdf, Davier:2013sfa, Garzia:2018ctr}. Among the most notable examples are the lightest excitation of the proton, Roper resonance $N^*(1440)$, a hybrid-meson candidate $\pi_1(1600)$, the charmed-molecule candidate $\chi_{c1}(3872)$, and its cousin, the recently discovered tetraquark candidate $T_{cc}^+(3872)$ \cite{Roper:1964, Arndt:2006bf, E852:1998mbq, COMPASS:2018uzl, E705:1993pry, Belle:2003nnu, LHCb:2020fvo, LHCb:2021ten, LHCb:2021vvq, LHCb:2021auc}. Systematic analysis of these states requires understanding the complicated final state interactions and building robust multi-body reaction amplitudes that satisfy the grounding principles of quantum mechanics, such as unitarity and analyticity. These two principles are essential when determining resonances manifesting as pole singularities in the scattering amplitudes.

While the importance of such a framework in experimental analysis is generally recognized, further discussion is required to motivate its significance in lattice QCD. The major challenge for accessing scattering observables via lattice QCD is the necessary truncation of the space-time. Indeed, most modern lattice QCD calculations use periodic cubic volumes. By making the volume finite, one can not define asymptotic states and consequently directly determine scattering amplitudes.\footnote{In principle, one could define wave packets in a finite volume and approximately access scattering amplitudes in a finite-volume~\cite{Briceno:2020rar}, but this would require real-time correlations that are not currently accessible using standard lattice QCD techniques.} However, it is possible to construct an exact, non-perturbative relation between finite- and infinite-volume observables. It was first presented by L\"uscher~\cite{LUSCHER1991531, Luscher:1985dn, Luscher:1986pf} for a system composed of two scalar bosons. His formalism, and its generalization to arbitrarily complex two-body systems~\cite{Rummukainen:1995vs, Kim:2005gf, Briceno:2012yi, Hansen:2012tf, Briceno:2014oea}, have resulted in a rich field of lattice QCD studies of scattering systems~\cite{Dudek:2014qha, Alexandrou:2017mpi, Prelovsek:2020eiw, Brett:2018jqw, Woss:2019hse, Woss:2020ayi, Wilson:2019wfr, Wilson:2015dqa, Briceno:2017qmb, Andersen:2017una, Wilson:2014cna, Briceno:2016mjc, Gayer:2021xzv, Dudek:2016cru,  Moir:2016srx, Rendon:2020rtw, Silvi:2021uya}.\footnote{See Ref.~\cite{RevModPhys.90.025001} for a recent review.} We note that these formulations are correct when applied to energies above the two-particle threshold. They may break down below that energy, an issue we discuss further in the text.

Similarly to the two-body sector, one may obtain relations that constrain infinite-volume scattering observables involving three-particle states based on finite-volume (FV) quantities. The first relativistic formulation relating the FV spectrum and the purely hadronic three-particle scattering amplitude was derived in Refs.~\cite{Hansen:2014eka, Hansen:2015zga, Blanton:2020gha}. A key outcome of this work is that in a finite volume, one places constraints on an infinite-volume object known as the three-body $K$ matrix. It is a generally unknown, real, and ``smooth" function of kinematic variables that describes the short-distance three-body interactions and can be considered an analog of the two-body phase shift. It is related to physical scattering amplitudes via a set of integral equations. This formalism was originally developed by assuming all particles to be identical scalar bosons that do not couple to two-particle states. These restrictions have been slowly lifted in Refs.~\cite{Briceno:2017tce, Briceno:2018aml, Hansen:2020zhy, Blanton:2021mih, Blanton:2020jnm}. Alternative and equivalent forms of the formalism were proposed in parallel~\cite{Mai:2017vot, Mai:2017bge, Hammer:2017uqm, Hammer:2017kms, Doring:2018xxx, Jackura:2018xnx, Dawid:2020uhn, Muller:2021uur}. The distinguishing features of these differing formalisms are technical, and it was shown that all of them are equivalent versions of the same underlying mathematical structure satisfying conditions imposed by the $S$-matrix unitarity~\cite{Jackura:2019bmu, Briceno:2019muc, Blanton:2020gmf, Jackura:2022gib}. 

The first implementation of the formalism in lattice QCD studies focused on determining the three-body $K$ matrix from FV spectra for maximal isospin~$3\pi$~\cite{Horz:2019rrn, Blanton:2019vdk, Mai:2019fba, Culver:2019vvu, Fischer:2020jzp, Brett:2021wyd, Blanton:2021llb}, $3K$~\cite{Alexandru:2020xqf, Blanton:2021llb}, and mixed $\pi \pi K$ systems \cite{Draper:2023boj}. The first study to take all the steps from the analysis of the lattice QCD correlations to physical scattering amplitude was presented in Ref.~\cite{Hansen:2020otl}.\footnote{For a calculation of three-particle systems in a toy-model, lattice $\varphi^4$ theory, see Ref.~\cite{Garofalo:2022pux}.} Although part of the limitation of studying increasingly rich systems is computational, the primary challenge is more formal. One of the essential unresolved obstacles is a proper understanding of the relationship between the three-body $K$ matrix and the physical scattering amplitude. It requires solving a system of integral equations in terms of purely on-shell dynamical inputs. These objects have kinematic and dynamical singularities, which result in amplitudes of a complicated analytic structure. For this reason, FV formalism must be accompanied by amplitude analysis techniques, which is the focus of this work.

First steps towards solving the particular set of integral equations, namely those presented in Ref.~\cite{Hansen:2015zga}, were carried out in Ref.~\cite{Jackura:2020bsk}. In that work, the authors considered one of the most singular scenarios, where the two-particle subsystem develops an $S$-wave bound state (dimer), labeled as $b$. They studied the scattering in the $S$ wave in the total, three-particle angular momentum $J$, and for simplicity, fixed the three-body $K$ matrix to zero. This model is the continued focus of our study.\footnote{Although they do not introduce new singularities, higher partial waves require additional consideration. Inclusion of a non-zero $K$ matrix is straightforward after first solving the vanishing $K$ matrix case.} Using the Lehmann, Symanzik, and Zimmermann (LSZ) reduction formula, this simplified theory can be used not only for studying $3\varphi \leftrightarrow 3\varphi$ scattering, where $\varphi$ is a generic label for a scalar boson of mass $m$ but also $\varphi+b \leftrightarrow \varphi+b$ and $\varphi+b \leftrightarrow 3\varphi$.

This same theory was previously investigated using the finite-volume formalism in Ref.~\cite{Romero-Lopez:2019qrt}. By obtaining energies below the three-particle threshold,  $s_{3\varphi } \equiv (3 m)^2$, these energies can be associated with those of a two-particle system composed of $\varphi+b$ and mapped to infinite-volume amplitudes using the L\"uscher formalism. The results there include a determination of $\Mc_{\varphi b}$, the $\varphi+b \to \varphi + b$ amplitude for energies below $s_{3\varphi }$ but also below the $\varphi b$ threshold, $s_{\varphi b } \equiv (m_b+m)^2$, where $m_b$ is the mass of the dimer. Below this threshold, the authors presented strong evidence for the three-particle bound states. It is important to note that in this same kinematic region, one does not expect the L\"uscher formalism to be generally applicable~\cite{Raposo:2023nex}, and as a result, the amplitude presented in Ref.~\cite{Romero-Lopez:2019qrt} may suffer of systematic corrections below $s_{\varphi b}$.

Study of Ref.~\cite{Jackura:2020bsk} followed the Nystr\"om method \cite{10.1007/BF02547521, delves1988computational} to establish a systematically improvable, numerical procedure for solving the three-body integral equations. It found a perfect agreement between the obtained $\Mc_{\varphi b}$ with the results of Ref.~\cite{Romero-Lopez:2019qrt} for energies in the range $s_{\varphi b } \leq s\leq s_{3 \varphi }$. Here we extend this work to investigate the integral equations and their solutions for energies below $s_{\varphi b }$, as well as in the complex energy plane, including the nearest unphysical Riemann sheet. Such an extension is far from trivial, as the partial-wave projected equations suffer from singularities that complicate the analytic properties of the final amplitude. For instance, these can result in the left-hand cuts below the $\varphi b$ threshold that obscure the presence of the bound-state poles. As discussed further in the text, we reach the correct solution by implementing analytic continuation techniques, which include integration contour deformation.

We achieve great agreement with the three-body bound states found in Ref.~\cite{Romero-Lopez:2019qrt}. However, we also witness the tension between our and the FV result for the $\varphi b$ amplitude. We interpret this as evidence of the L\"uscher formalism breaking down in the presence of the nearest $u$-channel cut associated with the partial-wave projection of one-particle exchange (OPE) amplitude. This OPE cut is also a key source of complication for solving the three-body integral equations for arbitrary kinematics, and we discuss this in great detail.

Before presenting our strategy for solving the desired integral equation, it is worth briefly summarizing the key literature on the topic. The analytical structure of the relativistic three-body amplitudes was an area of substantial research within the $S$-matrix theory literature in the 1960's~\cite{Hwa:1964xyz, Holman:1965prb, Grisaru:1966xyz} but also in the modern three-body approaches~\cite{Jackura:2018xnx}. Energy- and momentum-space contour deformations in the three-body integral equations have been employed as a solution tool necessary for reaching the unphysical energy domain~\cite{ Rubin:1966zz, Rubin:1967zza, Brayshaw:1968yia, Glockle:1978zz, Orlov:1984, Eichmann:2019dts, Sadasivan:2020syi}. The first analysis of the OPE cuts in the non-relativistic three-body system was performed by Rubin, Sugar, and Tiktopoulos in 1966 \cite{Rubin:1966zz}. It was considerably expanded by Brayshaw in 1968 \cite{PhysRev.167.1505, Brayshaw:1968yia}. In 1978, Gl\"ockle performed an analytic continuation of the non-relativistic, homogeneous Fadeev equation to describe poles of the three-neutron $^1S_0$ interaction~\cite{Glockle:1978zz}. He presented a procedure for avoiding the poles/ cuts of the non-relativistic OPE propagator via complex momentum contour deformation. It allowed him to trace trajectories of the $S$-matrix poles with evolving strength of the separable Yamaguchi potential, which was used as a model for two-body interactions between nucleons. In this work, we closely follow the ideas of Brayshaw and Gl\"ockle.\footnote{For parallel efforts in studying analytic properties of amplitudes in the context of Dyson-Schwinger equations and three-point functions, we point the reader to Refs.~\cite{Eichmann:2019dts} and~\cite{Huber:2022nzs}, respectively. An alternative relativistic description of three-boson bound states, known as Bethe-Salpeter-Fadeev equations, is described in Ref.~\cite{Ydrefors:2020duk} and references therein.} More recent efforts to compute three-body, relativistic amplitudes include those presented in Refs.~\cite{Sadasivan:2020syi, Sadasivan:2021emk}, where the authors studied the $a_1(1260)\to 3\pi$ resonance channel.

This work is organized in the following way. First, in Sec.~\ref{sec:integral-equation}, we summarize the formalism of interest; in particular, the building blocks of the relativistic three-body scattering equations. We focus on the system of Ref.~\cite{Jackura:2020bsk}, but the discussion applies to other physical scenarios. In Sec.~\ref{sec:analytic-structure}, we analyze the analytic properties of the building blocks of the integral equation, providing useful numerical examples. Next, in Sec.~\ref{sec:analytic-continuation}, we discuss the analytic properties of the solutions of the integral equation, namely the scattering amplitudes, and discuss their analytic continuation to the complex plane of the total energy of the system, including the unphysical Riemann sheets, where resonances and virtual states reside. Section~\ref{sec:results} starts with an outline of the solution procedure. We refer the readers interested in a practical implementation of the integral equations to this part of our work. Then, we present numerical results for the $\Mc_{\varphi b}$ for a wide range of kinematical variables. We show evidence of three-body bound states, which agree with those found in Ref.~\cite{Romero-Lopez:2019qrt}. Furthermore, we discuss the discrepancy of that finite-volume study with the $\Mc_{\varphi b}$ amplitude below the $s_{\varphi b}$ threshold due to the neglected left-hand cut. In Sec.~\ref{sec:conclusions}, we provide a summary of our findings. Some of the more pedagogical and technical aspects of the discussion are relegated to three appendices, Appendices~\ref{app:B}, \ref{app:A}, and \ref{app:C}. In particular, App.~\ref{app:C} contains concrete numerical routines applicable in studies of general three-body scattering reactions.

\section{Relativistic three-body equation}
\label{sec:integral-equation}

To ensure the self-sufficiency of this work, we review the necessary equations presented in Refs.~\cite{Hansen:2015zga, Jackura:2020bsk} for describing the on-shell scattering amplitude of three identical spinless bosons of mass $m$. We label the corresponding particles by ``$\varphi$". The $3 \varphi \to 3\varphi$ scattering occurs with the center-of-mass (CM) energy $E$. The corresponding total invariant mass squared is $s = E^2$. In the initial and final three-body state we choose a particle that we call an initial/ final \emph{spectator}. Their momenta are denoted by $\k$ and $\p$, respectively. The other two hadrons, associated with the given spectator, form a \emph{pair}. Their state is projected to a definite angular momentum, and here, we restrict ourselves to the $S$-wave case only.

The scattering process is described by the three-body amplitude $\Mc_3$, which is defined to be symmetric under the interchange of individual particles in the initial and final states. In this work, we consider the \emph{unsymmetrized} version labeled $\Mc^{(u,u)}_3$, which can be understood to describe a quasi-two-body spectator-pair reaction. The fully symmetric $\Mc_3$ is obtained by summing $\Mc^{(u,u)}_3$ over the nine choices of possible spectator momenta. The $\Mc^{(u,u)}_3$ amplitude can be written in terms of two other amplitudes, 
    \begin{align}
    \label{eq:Def.M3uu}
    \Mc^{(u,u)}_3( \p,\k) = \Dc^{(u,u)}(\p, \k) + \Mc_{\text{df},3}^{(u,u)}( \p,\k ) \, .
    \end{align}
\begin{figure}[t!]
    \centering
    \includegraphics[ width=0.45\textwidth]{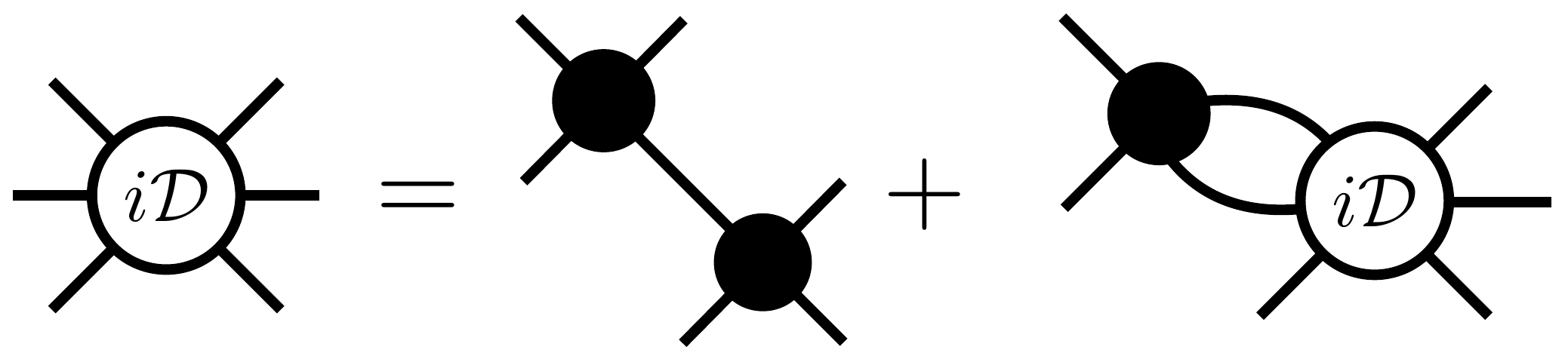}
    \caption{ Diagrammatic representation of ladder equation for the $\Dc$ amplitude defined in Eq.~\eqref{eq:Ddef}. Black circles represent the on-shell $2\to 2$ amplitude $\Mc_2$, introduced below that equation. Slanted lines represent amplitudes for one particle exchange, given by Eq.~\eqref{eq:G}. As made evident in Eq.~\eqref{eq:Ddef}, the building blocks of the integral equation are on their mass shell.}
    \label{fig:ladder}
\end{figure}
The first term of Eq.~\eqref{eq:Def.M3uu}, $\Dc^{(u,u)}$, depicted in Fig.~\ref{fig:ladder}, includes all pair-wise interactions in the absence of a pure three-body interaction. It is often referred to as the \emph{ladder} amplitude. The second term, amplitude $\Mc^{(u,u)}_{\text{df},3}$, includes all contributions that arise in the presence of a short-range three-body interaction. The short-distance dynamics is encoded in the relativistic three-body $K$ matrix, $\Kc_{\text{df},3}$. Although this separation between long- and short-range three-body interactions is scheme-dependent, $\Kc_{\text{df},3}$ is defined to assure that the resultant $\Mc^{(u,u)}_3$ is scheme-independent. In this work we assume that the three-body $K$ matrix is zero, which leads to $\Mc^{(u,u)}_{\text{df},3} = 0$. Having determined $\Dc^{(u,u)}$, one can include a non-zero $\Kc_{\text{df},3}$ contribution by solving equation for $\Mc^{(u,u)}_{\text{df},3}$.

\subsection{Ladder amplitude}

The ladder amplitude is defined by the integral equation,
    \begin{equation}
    \label{eq:Ddef}
    \Dc^{(u,u)}(\p, \k) = -\Mc_2(p) \, G(\p, \k) \, \Mc_2( k) - \Mc_2( p) \int \! \frac{d \q}{(2 \pi)^3 2 \omega_{q}}  \, G(\p, \q) \, \Dc^{(u,u)}(\q, \k) \, .
    \end{equation}
In general, $\Mc_2$ is a diagonal matrix in the pair's angular momentum space. Here we truncated it just to the $S$-wave element. It represents $2 \to 2$ scattering amplitude describing interactions among two particles in the initial and final pair. Their invariant mass squared, $\sigma_p$, is fixed by the momentum of the spectator, 
    \beq
    \label{eq:sigma_p}
    \sigma_p = (\sqrt{s} - \omega_p)^2 -  p^2 \, ,
    \eeq
 where $\omega_p = \sqrt{m^2+p^2}$, and $p = \lvert \p \rvert$ is the momentum magnitude. Two-body invariant mass squared can be used to express the CM spectator's momentum,
    \beq
    \label{eq:momentum_invs}
    p = \frac{ \lambda^{1/2}(s, \sigma_{p}, m^2) }{2 \sqrt{s} } \, ,
    \eeq
where $\lambda(x,y,z)$ is the triangle function, $\lambda(x,y,z) = x^2 + y^2 + z^2 - 2xy - 2yx - 2 zx$. The exchange propagator $G$, which describes the long-range interactions between the intermediate pair and spectator, is defined by
    \begin{equation}
    \label{eq:G}
    G(\p, \k) = \frac{H(p, k)}{b_{pk}^2 - m^2 + i \epsilon} \, ,
    \end{equation}
where $b^2_{pk} = (\sqrt{s} - \omega_p - \omega_{k})^2 - (\p + \k)^2 $, and $H( p, k)$ is a cut-off function necessary to ensure finitude of the integral in Eq.~\eqref{eq:Ddef}. In this work, we consider two classes of cut-off functions. The first is the smooth cut-off function presented in Ref.~\cite{Hansen:2015zga}, which we write explicitly in Sec.~\ref{sec:analytic-structure}. The second is a hard cut-off function that is equal to $1$ up to a maximum magnitude of the momentum, which we label as $q_{\rm max}$. 

Instead of considering the unsymmetrized ladder amplitude $\Dc^{(u,u)}$, it is beneficial to define the amputated amplitude $d(\p, \k)$,\footnote{For notation's simplicity, we drop the upper $(u,u)$ label in $d$, remembering that it is an unsymmetrized object.}
    \begin{align}
    \label{eq:d_def}
    \Dc^{(u,u)}(\p, \k) = \Mc_2( p ) \, d(\p, \k) \, \Mc_2( k) .
    \end{align}
in which one removes singularities in $\p$ and $\k$ associated with the $\Mc_2$. The amputated ladder amplitude satisfies an analogous integral equation,
    \begin{equation}
    \label{eq:d_intdef}
    d(\p, \k) = - G(\p, \k) - \int \frac{d \q}{(2 \pi)^3 2 \omega_{q}} \, G(\p, \q) \Mc_2( q) \, d(\q, \k)  \, .
    \end{equation}
It still depends on $\mathcal{M}_2$, which now enters the integral equation kernel. In the presence of a two-body bound state, $\mathcal{M}_2$ has a physical pole. Reference~\cite{Jackura:2020bsk} showed how to deal with such a singularity appearing in the integral equation when one considers physical energies $E$. 

Just as in Ref.~\cite{Jackura:2020bsk}, in addition to assuming that the two-body subsystem is well described by the $\ell = 0$ partial wave only, we also employ the partial wave projection in the total angular momentum, $J$, as defined in Eq.~(10) of Ref.~\cite{Jackura:2020bsk}, and we only consider the $J = 0$ scattering process. We denote the resultant amplitude with a subscript $S$, and it satisfies,
    \begin{align}
    \label{eq:d_Sproj}
    d_{S}(p, k)
    &=
    - G_{S}(p, k)  
    - \int_0^{q_{\rm max}} \! \frac{d q \, q^2}{ (2\pi)^2 \, \omega_{q}} \, G_{S}(p, q) \, \Mc_2( q) \, d_{S}(q, k) \, ,
    \end{align}
where we have introduced the $S$-wave projection of the OPE,
    \begin{align}
    \label{eq:Gs_proj}
    G_{S}(p, k)
    &= \int\limits_{-1}^1 dx \, 
    \frac{H(p, k)}{z(p, k) + i \epsilon - 2 p k x}
    \nonumber \\[5pt]
    & = - \frac{H(p, k)}{4pk} \, \log\left( \frac{z(p, k) + i \epsilon - 2pk}{z(p, k) + i \epsilon + 2pk} \right) \, ,
    \end{align}
with $z(p, k) = (\sqrt{s}-\omega_{k} - \omega_p)^2 - k^2 - p^2 - m^2$ and $x$ being the cosine of the scattering angle between final and initial spectators momenta. As described above, the upper bound of the integral in Eq.~\eqref{eq:d_Sproj}, which we label as $q_{\rm max}$, is fixed by the maximum value of $q$ for which the cut-off function $H$ has support. In our case, it is $q_{\rm max} = (s-m^2)/2\sqrt{s}$. It is useful to combine objects under the integral under one name, and define the \emph{integration kernel},
    \beq
    \label{eq:K_Sproj}
    K(p, q) = \frac{q^2}{(2\pi)^2 \omega_q}
    \, G_S(p, q) \,  \Mc_2(q) \, .
    \eeq
With this, we can rewrite Eq.~\eqref{eq:d_Sproj} as, 
    \begin{align}
    \label{eq:d_Sproj_kern}
    d_{S}(p, k)
    &=
    - G_{S}(p, k)  
    - \int_0^{q_{\rm max}} \! {d q } \, K(p, q)  \, d_{S}(q, k) \, ,
    \end{align}
In the remainder of this work, we consider this form of the ladder equation. Once one has obtained a numerical solution for $d_S$ using Eq.~\eqref{eq:d_Sproj}, it is possible to determine the $S$-wave projection of $\Dc^{(u,u)}$ using Eq.~\eqref{eq:d_def}, 
    \begin{align}
    \label{eq:DS}
    \Dc^{(u,u)}_S(p, k) 
    =    
    \Mc_2( p ) \,
    d_{S}(p, k) \,
    \Mc_2( k) \, .
    \end{align}

Partial-wave projection of the exchange propagator replaces the pole singularity in $\q$ with logarithmic branch cuts in $q$. Since these cuts play an important role in the process of the analytic continuation of Eq.~\eqref{eq:d_Sproj}, we delay their discussion to Sec.~\ref{sec:analytic-structure}. We just remark that having a non-zero value of $i\epsilon$ is necessary to define the integral equation in Eq.~\eqref{eq:d_intdef}; it follows from the $u$-channel pole shift in the OPE propagator, Eq.~\eqref{eq:G}, before the partial wave projection. In principle, the solution of the ladder equation is first obtained for finite $\epsilon$, and then the $\epsilon\to 0$ limit is taken. However, if the total invariant mass squared $s$ is complex, in practice we can set $\epsilon=0$ before solving for $d_S$. In doing so, one has to remember that $i\epsilon$ prescription defines a direction from which singularities of OPE are passed through by the integration contour in the first line of Eq.~\eqref{eq:Gs_proj}. Finally, this equation only holds when all orbital angular momenta have been set to zero. However, for any other amplitude with non-zero values of $J$ and the external pair's angular momenta, the pole structure of the OPE amplitude is the same. After partial-wave projection, the simple logarithm above would be replaced with a linear combination of Legendre functions of the second kind and non-singular functions. Given that the Legendre functions have the same singular points as the logarithm, the method for analytic continuation presented in this work applies to any partial wave.

\subsection{Bound-state--spectator scattering}
\label{sec:IE_BS}

We focus on a representative example of three-body scattering by considering a system where the two-body subsystem can become bound. Similarly to Ref.~\cite{Jackura:2020bsk}, we label the bound state of a pair as ``$b$". Although a toy model, it is a case of physical significance. Application of the relativistic three-body formalism to this system has hinted at emergent discrete scaling invariance \cite{Romero-Lopez:2019qrt, Jackura:2020bsk}, an underlying characteristic of Efimov systems \cite{Efimov:1970zz, Naidon:2016dpf}. We will explore this aspect of the model in the upcoming article, focusing here on the extension of results of Ref.~\cite{Jackura:2020bsk} to complex energy plane and verification of the finite-volume formalism.

In this model, the two-body amplitude $\Mc_2$ has a real pole in variable $\sigma_q$, below the two-particle threshold, $\sigma_q = (2m)^2$. We introduce this bound state by representing the on-shell, $S$-wave amplitude,
    \beq
    \label{eq:M2_general}
    \Mc_2(q) = \frac{1}{\Kc_2^{-1}(q) - i \rho(q)} \, ,
    \eeq
in the leading order (LO) effective range expansion (ERE). Namely, we take $\Kc_2(q) = - (16 \pi \sqrt{\sigma_q} )\, a$, where $a$ is the two-body scattering length and $\rho(q)$ is the two-body phase space for identical particles,
    \begin{equation}
    \label{eq:2p1.phase_space}
    i \rho(q) =  - \frac{1}{32 \pi \sqrt{\sigma_q}} \sqrt{4 m^2 - \sigma_q} \, .
    \end{equation}
For $a>0$, the system acquires a pole below the threshold in the first  $\sigma_q$ Riemann sheet. It corresponds to an imaginary relative momentum of the two-particle subsystem equal to $ i\kappa = \frac{i}{a} $. The total invariant mass squared of the bound state is then,
    \beq
    \label{eq:bound-state-mass}
    m_b^2 \equiv \sigma_b = 4 \left(m^2 - \kappa^2 \right) \, .
    \eeq
It corresponds to a relative bound-state--spectator momentum in the total CM frame,
    \beq
    \label{eq:bound-state-momentum}
    q_{b} = \frac{\lambda^{1/2}(s, \sigma_b, m^2)}{2 \sqrt{s}} 
    =
    \frac{
    \sqrt{s-s_{\varphi b}}
    \sqrt{s-(m_b-m)^2}
    }{2 \sqrt{s} } \, .
    \eeq
Finally, one finds that residue of $\Mc_2$ at the pole is $-g^2$, where $g$ is the $b\to2\varphi$ coupling given by,
    \begin{align}
    g = 8\sqrt{ 2 \pi \sqrt{\sigma_b}\,\kappa  } \, .
    \end{align}

As discussed in Ref.~\cite{Jackura:2020bsk} in detail, continuing external momenta of $\Dc^{(u,u)}$ to the value $q_b$ leads to factorization of the poles associated with the external two-body bound states. The residuum at the double-pole becomes proportional to the $S$-wave spectator--bound-state scattering amplitude $\Mc_{\varphi b}(s)$. The three-body amputated ladder amplitude is related to the $\varphi b$ through,
    \begin{align}
    \label{eq:M-phi-b-limit}
    \Mc_{\varphi b}(s)
    &=
    g^2 \lim_{p, \, k \, \to q_b}
    \, d_{S}( p, k) \, .
    \end{align}
We note that by continuing to other values of external momenta, one can also obtain three-to-three, $3\varphi \to 3\varphi$, or two-to-three, $\varphi b \to 3\varphi$, amplitudes. Reference~\cite{Jackura:2020bsk} explained how to evaluate these amplitudes for energies along the real axis above the bound-state--spectator threshold.

Between the $\varphi b$ and $3\varphi$ thresholds, similarly to $\Mc_2$ in Eq.~\eqref{eq:M2_general}, the $\Mc_{\varphi b}$ amplitude can be parametrized in the $K$-matrix form,
    \begin{align}
    \label{eq:phi-b-Kmat}
    {\Mc}_{\varphi b} (s) &= \frac{1}{ \Kc_{\varphi b}^{-1}(s) - i  \rho_{\varphi b}(s) } \, ,
    \end{align}
where $\rho_{\varphi b}$ is the phase space between the bound state and the spectator,
    \beq
    \rho_{\varphi b}(s) = \frac{q_b}{8 \pi \sqrt{s}} \, .
    \eeq
The bound-state--spectator $K$ matrix, $\Kc_{\varphi b}(s)$ is real between the $\varphi b$ and $3\varphi$ thresholds but can potentially acquire an imaginary part below $s_{\varphi b}$. Using Eq.~\eqref{eq:phi-b-Kmat}, one defines the $\varphi b$ phase shift,
    \beq 
    \label{eq:cot_delta}
    q_b \cot\delta_{\varphi b} = 8 \pi \sqrt{s} \, \Kc_{\varphi b}^{-1}(s) = 8 \pi \sqrt{s} \, \mathcal{M}_{\varphi b}^{-1}(s) + i q_b \, .  
    \eeq 
In Sec.~\ref{sec:results}, we provide numerical solutions for the $\varphi b \to \varphi b$ amplitude below the ${\varphi b}$ threshold and in the complex $s$ plane. We use Eq.~\eqref{eq:cot_delta} to define the analytic continuation of the two-body $K$ matrix below the $\varphi b$ threshold.

\subsection{Three-body bound and virtual states}    

One of the goals of this work is the computation of the positions of the three-body bound states, which manifest themselves as poles on the real axis below the $s_{\varphi b}$ threshold. Close to the pole, denoted by $s_b$, the amplitude factorizes,
    \beq
    d_S(p, k)= - \frac{\zeta(p) \, \zeta^*(k)}{s-s_b} + \dots \, ,
    \label{eq:dS_pole}
    \eeq
 where $\zeta$ is called the \emph{vertex function} and constitutes the momentum-dependent residue of the pole. From Eq.~\eqref{eq:DS}, it is evident that if $d_S$ has a pole in $s$, $\Dc_S^{(u,u)}$ must as well. Writing the residue of the latter as $-\Gamma(p) \, \Gamma^*(k)$, one finds these satisfy, 
    \begin{align}
    \Gamma(p)=
    \zeta(p) \, \Mc_2(p) \, .
    \label{eq:Gamma_vertex}
    \end{align}
This residue can be understood as the coupling between the three-body bound-state and the $3 \varphi$ scattering states. We note that, by definition, $\Gamma(p)$ describes a scattering process that has not been symmetrized with respect to the choice of external pairs, but we keep the $(u)$ label implicit. Finally, the vertex function of the $\Mc_{\varphi b}$ amplitude is 
    \begin{align}
    \Gamma_{\varphi b} = g \,
    \zeta(q_b) \, .
    \label{eq:phi_b_vertex}
    \end{align}

Inserting Eq.~\eqref{eq:dS_pole} into Eq.~\eqref{eq:d_Sproj_kern} leads to the homogeneous ladder equation for the residue,
    \beq
    \label{eq:homo-eq}
    \zeta(p) = -
    \int\limits_{0}^{q_{\rm max}} dq \, K(p, q) \, \zeta(q) \, ,
    \eeq
This equation is satisfied at the three-body bound-state invariant mass squared $s=s_b$. As a result, one might use it to solve for the bound-state location. Assuming that $\zeta$ is non-zero, this only has a solution if the following determinant condition is satisfied,
    \beq
    \label{eq:determinant-condition}
    \det \Big[ \one + K \Big] = 0 \, ,
    \eeq
where the determinant is calculated in the $(p,q)$ momentum space. In other words, Eq.~\eqref{eq:determinant-condition} serves as a quantization condition for the three-body bound state.

To determine the residue itself, one solves the generalized eigenvalue problem,
    \beq
    \label{eq:homo-eigen}
    \eta(s_b) \zeta(p) = -
    \int\limits_{0}^{q_{\rm max}} dq \, K(p, q) \, \zeta(q) \, ,
    \eeq
where one treats $s$ as the external parameter evaluated at $s_b$. For $\eta(s_b) = 1$, Eq.~\eqref{eq:homo-eigen} coincides with Eq.~\eqref{eq:homo-eq}, and the corresponding eigenvector, $\zeta$, is the sought vertex function~\cite{Glockle:1978zz}. Numerically, one solves the homogeneous equation similarly to the inhomogeneous one, i.e., by discretizing the momenta $(p,q)$, solving the eigenvalue problem, and finding the value of $\eta$ closest to $1$.

We note it is also possible to find the position of the three-body bound-state pole and its residue by solving the inhomogeneous ladder equation for $d_S(p, k)$, Eq.~\eqref{eq:d_Sproj}, for a range of energies and searching for the pole explicitly in the complex-valued amplitude.
    
To study virtual states or resonance poles, one needs to continue the amplitude in Eq.~\eqref{eq:M-phi-b-limit} to the unphysical Riemann sheet, which is continuously connected to the first one through the unitarity cut. For the system under study, the relevant branch cut is due to the $\varphi b$ threshold. Using Eq.~\eqref{eq:phi-b-Kmat}, we can analytically continue the amplitude to the second sheet,
    \beq
    \label{eq:phi-b-second-sheet}
    \Mc_{\varphi b}^{\text{II}}(s) = \frac{\Mc_{\varphi b}(s)}{1 + 2 i \rho_{\varphi b}(s) \Mc_{\varphi b}(s) } \, .
    \eeq
From this, it is easy to see that resonance or virtual state poles are found by using the condition,
    \beq
    \label{eq:5-second-sheet-resonance}
    1 + 2 i \rho_{\varphi b}(s) \Mc_{\varphi b}(s) = 0 \, .
    \eeq
If one is interested exclusively in the virtual states, the knowledge of $\Mc_{\varphi b}$ on the first sheet below the $s_{\varphi b}$ threshold is sufficient for their determination.

\section{Analytic properties of the integration kernel}
\label{sec:analytic-structure}

Having recollected all the components of the integral equation and reviewing strategies for determining the three-body bound-state poles, we proceed to discuss the analytic properties (singularity structure) of the ladder equation and its integration kernel.

We restrict our attention to the integral equation as expressed in Eq.~\eqref{eq:d_Sproj_kern}, i.e., in terms of the spectator momenta. Alternatively, one can write it using the external pairs' invariant masses. The resulting amplitudes in momentum-space, $d_S(p,k)$, and invariant space, $d_S(\sigma_p,\sigma_k)$, are equivalent, but two forms of the integral equation can offer different types of insight into the analytic structure of the integration kernel. We discuss this in App.~\ref{app:B}.

In the following paragraphs, we outline the singularities of the components of Eq.~\eqref{eq:d_Sproj_kern}. It contains three objects: the OPE term $[G_S]$, the kernel $[K]$, and the subsequent solution $[d_S]$. The kernel depends on $G_S$, the two-body amplitude $\Mc_2$, and the Jacobian. Below we discuss each one of these in reverse order. The properties of $d_S$ emerge from those of $G_S$ and the integration of the kernel. We discuss them separately in Sec.~\ref{sec:analytic-continuation}. 

Each object depends on the invariant mass squared $s$ and two of the spectator momenta $(q,k)$. In general, their analytic properties in one variable, e.g., placement of pole and branch-point singularities in the complex $q$ plane, depends on the values of the other two, $(k,s)$. As these kinematic parameters change, e.q., $s \to s'$, singularities can approach and cross the real $q$ axis in the integration interval $[0, q_{\rm max}]$. Such a crossing signals the emergence of singularities of $\Mc_{\varphi b}(s)$ in the complex $s$ plane. To evaluate the amplitude at the new value of total invariant mass, $s'$, one must understand the nature of the resulting $s$-plane singularity and whether it can be avoided. If possible, it is accomplished by analytic continuation, which is equivalent to the $q$-plane integration path deformation. In App.~\ref{app:A}, we provide a basic, pedagogical introduction to these concepts and a collection of helpful references.

Because all functions entering the kernel are symmetric under a parity transformation $q \rightarrow -q $, for each complex singularity at point $q$, there is a corresponding ``copy" at $-q$. It is easy to see in the Jacobian, which contains the single-particle energy, $\omega_q=\sqrt{q^2+m^2}$. It has two imaginary branch points starting at $q = \pm i m$. We orient the associated branch cuts along the imaginary axis, and they go to $\pm i \infty$, respectively.

\subsection{Singularities of the two-body amplitude}

\begin{figure}[t]
    \centering
    \includegraphics[width=0.95\textwidth]{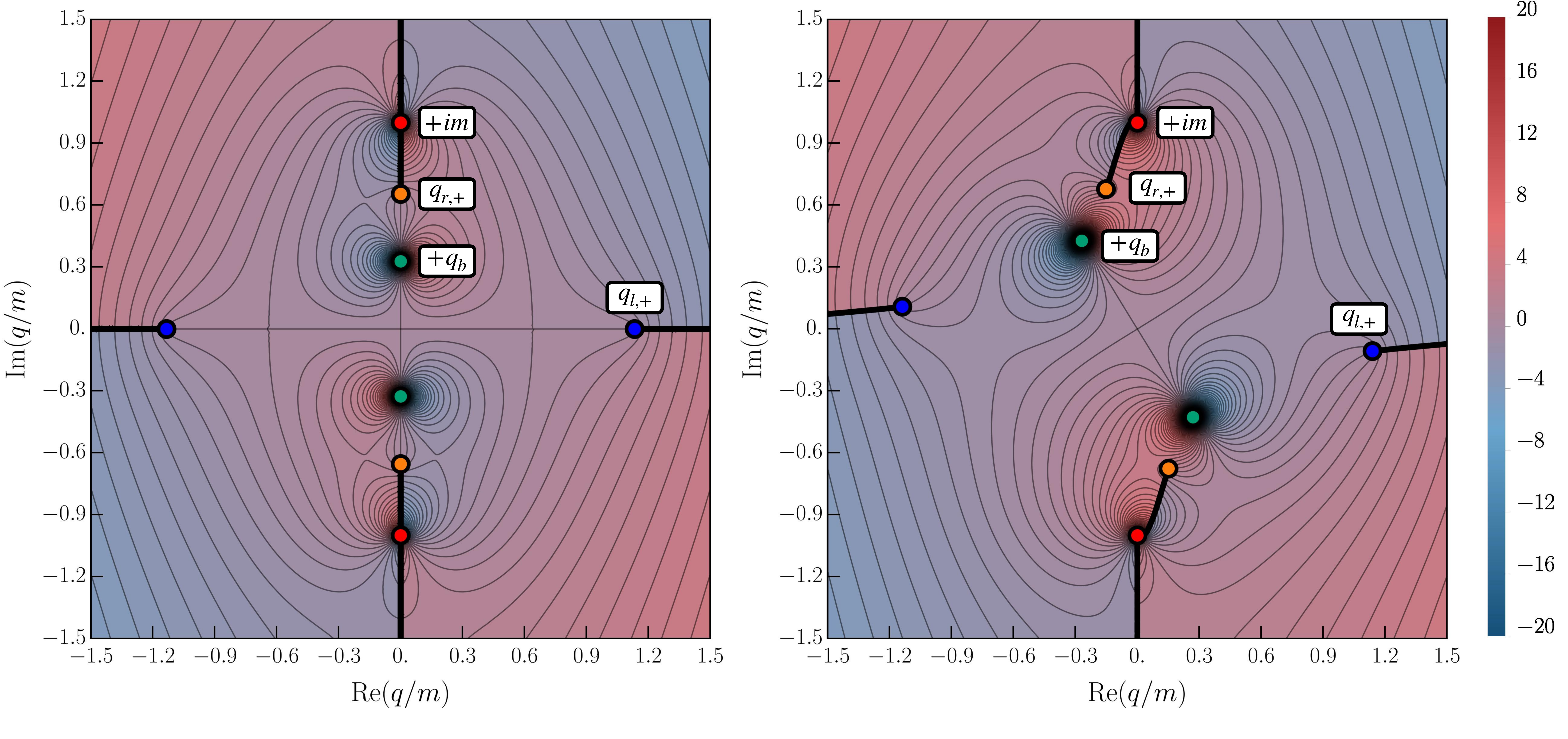}
    \caption{Plot of the imaginary part of $q^2 \Mc_2(q)/ (2\pi)^2 \omega_q$ in units of $m$, for $ma=2$ and \textbf{(left)} $s/m^2=7$, \textbf{(right)} $s/m^2=7-i$, with the singularity structure highlighted. Branch cuts are represented by black lines, while branch points by colored points. We explicitly label the upper half-plane copies. Branch points of $\omega_q$ are shown in red, branch points $q_{r,\pm}$ in orange, and branch points $q_{l,\pm}$ in blue. Two copies of the pole singularity are depicted with green points.}
    \label{fig:M2-cuts}
\end{figure}

The two-body amplitude, $\Mc_2(q)$, has a pole and branch points. We first discuss the latter. This amplitude has three pairs of branch points, associated with the momentum dependence of $\rho(q)$. Since it depends on $q$ through $\sigma_q$, it must have the same branch cuts as the energy $\omega_q$, which enters Eq.~\eqref{eq:sigma_p}. Moreover, from Eq.~\eqref{eq:2p1.phase_space}, we see it is singular when $\re \sigma_q < 0$ and $\re \sigma_q > 4 m^2$ and $\im \sigma_q = 0$. These conditions define the unphysical left-hand cut and the right-hand cut required by unitarity, respectively. They translate into $s$-dependent or ``movable" branch points in the $q$ plane,
    \beq
    \label{eq:M2-cut-one}
    \text{[right-hand cut]}~~~
    \sigma_q = 4m^2 ~~~\Leftrightarrow ~~~ q_{r,\pm} = \frac{\lambda^{1/2}(s, 4m^2, m^2)}{2 \sqrt{s}} = \pm \left( \frac{\sqrt{s-m^2} \sqrt{s-(3m)^2}}{2 \sqrt{s} } \right) \, ,
    \eeq
and
    \beq
    \label{eq:M2-cut-two}
    \text{[left-hand cut]}~~~
    \sigma_q = 0 ~~~\Leftrightarrow ~~~ q_{l,\pm} = \frac{\lambda^{1/2}(s, 0, m^2)}{2 \sqrt{s}} = \pm \left( \frac{s-m^2}{2\sqrt{s}} \right) \, .
    \eeq
For the real total invariant mass, $m^2 < s < (3m)^2$, the cuts starting at $q_{r,\pm}$ are aligned with the imaginary axis and go to $\pm i \infty$, respectively. The cuts starting at $q_{l,\pm}$ are aligned with the real axis and go to $\pm \infty$, respectively. For complex $s$, they are curved and described by complicated equations relating $\im q$ and $\re q$. The presence of the $q_{r,\pm}$ branch point is required by the unitarity of $\Mc_2$. The occurrence of $q_{l,\pm}$ is a feature of the particular model we study. One can remove the associated unphysical cut by dispersing the $\Mc_2$ amplitude and ensuring its analyticity, as typically done in the so-called FVU approach~\cite{Mai:2017vot, Sadasivan:2020syi}. 

We note that $q_{l,\pm}$ and $q_{r,\pm}$ are expressed in terms of $\lambda^{1/2}$, which is a function of $s$ with cuts in that variable. Both points have an inverse square root singularity at $s=0$, which we ignore, as we do not consider $s<m^2$ in this work. In addition, $q_{r,\pm}$ has two branch points at real $s = m^2$ and at the three-body threshold $s= s_{3\varphi} = (3m)^2$. Commonly one makes corresponding cuts of $\lambda^{1/2}$ to run between these two points or have them go to $-\infty$ and $+\infty$, respectively. Choosing the latter option makes the $q_{r,\pm}$ points evolve smoothly when changing between positive and negative values of $\im s$ for $\re s < s_{3\varphi}$. Choosing the former option leads to a switch $\im q_{r,+} \leftrightarrow \im q_{r,-}$ when changing the sign of $\im s$, while the real parts of $q_{r,\pm}$ are symmetric under complex conjugation of $s$. It is a general property of the spectator's momentum defined at a fixed value of the corresponding pair's invariant mass, Eq.~\eqref{eq:momentum_invs}. We use this definition. Regardless of this choice, the two resulting branch points of $\Mc_2$ remain ``parity copies" of each other in the $q$ plane. From the point of view of analytic continuation of the $d_S$ amplitude, we try to determine the presence of the branch points that could potentially cross the integration path. Thus, thanks to the parity symmetry property of $\Mc_2$, it is not ultimately important whether we label these branch points $q_{l,+}$ or $q_{l,-}$. We show all singularities of the Jacobian and the two-body amplitude in Fig.~\ref{fig:M2-cuts} for two example values of $s$ and scattering length $ma=2$.

As mentioned in the previous section, $\Mc_2$ develops a pole at $\pm q_b$. Similarly to the branch points, it depends on the total invariant mass via the triangle function. For complex values of $s$, momentum $q_b$ has a cut between $(\sqrt{\sigma_b} - m)^2$ and $s_{\varphi b}$. When crossed, $\im q_b \to - \im q_b$, while $\re q_b$ remains unchanged. Thus, under complex conjugation of $s$, two parity copies of $q_b$ transform into each other.

\subsection{Cut-off function and potential essential singularity}

The most interesting contributions to the singularity structure of the integration kernel come from the $S$-wave OPE amplitude, Eq.~\eqref{eq:Gs_proj}. Before we discuss its logarithmic part, let us first analyze the analytic properties of the cut-off function $H(p,q)$, included in $G_S$. Here, we explore two types of regularization. One is the smooth cut-off defined in Ref.~\cite{Hansen:2015zga}, $H(p, q) = J(\sigma_p/ 4 m^2 )  J(\sigma_q/ 4 m^2 )$, where
    \begin{align}
    \label{eq:cut-off}
    J(x) & =
    \begin{cases}
    0 \,, & x \le 0 \, , \\
    \exp \left( - \frac{1}{x} \exp \left [-\frac{1}{1-x} \right] \right ) \,, 
    & 0<x \le 1 \, , \\
    1 \,, & 1<x \, .
    \end{cases}
    \end{align}
This function equals unity in the physical region $\sigma_q > 4m^2$ and smoothly transitions to zero at $\sigma_q = 0$. The other choice is the hard cut-off $H(p,q) = \theta(p)\theta(q_{\rm max}-p) \, \theta(q) \theta(q_{\rm max}-q)$.

Both functions are originally defined for real values of momenta. Since, in the process of analytic continuation, we will perform integration over complex variables, they have to be generalized to the complex plane—possibly without introducing additional singularities. Restriction of the integration range is implemented by requiring that the complex-momentum integration contour $\Cc$ has fixed endpoints, $q=0$ and $q=q_{\rm max}$. For the hard cut-off, we take $H(p,q) = 1$ in the whole complex plane, which is the unique analytic extension of the constant function. For the smooth cut-off, we extend,
    \beq
    J(x) \to J(z) = \exp \left( - \frac{1}{z} \exp \left [-\frac{1}{1-z} \right] \right ) \, ,
    \eeq
for all complex $z$; removing conditions that make $J$ constant for certain values of its argument. This function is analytic everywhere except for $z=0$ and $z=1$, for which it develops essential singularities. In the language of the complex momentum variables, those essential singularities coincide with branch points of $\Mc_2$, $q_{l,\pm}$, and $q_{r,\pm}$, respectively. We note that, since $q_{r,\pm} = 0$ for $s = s_{3\varphi}$, one can not use the smooth cut-off when performing analytic continuation above the three-particle threshold, as the collision of the essential singularities with the integration endpoint could induce an unphysical right-hand cut structure of the $d_S$ amplitude. It points to a serious tension between finite- and infinite-volume counterparts of the formalism: one requires a smooth cut-off of the form \eqref{eq:cut-off} in the rigorous derivation of the three-body quantization condition; however, it can not be used when identifying properties of resonances. In the bound-state--spectator system, we avoid this problem by considering $\re s < s_{3\varphi}$. In Sec.~\ref{sec:results}, we present results for both the smooth and hard cut-off cases.

\subsection{Logarithmic singularities of the OPE amplitude}

Apart from the potential singularities associated with the regularizing functions, the $S$-wave OPE amplitude has logarithmic discontinuities that can manifest themselves both in the kernel and the inhomogeneous term of the ladder equation. The analytic representation of the cuts is obtained most simply from the integral representation of the $G_S$, i.e., the first line of Eq.~\eqref{eq:Gs_proj}. They are produced when the pole of the propagator crosses the integration path in the $x$ variable,
    \beq
    \label{eq:OPE-pole-condition}
    z(p,k) + i \epsilon + 2 p k x  = 0 \, .
    \eeq
Solving the above condition yields an explicit parametrization of the cuts,
    \beq
    \label{eq:OPE-cut-parametrization}
    p_{\text{cut},\pm}(s,k,x) = \frac{ k x \, (\beta_1 + i \epsilon) \pm \sqrt{\beta_0} \sqrt{(\beta_1 + i \epsilon)^2 - 4 m^2 \beta_x} }{ 2 \beta_x } \, ,
    \eeq
where the ``$\pm$" sign refers to two parity copies of the cut and the parameter $x \in [-1,1]$. Here, we defined a function of $s$ and $k$,
    \beq
    \label{eq:beta-function}
    \beta_x \equiv \beta_x(s,k) = (\sqrt{s} - \omega_k)^2 - x^2 k^2 \, .
    \eeq
We note that $\beta_{\pm1}(s,k) = \sigma_k$.
Equation \eqref{eq:OPE-cut-parametrization} is the master formula describing the analytic structure of the OPE amplitude for arbitrary kinematics, and has been previously studied under different guises both in the non-relativistic \cite{Brayshaw:1968yia, Glockle:1978zz, Orlov:1984} and relativistic~\cite{Jackura:2018xnx, Sadasivan:2020syi, Sadasivan:2021emk} three-body approaches. Considered as a function of $s$ and $k$, it has an analytic structure of its own with various square-root branch points\footnote{For example, due to the analytic properties of the triangle function, for various differing values of $s$ and $k$, points $p_+$ and $p_-$ in Eq.~\eqref{eq:mom-plus-minus} can transform into each other or their parity copies. An unambiguous definition of $p_\pm$ requires specification of the cut structure of the $p_{\rm cut,+}$ function, e.g., resulting from the condition $\sigma_k^2 - 4 m^2 \beta_x < 0$. As discussed below Eq.~\eqref{eq:M2-cut-two}, it does not affect the problem of analytic continuation.}. It is beyond the scope of this work to explain them all; instead, we focus on those features of the OPE singularities that affect the determination of the $\Mc_{\varphi b}(s)$ amplitude and the three-body bound-state and virtual-state poles.

\begin{figure}[t]
    \centering
    \includegraphics[width=0.95\textwidth]{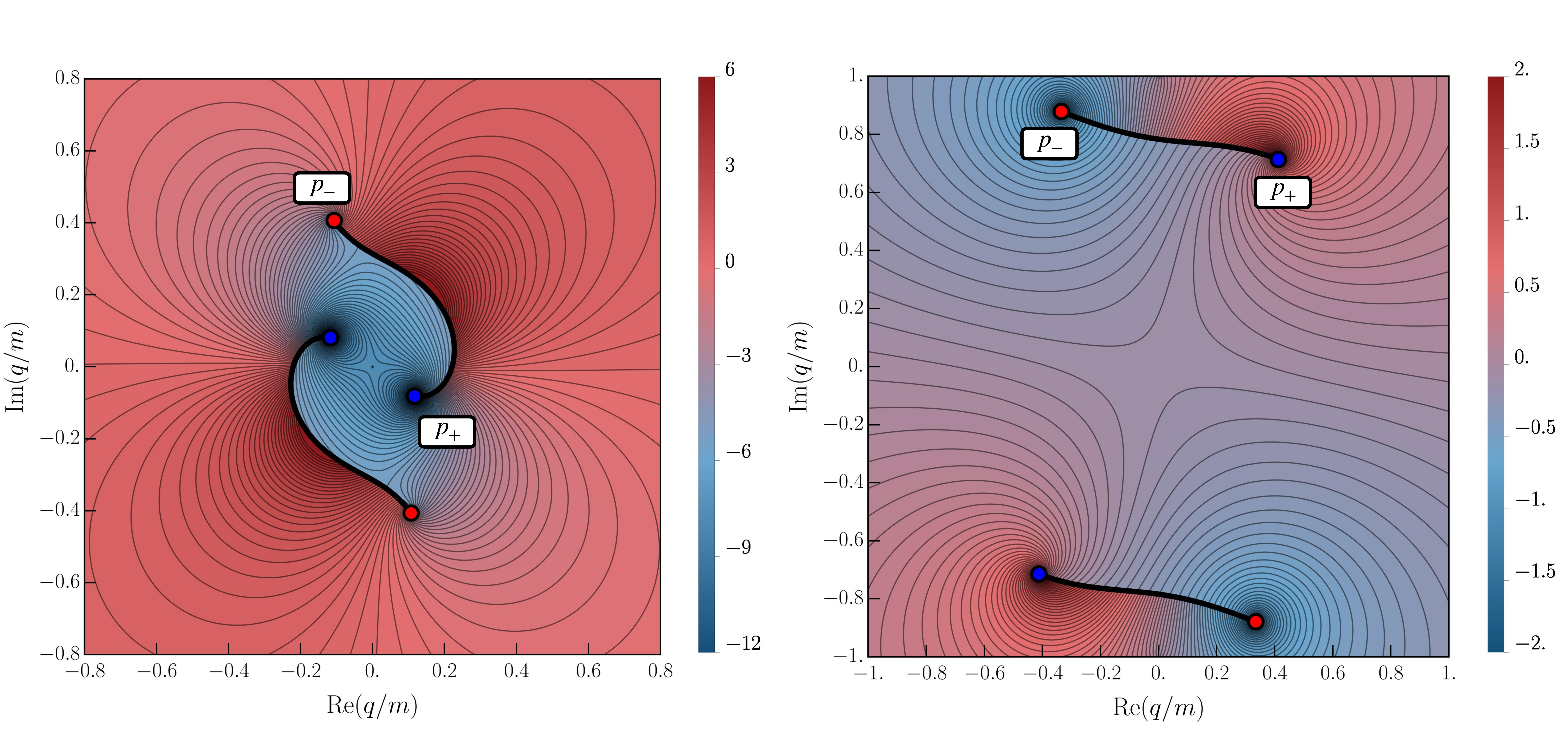}
    \caption{Contour plot of the imaginary part of $G_S(q,k)$ in the complex $q$ plane, in units of $1/m^2$. Here, we chose the hard cut-off, total invariant mass $s/m^2 = 8 - i$, and $p$ corresponding to a fixed $\sigma_k$: \textbf{(left)} $\sigma_k=\sigma_b$, where $ma=6$, and \textbf{(right)} $\sigma_k=2m^2$. The singularity structure is highlighted: branch cuts are represented by black lines and branch points by colored points. We explicitly label the positive-parity copies. Branch points $\pm p_+$ are shown in blue, and branch points $\pm p_-$ are in red.}
    \label{fig:OPE-cuts}
\end{figure}

In the following expressions, we set $\epsilon=0$ unless explicitly stated otherwise. A cut runs between the two associated branch points, whose positions are obtained by setting $x=\pm 1$, e.g.,
    \beq
    \label{eq:mom-plus-minus}
    p_\pm = p_{\text{cut},+}(s,k,\pm 1) = 
    \frac{\lambda^{1/2}(s,\sigma_{\pm},m^2)}{2 \sqrt{s}} \, ,
    \eeq
where $\sigma_\pm$ is a function of $s$ and $k$, describes the position of the branch points in the $\sigma_p$ plane, and is derived in App.~\ref{app:B}. The other two branch points are $-p_{\pm}$. The above expressions hold universally for real and complex values of $s$ and $k$. 

Finally, we note two key properties of $G_S$ that are useful in an upcoming discussion of the OPE cuts in various variables, namely,
    \beq
    \label{eq:G-property}
    G_S(p,k; s) = G_S^*(p^*,k^*;s^*) \, , ~~~ G_S(p,k; s) = G_S(k,p;s)  \, ,
    \eeq
where we wrote the $s$ dependence explicitly. For example, we see that cuts in $k$ for fixed $(p,s)$ are given by the equations analogous to the ones derived in this subsection.

\begin{figure}[b]    \includegraphics[width=0.99\textwidth]{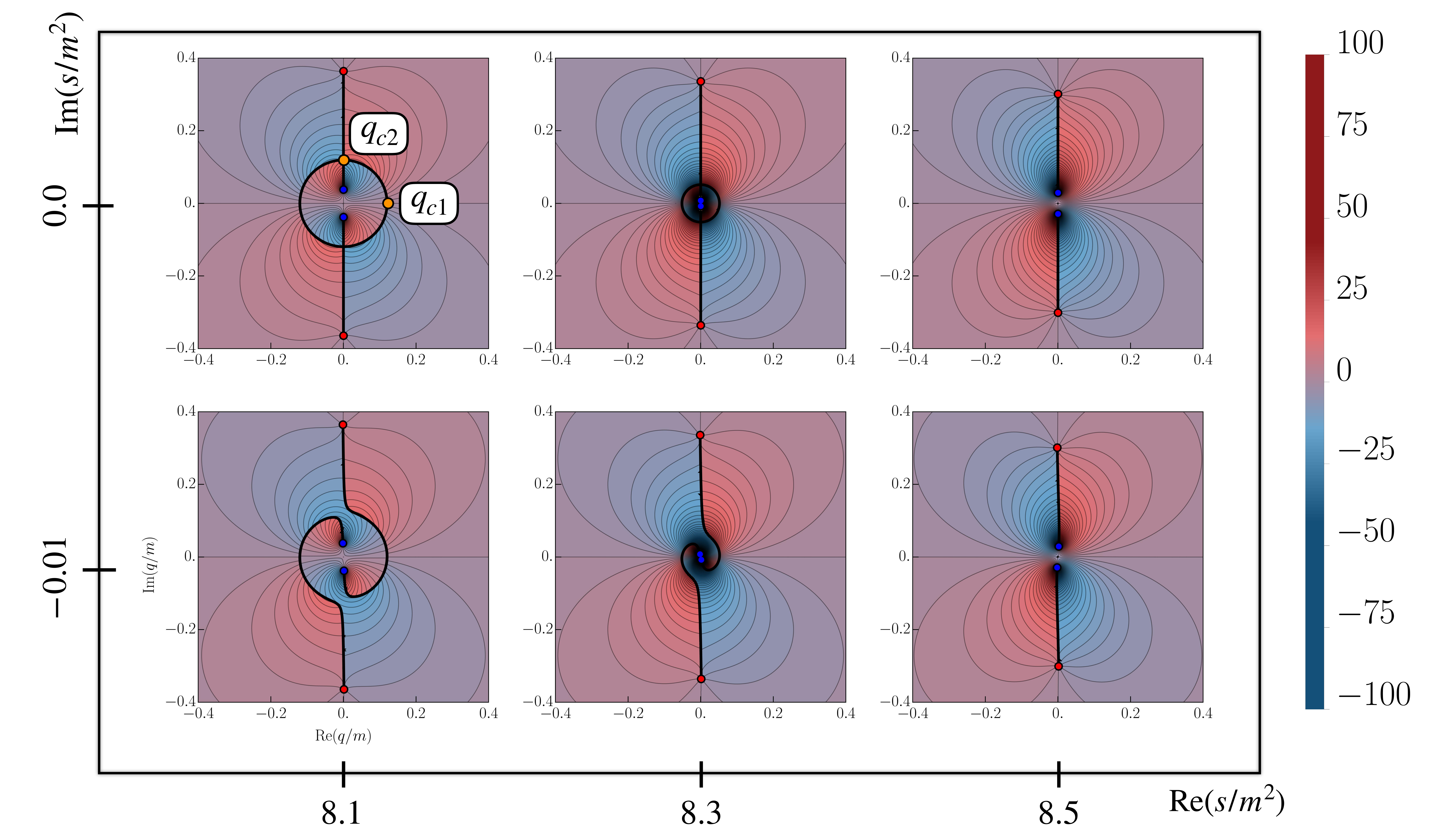} 
\caption{Cut structure of the imaginary part of the OPE, $\im G_S(q, k)$, in the complex momentum $q$ plane, in units of $1/m^2$. We set $\sigma_k =\sigma_b$ for $ma=6$ and use the hard cut-off. Total invariant mass changes from $\re s/m^2=8.5$ (most right panel) to $\re s/m^2=8.1$ (most left panel). We present two example values of $\im s/m^2$. The top panels exhibit the emergence of the circular cut, which encloses the origin of the complex plane. On the left panel, we highlight the positions of the $q_{c1}$ and $q_{c2}$ points. For non-zero $\im s$ (bottom panels), the circular cut ``opens". This creates a slit through which the $q=0$ point can be accessed by the integration contour.
}
\label{fig:circular-cut-1}
\end{figure}

\subsection{Illustrative example of the OPE cuts}

An example illustration of the OPE cuts is presented in Fig.~\ref{fig:OPE-cuts}. We focus on two cases of practical interest, i.e., real-valued external pair invariant masses, $\sigma_k = \sigma_b$ and $\sigma_k = 2m^2$. Of the two choices, the former corresponds to our $\Mc_{\varphi b}(s)$ calculation, while the latter is considered, e.g., when solving for the bound-state pole positions and vertex functions in the homogeneous equation, Eq.~\eqref{eq:homo-eq}. As we can see, for some values of kinematic parameters, the cuts cross the real $q$ axis, requiring deformation of the real integration interval $[0, q_{\rm max}]$, as discussed in the subsequent section. Additionally, as shown in Fig.~\ref{fig:circular-cut-1}, an unusual structure, known as the \emph{circular cut}\footnote{Name ``circular" can be misleading as the branch cut does not necessarily form a perfect circle for all values of $s$ and $k$. We handle the cut under the approximation that it is a circle and not an ellipse because for considered kinematics the difference is negligible.}, emerges when the (real-valued) $s$ decreases from the physical value below the point $s_{\circ}$ given in Eq.~\eqref{eq:s-cut-condition-2}. Both copies of the OPE branch cut assume the shape of two short lines attached to a semi-circle. For decreasing $s<s_{\varphi b}$, they become connected along the imaginary $q$ axis. Points $\pm p_+$ approach and touch the origin of the complex plane and then ``bounce" back, moving away from the $q=0$ point along the imaginary axis as the circular cut grows. 

Assuming values of $s$ and $\sigma_k$ are real we determine positions where the circle coincides with complex plane axes. The real one is passed at,
    \beq
    \label{eq:mom_c1}
    q_{c1} = \frac{\lambda^{1/2}(s,\sigma_{c1},m^2)}{2 \sqrt{s}} \, .
    \eeq
Since the cut consists of two parity copies that are ``glued" together when $\im s=0$, the imaginary axis is not crossed by the cut, but approached as $\im s \to 0$, the semi-circle starting at,
    \beq
    \label{eq:mom_c2}
    q_{c2} = \frac{\lambda^{1/2}(s,\sigma_{c2},m^2)}{2 \sqrt{s}} \, .
    \eeq
The two-body invariant masses $\sigma_{c1}$ and $\sigma_{c2}$ are derived in App.~\ref{app:B} and both depend on variables $s$ and $k$. As usual, points $q_{c1}$ and $q_{c2}$ have their corresponding parity copies. Knowledge of the functional form of these points is useful for determining the appropriate integration contour that leaves the neighborhood of the $q=0$ point without crossing any cuts. This is discussed in App.~\ref{app:C}, where we also derive their generalization for complex values of $s$ and $\sigma_k$.

Opening of the circular cut is shown in the bottom panel of Fig.~\ref{fig:circular-cut-1}, where the $\im s=-10^{-2}$, and $k=q_b$ case is presented. An equivalent branch cut structure is obtained for purely real $s$ but non-zero, positive $\epsilon$. For $\im s >0$, the analytic structure of $G_S$ is obtained by a reflection of the $\im s < 0$ cuts with respect to the real $q$ axis. Indeed, the complex conjugation of $s$ leads to reflection $p \to p^*$ in the argument of $G_S$ as can be seen from Eq.~\eqref{eq:G-property}, and the following transformation,
    \beq
    \label{eq:circular-cut-discontinuity}
    G_S(p,k; s^*) \to G_S(p,\re k-i \im k; s^*) = G_S^*(p^*,(\re k-i \im k)^*;s) = G_S^*(p^*,k;s) \, ,
    \eeq
where, in the first transformation, we used the property of the spectator's momentum, $\im k \to -\im k$, under complex conjugation of the total invariant mass, $s \to s^*$, which holds for $k$ defined for a fixed $\sigma_k$ and $s\leq(\sqrt{\sigma_k}+m)^2$.

As seen on the right panel of Fig.~\ref{fig:OPE-cuts}, the circular cut is not present for all values of $\sigma_k$. We present an example position of the OPE amplitude cuts for $\sigma_k = 2m^2$.

Finally, as a side remark, let us observe that the position of the OPE branch cuts is arbitrary and can be chosen in various ways leading to a different definition of $G_S$. It can be introduced by considering contour deformation in the $x$ variable in Eq.~\eqref{eq:Gs_proj}. An integration path starting at $x=-1$ and ending at $x=1$ but going into the complex $x$ plane gives the same branch points but a different cut structure of the OPE amplitude. It might allow one to ``open" the circular cut for those values of $(s,k)$ for which it is ``closed" when the regular $[-1,1]$ integration interval is chosen to define $G_S$. Although it is useful, we do not explore this procedure further.

\section{Analytic continuation of the amplitude}
\label{sec:analytic-continuation}

The original ladder equation, Eq.~\eqref{eq:d_Sproj_kern}, is defined in the physical kinematical region. In the model of the bound-state--spectator scattering, it is given by the condition $s \geq s_{\varphi b }$. In this case, all the variables describing the amplitude: external momenta $(p,k)$, total invariant mass squared $s$, and the integration variable $q$ are real. The solution of the integral equation for these energies is explored in Ref.~\cite{Jackura:2020bsk}. After the discussion of the previous sections, we are ready to extend the results of this work by studying energies below the $\varphi b$ threshold and complex values of the total invariant mass $s$. As predicted in Ref.~\cite{Romero-Lopez:2019qrt}, one expects the presence of the three-body bound states there, and verification of this result is one of our aims.

The $\Mc_{\varphi b}$ amplitude is obtained from $d_S(p,k)$ by continuing the external momenta to the relative $\varphi b$ momentum $q_b$. Let us observe that for real $s < s_{\varphi b}$, this point becomes purely imaginary. Thus, the analytic continuation of $d_S(p,k)$ in $s$ naturally forces one to continue $d_S(p,k)$ in the momentum arguments as well. Amplitude $d_S(p,k)$ becomes a multi-variable complex function that develops singularities in each of the three arguments ($p,k;s$); with their presence in one variable usually depending on the values of the other two. Therefore, one should not study the formation of the pole in $s$ independently from the analytic behavior of $d_S$ in the $(p,k)$ variables. For this reason, we devote an entire section to the analysis of the analytic properties of $d_S$.

To simplify our discussion, we narrow our interest mostly to the $\Mc_{\varphi b}$ amplitude. However, the methods described below apply to more general cases. They were originally described by Brayshaw in Refs.~\cite{PhysRev.167.1505, Brayshaw:1968yia}. We simplify and modify some aspects of his discussion, as explained in Sec.~\ref{sec:domain-of-non}. In particular, we have to extend Brayshaw's method to incorporate the two-body bound-state case, which leads to the appearance of the circular cut. It is done following the work of Gl\"ockle in Ref.~\cite{Glockle:1978zz}. When appropriate, we present additional extensions of Brayshaw's and Gl\"ockle's methods that are necessary for the system under study. 

\subsection{Overview of singularities of the bound-state--spectator amplitude}
\label{sec:analytic-continuation-A}

Before turning to the analysis of the integral equation, it is useful to discuss the expected analytic structure of the solution and its origin. The amplitude $\Mc_{\varphi b}(s)$ inherits its singularities from the two terms on the right-hand side of Eq.~\eqref{eq:d_Sproj_kern}. First, it has explicit singularities of $G_S(q_b, q_b)$. Second, it has singularities of the integral term, considered here as a function of $s$. These can be either explicit or emerge from the collision of the $s$-dependent singularities of the integrand in the complex $q$ plane with the integration contour, as summarized in App.~\ref{app:A}.

From Eq.~\eqref{eq:Gs_proj}, evaluated at identical external momenta, $p,k=q_b$, we find that the function $G_S(q_b,q_b)$ has a cut in the complex $s$ variable that connects two branch points,
    \beq
    \label{eq:sl1}
    s_{L1} = \frac{(m^2 - \sigma_b)^2}{m^2} \, ,
    \eeq
and
    \beq
    \label{eq:sl2}
    s_{L2} = m^2 + 2 \sigma_b \, .
    \eeq
We refer to this as the ``short" OPE cut. As explained in Ref.~\cite{Jackura:2018xnx}, for certain values of external momenta it can occur in the physical region, i.e., when it is allowed for a pair to decay, and then corresponds to the real particle exchange. However, in our model, these two points are found below the $s_{\varphi b}$ threshold. We note that for $\sigma_b =4m^2$, they both coincide with the $3\varphi$ threshold. The ``short" cut is the only singularity contributed by the inhomogeneous term to the $\Mc_{\varphi b}$ amplitude.

Considering the second term of the integral equation, the right-hand cut structure of $\Mc_{\varphi b}(s)$ is fixed by the presence of the pole in $\Mc_2$ at $q=q_b$. Namely, as implied by Eq.~\eqref{eq:bound-state-momentum}, for real $s<s_{\varphi b}$, both copies of the pole are located on the imaginary $q$ axis. In the limit $s \to s_{\varphi b}$, they approach the origin of the complex momentum plane and collide with the lower limit of the integration. It leads to the emergence of the unitarity cut of $\Mc_{\varphi b}$ at $s_{\varphi b}$. For increasing $s$, both copies of the pole travel along the real axis in opposite directions. We note that branch points $q_{r,\pm}$ follow this behavior, colliding with $q=0$ point at $s=s_{3\varphi}$. It results in the logarithmic branch point of the amplitude, corresponding to the opening of the three-body channel. For increasing $s>s_{3\varphi}$, points $q_{r,\pm}$ continue their motion along the real axis in opposite directions, with the cut running between them.

\begin{figure}[t]
    \centering
    \includegraphics[width=0.9\textwidth]{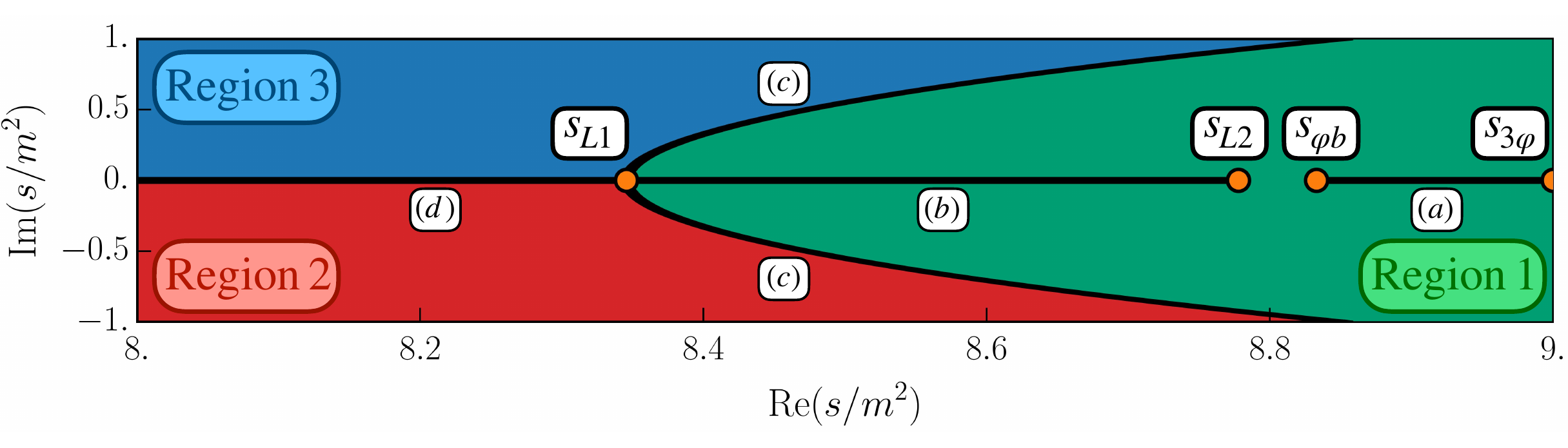}
    \caption{First-sheet singularity structure of $\Mc_{\varphi b}(s)$ for $ma=6$. Branch points discussed in the text are presented as orange points, while associated cuts are shown as black lines. In particular: (a) unitarity cut starts at $s_{\varphi b}$ and goes to the right, (b)  the ``short" OPE cut is placed between $s_{L1}$ and $s_{L2}$, (c) cuts associated with the condition $\im p_+ =0$ start at $s_{L1}$ and take rotated parabola-like shape, (d) cut associated with the presence of the circular cut starts at $s_{L1}$ and goes to the left. To uncover the three-body bound-state pole positions, one has to continue the amplitude from Region 1 to Region 2 and rotate the cut aligned with the real axis to the complex plane. Here, we do not indicate the position of a potential three-body bound-state pole that resides anywhere below $s_{\varphi b}$.}
    \label{fig:mphib-strucutre}
\end{figure}

In addition to these structures, the second term of the equation develops unphysical singularities below $s_{\varphi b}$ when the $s$-dependent cuts of OPE collide with the integration interval. Setting $p=0$ in Eq.~\eqref{eq:OPE-pole-condition}, and solving for $s$ yields,
    \beq
    \label{eq:s-cut-condition-1}
    s_\circ  = (m + 2 \omega_k)^2 \, ,
    \eeq
or equivalently,
    \beq
    \label{eq:s-cut-condition-2}
    s_\circ = \frac{(m^2 - \sigma_{k})^2}{m^2} \, .
    \eeq
It is the condition for the branch points $p_{\pm}$ to coincide with the origin of the complex plane for fixed values of $(s,k)$.
At this value, the integral equation solution develops a branch point and an associated cut in the upper-half complex $s$ plane (and its complex-conjugate copy in the lower half). It is given by the conditions $\im p_+ = 0$ and $\re p_+> 0$, which describe the collision of the OPE $q$-plane cuts with the integration interval and translate into an involved relationship between the real and imaginary part of the total invariant mass. For $k=q_b$, it describes half of a rotated parabola-like shape. We note that for $\sigma_k=\sigma_b$, Eqs.~\eqref{eq:s-cut-condition-2} and~\eqref{eq:sl1} become identical. The presence of the $s$-plane cut can be understood as corresponding to two possible ways in which the deformed integration contour circumvents $q$-plane cut of $d_S(q,k)$ that was inherited from the inhomogeneous term of the equation. It can be passed either from the left or the right, the difference equal to the integral of the integration kernel $K(p,q)$ with discontinuity of $G_S(q,k)$ along the cut.

Moreover, the second term of the ladder equation has an explicit singularity implied by the presence of the circular branch cut. The circular cut enters the integrand through the amplitude $d_S(q,k)$ evaluated at $k=q_b$ since it inherits it from the first term of the integral equation. It leads to the left-hand discontinuity of $\Mc_{\varphi b}$ along the real $s$ axis since the cuts of $d_S(q,k)$ in the $q$ complex plane are reflected with the complex conjugation of $s$, as explained by Eq.~\eqref{eq:circular-cut-discontinuity}. This discontinuity starts at $s_\circ$, meaning there are four cuts in total emerging out of this point, all having different origins. It is an atypical feature of the three-body integral equations we solve. They were derived without considering the analyticity of the resulting amplitudes, which in turn happen to have complicated unphysical singularities below the threshold \cite{Jackura:2018xnx, Dawid:2020uhn}.

We present and summarize the analytic structure of $\Mc_{\varphi b}$ in Fig.~\ref{fig:mphib-strucutre}. We note it can also develop three-body poles on the real axis, where the left-hand cuts are present. Nothing can be inferred about their positions beforehand, and the integral equation has to be solved to identify their presence. They might necessitate the rotation of the cuts obscuring the bound-state physics to the complex plane. From this point of view, it is advantageous to consider Eq.~\eqref{eq:s-cut-condition-2} as a condition for $\sigma_k$ evaluated at fixed $s$. The circular cut disappears when,
    \beq
    \label{eq:s-cut-condition-3}
    \sigma_{k} \leq m (m + \sqrt{s}) \, .
    \eeq
In particular, for $\sigma_k \leq 2 m^2$ the circular cut does not occur for any $s>m^2$, which is the lowest value of the total invariant mass we consider. Thus, when $\sigma_b \leq 2 m^2$, or correspondingly $ma \leq \sqrt{2}$, the left-hand cuts of the $\Mc_{\varphi b}$ amplitude travel far to the left. Then, since we expect the bound-states to lie close to the $s_{\varphi b}$ threshold, the poles should not overlap with the cuts, simplifying the extraction of their positions. Evaluation of the $d_S(p,k)$ amplitude both for $\sigma_k = \sigma_b$ and $\sigma_k = 2m^2$ at different values of $a$ is a natural way to verify that the cut rotation procedure does not introduce numerically significant systematic errors and leads to correct bound-state pole positions. It is shown in Sec.~\ref{sec:results}.

\subsection{Extrapolation of the integral equation}
\label{sec:extrapolation}

Let us consider $d_S(p,k)$ as a function of $p$ for fixed $s$ and $k$. As already noted, the integral equation, Eq.~\eqref{eq:d_Sproj_kern}, contains two terms, the inhomogeneous one, which consists of the OPE amplitude $G_S$, and the homogeneous one, which is an integral of the kernel $K(p,q) \, d_S(q,k)$ over the intermediate spectator's momentum, $q$. Let us assume that $d_S(p,k)$ is known for real values of the outgoing momentum, $p \in [0,q_{\rm max}]$. One can use this knowledge in the homogeneous term of the equation, where the integration over $q$ is performed in the same interval, to obtain $d_S(p,k)$ at other values of $p$. Indeed, the right-hand side of the equation depends on $p$ through $G_S(p,k)$ in the first term and $K(p,q)$ in the second one. Since these functions are known analytically, it is possible to extrapolate $d_S(p,k)$ from the real axis to the complex $p$ plane simply by using a complex value of the left-hand momentum argument in both terms.

However, not all complex values of $p$ are accessible with this method. In particular, we are interested in extrapolating the amplitude to the point $p=q_b$. The extrapolation region and its potential extensions are determined from the singularity structure of the integral equation. Using the result of Eq.~\eqref{eq:OPE-cut-parametrization} inside Eq.~\eqref{eq:Gs_proj}, we rewrite the ladder equation as,
    \begin{align}
    \label{eq:d_S_explicit}
    d_{S}(p, k)
    =&
    - \int\limits_{-1}^1 dx \, 
    \frac{H(p, k)}{4 \beta_x(s,k) [p - p_{\text{cut},+}(s,k,x)] [p - p_{\text{cut},-}(s,k,x)]} \\ \nonumber
    & - \int\limits_0^{q_{\rm max}} \frac{d q \, q^2}{ (2\pi)^2 \, \omega_{q}} \int\limits_{-1}^1 dx \, \frac{H(p, q) \Mc_2(q) }{4 \beta_x(s,q) [p - p_{\text{cut},+}(s,q,x)] [p - p_{\text{cut},-}(s,q,x)]}  \, d_{S}(q, k) \, .
    \end{align}
It allows us to clearly identify the singularities of $d_S(p,k)$ in the $p$ variable. The amplitude $d_S(p,k)$ depends on the momentum $p$ through its explicit presence in the cut-off function $H(p,q)$ and the denominators of the two terms of the ladder equation. They are singular when $p$ coincides with poles at $p_{{\rm cut},\pm}$. In the first term, at fixed $k$, the collision points constitute a cut parametrized by $x$, as described in Sec.~\ref{sec:analytic-structure}. This explicit singularity is inherited by $d_S(p,k)$ on the left-hand side of the ladder equation.

\begin{figure}[t]
    \centering
    \includegraphics[width=0.95\textwidth]{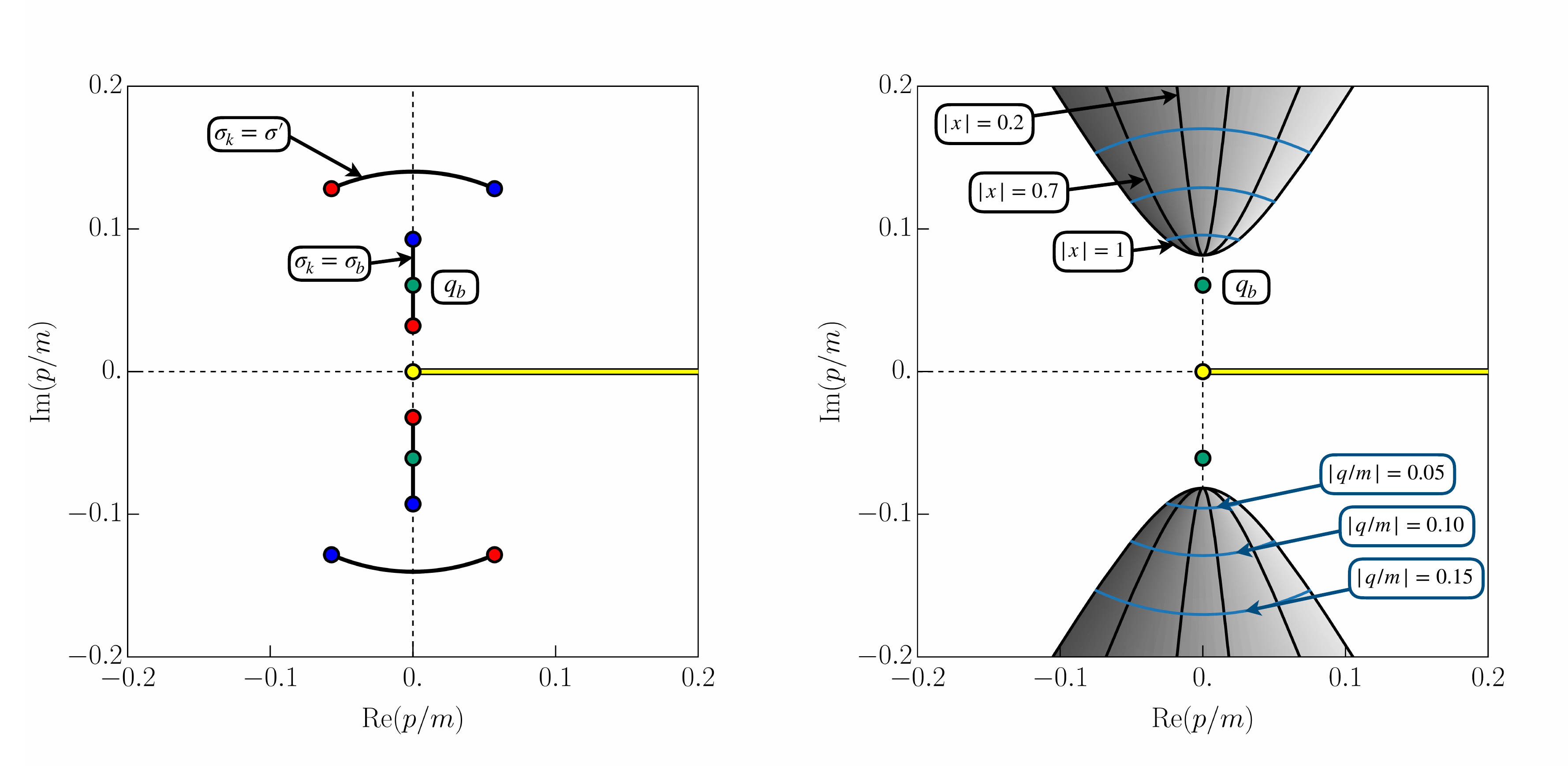}
    \caption{\textbf{Left:} Cuts (black lines) of the inhomogeneous term of the ladder equation, $G_S(p,k)$, for $s/m^2 = 8.96$ and two different choices of external momentum $k$: one corresponding to fixed $\sigma_k =\sigma_b \approx 3.984m^2$ ($ma=16$), and one to fixed $\sigma_k = \sigma' = 3.934 m^2$. Threshold energy is $s_{\varphi b}/m^2 \approx 8.977$, while $s_\circ/m^2 \approx 8.906$. For both values of $\sigma_k$ the cut does not cross the integration interval (yellow line). However, for $\sigma_k = \sigma_b$ it coincides with the point $p = q_b$. \textbf{Right:} Domain of non-analyticity, $\bar{\Rc}$, (shaded area) of the homogeneous term of the ladder equation for $s/m^2 = 8.96$ and $q \in [0, q_{\rm max}]$. Curves corresponding to fixed values of $x$ are shown in black, while curves corresponding to fixed values of $q$ are shown in blue. For the considered value of $s$, we find the point $p = q_b$ outside of the $\bar{\Rc}$.}
    \label{fig:ancon1}
\end{figure}

In the second term, the OPE poles occur for all values of $x \in [-1,1]$ and $q \in [0, q_{\rm max}]$. It is useful to consider them from two points of view: as cuts parametrized by $x$, emerging for all different values of $q$, or, equivalently, as cuts parametrized by $q$, emerging for all possible values of $x$; see the right panel of Fig.~\ref{fig:ancon1}. These curves cover a region in the complex $p$ plane in which the extrapolated solution $d_S(p,k)$ is not analytic. Following Gl\"ockle, we call this area a \emph{domain of non-analyticity} and denote it by $\bar{\Rc}$\footnote{In the language of Brayshaw \cite{Brayshaw:1968yia}, it is called $\bar{R}(W,z=-1)$, where in our relativistic notation, $W = \sqrt{s} - 3m$ and $z = x$. Brayshaw observes that for real $q$, the constant-$z$ curves, $C(W,z)$, can be ordered by the value of $z$, and the $C(W,-1)$ is the boundary of region that contains all of them. The relativistic ladder equation exhibits analogous property.}. The rest of the complex plane is called \emph{domain of analyticity} and is denoted by $\Rc$. In the following discussion, we do not consider the presence of the branch cut singularities of $\omega_q$ and $\Mc_2$ in the homogeneous term. Since for real $s$, $\im q_{\pm,r} > q_b$, and we are interested in continuing $p \to q_b$, we can ignore the regions $\im p, \im q \leq \im q_{\pm,r}$. In other words, for the kinematics of interest, they are far from the path of integration and the complex $p$ region of interest. 

In Fig.~\ref{fig:ancon1}, we present an example position of these structures for a fixed total invariant mass $s_{\varphi b} > s > s_\circ$ and momentum $k$ corresponding to a fixed $\sigma_k < 4m^2$. This particular choice of kinematical variables produces a relatively simple set of singularities of $d_S(p,k)$. Let us consider first the $\sigma_k=\sigma'$ case, for which the non-analytic regions neither cross the integration interval nor contain the extrapolation point of interest $p=q_b$. As noted in the introduction, the numerical solution of the ladder equation is obtained using the Nystr\"om method, i.e., via discretization of the momenta and solution of the resulting matrix equation as explained in App.~\ref{app:C}. It requires fixing the value of argument $k$ and evaluation of two remaining momentum variables, $p$ and $q$, on the real integration contour $\Cc$. Since the integration path (yellow line) is not crossed by any singularity, we can safely evaluate $p$ there and achieve the desired solution. Once $d_S(p,k)$ is known on the real axis, one can extend it to those complex values of $p$, which lie outside of $\bar{\Rc}$ (shaded area).

As can be seen from the ladder equation itself, the domain of non-analyticity does not depend on the variable $k$. Thus, we reach similar conclusions in the second illustrated case, $\sigma_k=\sigma_b$ (equivalently $k=q_b$), with one exception. For this value of the incoming spectator's momentum, the cut of the inhomogeneous term coincides with the $p=q_b$, as can be seen on the left panel of Fig.~\ref{fig:ancon1}. Its presence does not prevent one from solving the equation, as it does not cross the integration interval. It corresponds to a cut of the amplitude $d_S(p=q_b,k=q_b)$ in the complex $s$ plane, inherited from the inhomogeneous term in the ladder equation. This is the explicit cut of $G_S(q_b,q_b)$ discussed in Sec.~\ref{sec:analytic-continuation-A}, running between $s_{L1}$ and $s_{L2}$. From the complex $p$ plane point of view, the emergence of the $s$-plane cut is understood by studying the behavior of the $p$-plane cut of $G_S(p,q_b)$ for small non-zero values of $\im s$. Adding a small positive/ negative imaginary part to $s$ moves the cut to the left/ right of the $p=q_b$ point, leading to a discontinuity in $d_S(p=q_b,q_b)$ along the real $s$ axis.

\subsection{Continuation to the domain of non-analyticity}
\label{sec:domain-of-non}

From the above examples, we observe there exists an area of the complex $p$ plane that is not immediately accessible via straightforward extrapolation. Although in the cases discussed above, the desired extrapolation point $p=q_b$ lies outside of the $\bar{\Rc}$ region, it might travel to the domain of non-analyticity for other values of $s$. We discuss such a case in the following subsection. It is, therefore, useful to study the continuation of our solution into this region. There are two ways of extending the solution $d_S(p,k)$ from $\Rc$ to the domain of non-analyticity, $\bar{\Rc}$. 

{\bf Modification of the kernel (Brayshaw's Method):} In the first one, one includes the discontinuity of $G_S(p,q)$ in the kernel of the homogeneous part of the ladder equation. Namely, following Brayshaw, we redefine the ladder equation in the following way,
    \begin{align}
    \label{eq:d_S_continuation}
    d_{S}(p, k)
    &=
    - G_{S}(p, k)  
    - \int\limits_0^{q_{\rm max}} dq \, \bar{K}(p,q) \, d_{S}(q, k) \, ,
    \end{align}
where
    \begin{align}
    \bar{K}(p,q) &= \frac{q^2}{ (2\pi)^2 \, \omega_{q}} \, G_{S}(p, q) \, \Mc_2( q) \, , ~~~\text{if}~~~ p \in \Rc \, , \\
    & = \frac{q^2}{ (2\pi)^2 \, \omega_{q}} \, \left[G_{S}(p, q) + \Delta(p,q) \theta(q - q_{\text{cut}}) \right] \, \Mc_2( q) \, , ~~~\text{if}~~~ p \in \bar{\Rc} \, .
    \end{align}
Here we defined the discontinuity of the OPE amplitude,
    \beq
    \label{eq:OPE-discontinuity}
    \Delta(p,q) = -(2\pi i)\frac{H(p,q)}{4 p q} \, ,
    \eeq
and a momentum $q_{\text{cut}}$ for which the first constant-momentum cut of the integration kernel (blue lines in the right panel of Fig.~\ref{fig:ancon1}) crosses the external extrapolation momentum $p$. It is given by the condition, 
    \beq
    \label{eq:cont-condition}
    p = p_+(s, q_{\text{cut}}, x)
    \eeq
for some $x \in [-1,1]$. 

The modified kernel, $\bar{K}(p,q)$, is defined to be smooth for all values within the integration region. Because the discontinuity in the kernel was the origin of the area of non-analyticity, it should not be to surprising that Eq.~\eqref{eq:d_S_continuation} constitutes the analytic continuation of $d_S(p,k)$ to $p \in \bar{\Rc}$ except for points where $\Delta(p,q)$ is singular in this region. In particular, we note the explicit essential singularity of the smooth cut-off function $H(p,q)$ belongs to the domain of non-analyticity for $s \leq s_{\varphi b}$.

\begin{figure}[t]
    \centering
    \includegraphics[width=0.45\textwidth]{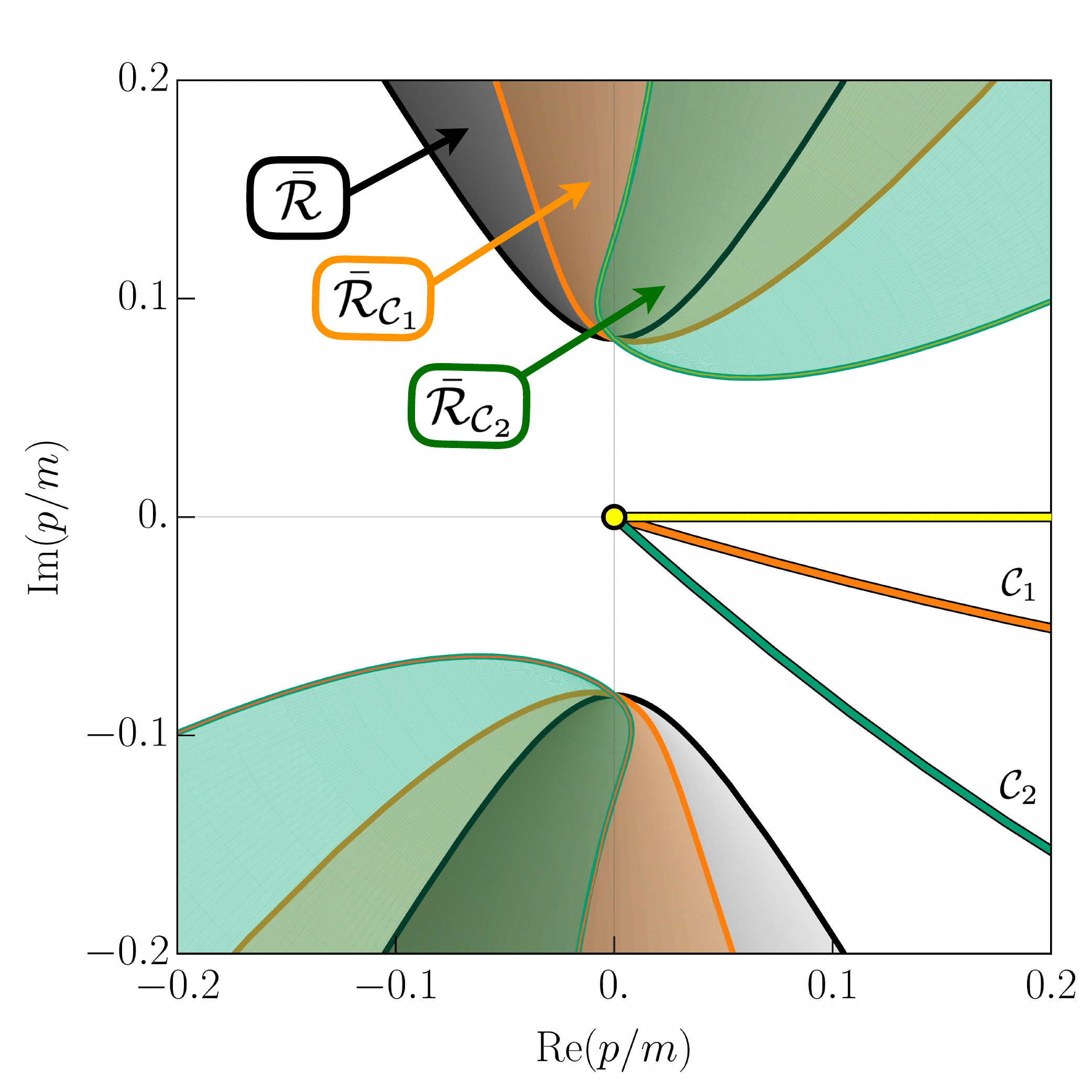}
    \caption{Domains of non-analyticity of the solution $d_S(p,k)$ for different choices of the complex integration path. We use $s/m^2 = 8.96$ and three different integration contours: the original interval $[0, q_{\rm max}]$ (yellow), contour $\Cc_1$ described by parametrization $\gamma(t) = q_{\rm max} (t + 0.3 \, i t (t-1))$, $t \in [0,1]$ (orange), contour $\Cc_2$ described by parametrization $\gamma(t) = q_{\rm max} (t + 0.9 \, i t (t-1))$, $t \in [0,1]$ (green). We highlight corresponding domains of non-analyticity. Their boundaries cross the imaginary axis at point $p_0$.}
\label{fig:dom-of-anal}
\end{figure}

In the equations above, we simplified the original method of Ref.~\cite{Brayshaw:1968yia}. There, see Eqs.~(37) to (41), the author changes the order of the $q$ and $x$ integration in the non-relativistic analog of our Eq.~\eqref{eq:d_S_explicit} and considers the continuation of the inhomogeneous term through the fixed-$x$ curves similar to the ones presented in the right panel of Fig.~\ref{fig:ancon1}. In other words, instead of adding discontinuity to the function $Z$ (a non-relativistic analog of $G_S$) along cuts understood as curves parametrized by $x$ for fixed $q$, he adds it along lines corresponding to fixed $x$ and parametrized by $q$. It allows for a clear geometric interpretation of the analytic continuation procedure since those curves are boundaries of the regions to which we continue the amplitude. 

However, this procedure leads to an expression for discontinuity that contains the integral of the ladder equation solution $X$ (a non-relativistic analog of our $d_S$) evaluated along fixed-$x$ curves; see Eq.~(39) therein. It makes the solution of the integral equation through the numerical Nystr\"om technique more difficult. 

In our work, we use the original order of integration and the fact that one can perform the integration over $x$ analytically since $d_S$ is independent of the scattering angle. These two ways of defining appropriate discontinuities and analytic continuation are mathematically equivalent. However, our method is not completely free of difficulties. The trade-off is that one loses the simple geometric interpretation of Brayshaw and has to solve Eq.~\eqref{eq:cont-condition} for the value of $q_{\text{cut}}$. We note that for purely imaginary $p$, it can be obtained by setting $x=0$, leading to, 
    \beq
    q_{\text{cut}} = \frac{\sqrt{ \sigma_k^2 - 4 m^2 (\sqrt{s} - \omega_k)^2 }}{2(\sqrt{s}-\omega_k)} \, .
    \eeq
For a general complex value of $p$, the simplest way to determine $q_{\text{cut}}$ is by solving Eq.~\eqref{eq:cont-condition} numerically. We come back to this issue in Sec.~\ref{sec:an-con-circ} when it becomes relevant for total invariant mass $s < s_{L1}$.

{\bf Contour deformation (\`a la Gl\"ockle):}  The alternative way of continuing the solution to $\bar{\Rc}$ is via the deformation of the integration contour. It is a method employed by Gl\"ockle in Ref.~\cite{Glockle:1978zz}. After the analysis of the previous paragraphs, in principle, the solution of the ladder equation is known not only for real $p$ but also for all $p \in \Rc$. Thus, one can generalize the integration path from the interval $[0,q_{\rm max}]$ to a complex curve $\Cc \subset \Rc$ that starts at $q=0$ and ends at $q_{\rm max}$. The contour deformation must itself be continuous, i.e., it can not cross any singularities of the integration kernel. Because the region $\bar{\Rc}$, defined by pole positions, $p_{{\rm cut},\pm}(s,q,x)$, becomes different when parametrized by $q \in \Cc$, the contour deformation leads to a new domain of non-analyticity, which we denote $\bar{\Rc}_\Cc$. We show an example of this behavior in Fig.~\ref{fig:dom-of-anal}. 

This way, one may continue $d_S(p,k)$ to all $p \in \bar{\Rc} \cap \Rc_{\Cc}$, by defining the analytic continuation of $d_S(p,k)$ as the solution of the ladder integral equation with the deformed contour $\Cc$. We note that the contour deformation procedure in general does not allow one to uncover the whole $\bar{\Rc}$ region with a single contour $\Cc$. For instance, evaluating Eq.~\eqref{eq:OPE-cut-parametrization} at $x=0$ and zero momentum, we find that the boundary of every $\Rc_{\Cc}$ crosses a purely imaginary point,
    \beq
    p_0 = p_{\text{cut},\pm}(s,q=0,x=0) = \frac{ \sqrt{(\sqrt{s} - 3m) (\sqrt{s}+m) } }{ 2 } \, .
    \eeq
    
Several regions $\Rc_{\Cc}$ might be needed to cover its vicinity. Moreover, the deformed integration path should not cross the new region of non-analyticity. Otherwise, it does not define the analytic continuation of the amplitude, and one can not achieve the solution via the Nystr\"om method, where both $p$ and $q$ must be evaluated on the integration contour. In addition to that, $\Cc$ should not cross singularities of the product $H(p,q) \Mc_2(q)$ and those singularities of $d_S(q,k)$ which are inherited from the inhomogeneous term of the equation. We call contours that satisfy these constraints \emph{self-consistent}.

\subsection{Analytic continuation in the presence of the circular cut}
\label{sec:an-con-circ}

\begin{figure}[t]
    \centering
    \includegraphics[width=0.95\textwidth]{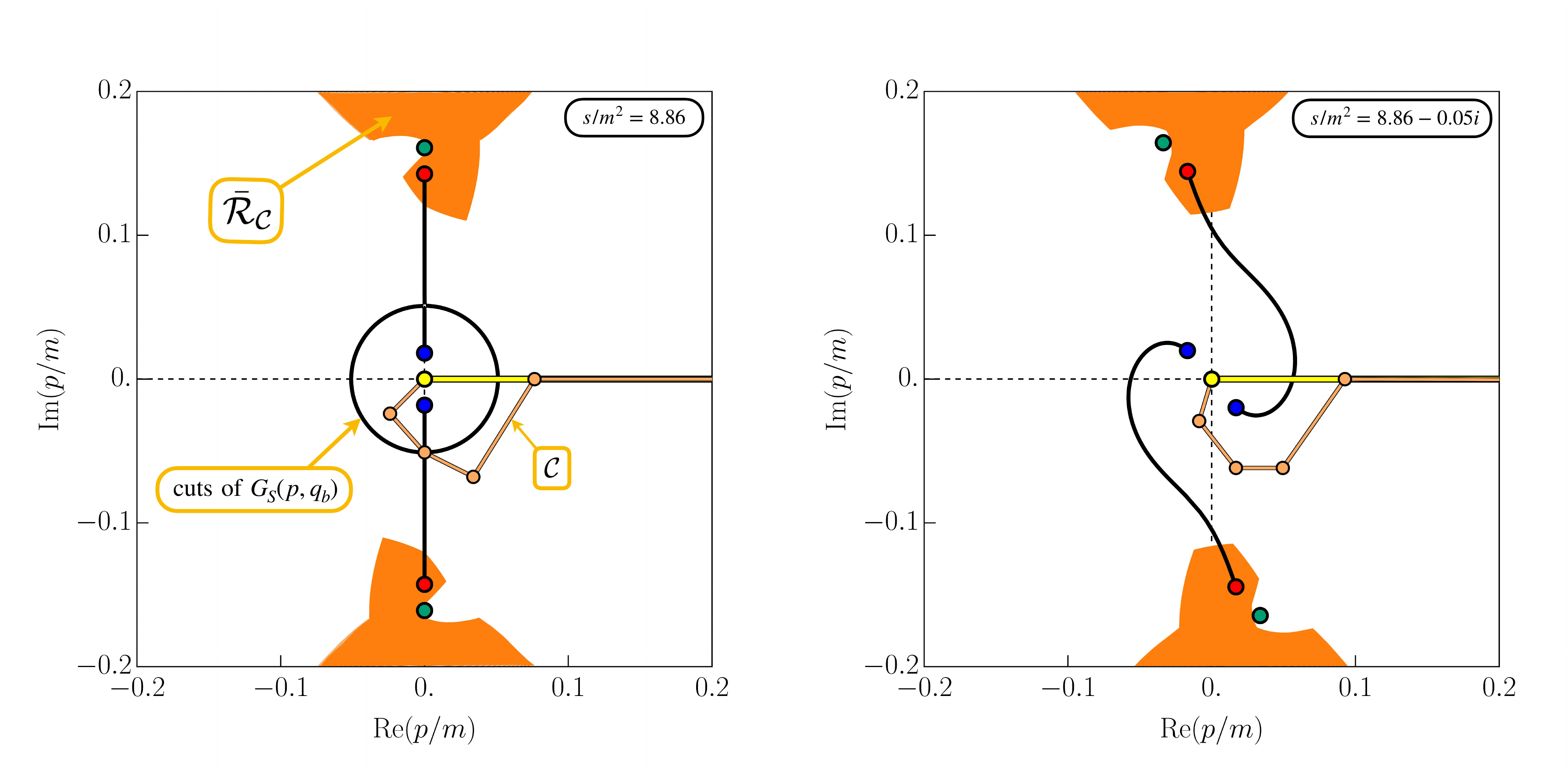}
    \caption{Singularities of both terms of the ladder equation for $s/m^2 = 8.86$ (left) and $s/m^2 = 8.86-0.05i$ (right). We set the external momentum to $k = q_b$ for $ma=16$. In this case, the threshold energy is $s_{\varphi b}/m^2 \approx 8.977$, while $s_\circ/m^2 \approx 8.906$. Black lines represent cuts of the inhomogeneous term, $G_S(p,k)$. The cut crosses the original integration interval (yellow), and it must be deformed (orange line). Here, we chose a piece-wise linear contour $\Cc$. The green dot represents the point $p=q_b$. The domain of non-analyticity, $\bar{\Rc}_{\Cc}$, of the homogeneous term corresponding to the orange integration contour is shown as an orange area. The integration contour was chosen such that they do not cross. For $\im s = 0$ point $p = q_b$ is found on the boundary of $\bar{\Rc}$. For $\im s < 0$ it is outside of the domain of non-analyticity for appropriate contour, while for $\im s > 0$ it belongs to $\bar{\Rc}_{\Cc}$.}
    \label{fig:an-con-D}
\end{figure}

In the above examples, the cut of the inhomogeneous part of the ladder equation does not cross the original integration path. This situation changes when one fixes $k=q_b$ and extends $s$ below the point $s_\circ$ or sufficiently deep into the complex plane, as exemplified in Fig.~\ref{fig:circular-cut-1}. The OPE amplitude develops a cut that coincides with the real $p$ axis. In this case, since $d_S(p,k)$ inherits this singularity, and thus it propagates to $d_S(q,k)$ in the homogeneous term, the integration contour deformation is no longer optional but required. We note that both Ref.~\cite{Brayshaw:1968yia} and~\cite{Glockle:1978zz}, which we followed so far, do not discuss this possibility.

As discussed in Sec.~\ref{sec:analytic-continuation-A}, when the OPE cut crosses the integration interval, the resulting amplitude $\Mc_{\varphi b}$ develops discontinuity in the $s$ variable. By deforming the integration contour, we analytically continue the amplitude from Region 1 to Region 2 and 3 through cuts denoted by \textbf{(c)} in Fig.~\ref{fig:mphib-strucutre}. As already noted, the integration contour can be deformed to circumvent the OPE cut in the complex $p$ plane, either from the right or left. One determines the integration path by fixing $\im s/m^2 = \pm \delta$, where $\delta>0$ is a small number, and steadily changing $\re s$ from $s_{\varphi b}$ below $s_{\circ}$. When the OPE amplitude branch cuts are positioned deep in the complex plane, the integration can be performed over the real axis. For decreasing $\re s$, the singularities approach and finally cross the real axis wrapping around the origin of the complex plane. The integration contour is deformed according to their trajectory.

\begin{figure}[t]
    \centering
    \includegraphics[width=0.98\textwidth, trim = {4 4 4 5}, clip]
    {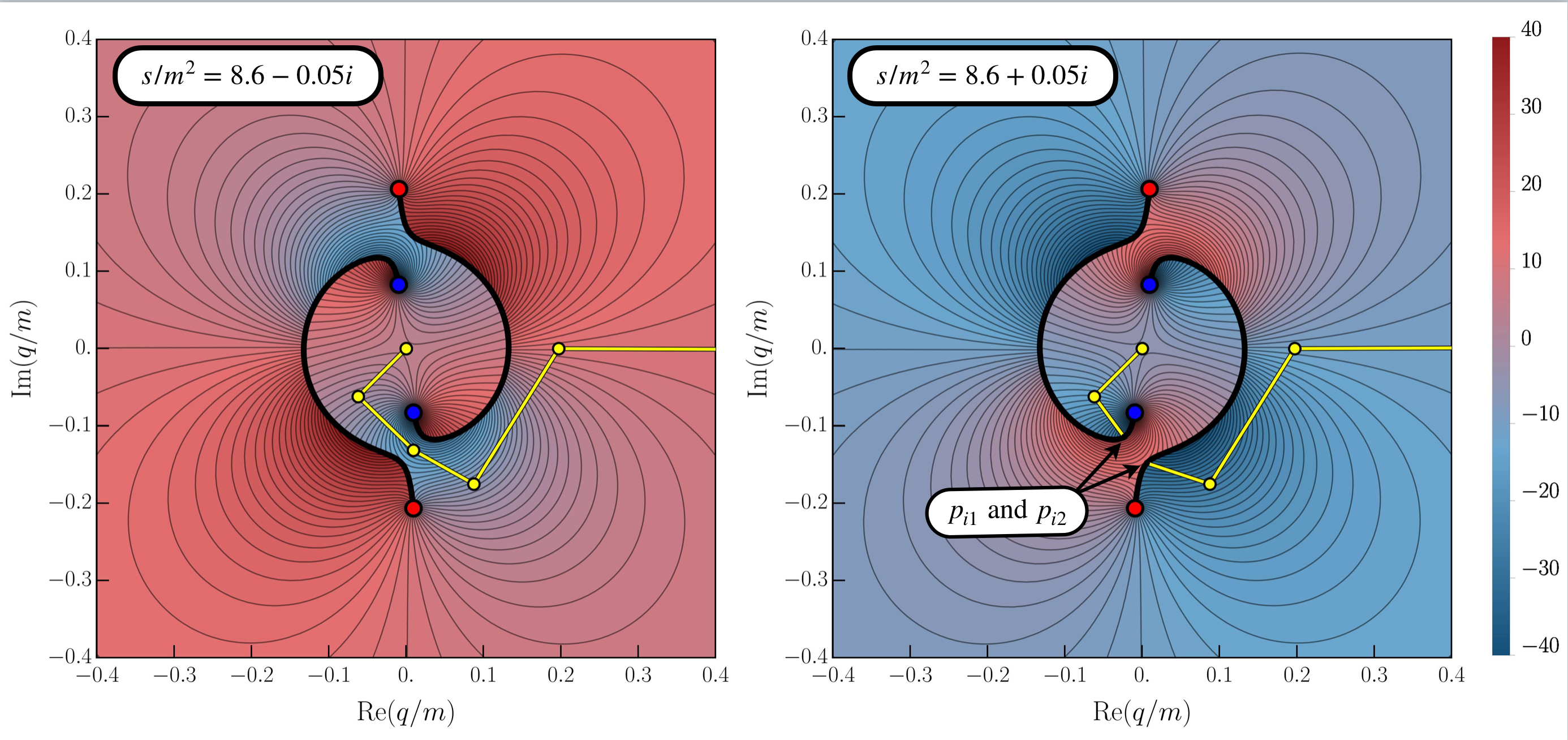}
    \caption{$\im G_S(q, k)$, in the complex momentum plane for $s/m^2 = 8.6 \pm 0.05 i$ and $k = q_b$ ($ma=16$). A hard cut-off is used and the amplitude is given in units of $1/m^2$. Negative $\im s$ is presented on the left panel, and positive on the right. Example contour for negative $\im s$ can not be continuously deformed to work for $\im s>0$ due to reflection of the cuts, therefore we continue it through the second Riemann sheet. It is presented as the disappearance and reappearance of the contour crossing the cuts on the right panel.}
    \label{fig:two-contours}
\end{figure}

In Fig.~\ref{fig:an-con-D}, we present singularities of the ladder equation and examples of appropriate contours for continuation below the $\re s = s_{\circ}$ point. We present two cases, for which $\im s/m^2 = 0$ (left panel) and $\im s/m^2 < 0 $ (right panel). For the zero imaginary part of $s$, the contour must pass through the lower-half copy of point $q_{c2}$, defined in Eq.~\eqref{eq:mom_c2}. For the non-zero imaginary part of $s$, the circle ``opens" and a contour is no longer constrained by this condition. The integration path $\Cc$ chosen to avoid cuts of the inhomogeneous term must also avoid the corresponding domain of non-analyticity $\bar{\Rc}_{\Cc}$. Presented contours allow for continuing the $\Mc_{\varphi b}$ from Region 1 to Region 2 in Fig.~\ref{fig:mphib-strucutre} through the lower-half copy of cut (c). 

By reflecting the presented singularities with respect to the $\re p$ axis, one obtains the singularity structure of the equation for positive values of $\im s/m^2$. Thus, by analogy, to analytically continue the amplitude from Region 1 to Region 3, one has to use a deformed integration path that is a complex conjugation of the one presented in Fig.~\ref{fig:an-con-D}.

From the perspective of the complex $p$ plane, the presence of the unphysical left-hand cut of $\Mc_{\varphi b}(s)$ starting at $s_\circ$ is a consequence of property~\eqref{eq:circular-cut-discontinuity} and the resulting discrepancy between the two choices of contours for $\im s \leq 0$ and $\im s > 0$ case. It might disable one from identifying the bound-state pole positions and residues below $s_\circ$ as they may overlap with the left-hand cut of the solution. To access the real axis below $s_\circ$, one has to continue the amplitude from the lower half-plane through the cut to the upper half-plane. Equivalently, one can say that the left-hand $s$-plane cut has to be rotated away from the real axis.

However, we note no smooth contour deformation allows for the analytic continuation across the $\re(s) \leq s_\circ$ line. When we increase $\im s$ from negative to positive values, the circular cut closes (at $\im s = 0$) and then opens in a manner that makes the integration contour $\Cc$ cross the OPE cuts in the complex $p$ plane twice, see the right panel of Fig.~\ref{fig:two-contours}. It is not possible to use the Cauchy theorem in the usual manner to define a contour that avoids the singularities of the OPE and, at the same time, is a smooth deformation of the original contour $\Cc$.

{\bf Analytic continuation across the $\bm{{\rm Re}(s)}$ axis:} Nevertheless, the behavior of the OPE cuts suggests a natural way to extend the solution from negative to positive values of $\im s$. We define the analytic continuation of $d_S(p,q_b)$ from $\im s \leq 0$ to $\im s > 0$ in the following way. For $\im s=0 $, we fix an appropriate contour $\Cc_0$ that passes through the lower-half copy of $q_{c2}$ and is self-consistent. In principle, solution $d_S(p,q_b)$ for $p \in \Cc_0$ is defined using prescriptions of the previous sections and of App.~\ref{app:C}. For small $\im s>0$, the cuts of the inhomogeneous term are crossed twice by this contour. We call two intersection points $p_{i1}$ and $p_{i2}$ and define $\Cc'$ as the piece of the contour $\Cc_0$ starting at $p_{i1}$ and ending at $p_{i2}$. The inhomogeneous part of the ladder equation, which has one momentum fixed at $q_b$, has a discontinuity along the contour $\Cc_0$ at these two points. We can remove it by adding the OPE discontinuity to this part whenever the $p$ momentum is evaluated between them. It implies evaluation of the OPE amplitude on the second sheet associated with the OPE cuts whenever $p \in \Cc'$. 

The kernel appearing in the homogeneous term of the equation, $K(p,q)$, is evaluated with momenta $p$ and $q$, both of which are in $\Cc_0$, and does not have a discontinuity in this region. With this in mind, for $p \in \Cc_0$ and $\im s>0$, we define $d'_S(p,q_b)$ as,
    \begin{align}
    \label{eq:d_S_II}
    d'_{S}(p, q_b)
    =
    - \left[  G_S(p,q_b) + \Delta(p,q_b) \theta(p \in \Cc') \right]
    - \int_{\Cc_0} dq \, K(p,q) \, d'_{S}(q, q_b) \, .
    \end{align}
Here, $\theta(p \in \Cc')$ is a function equal to 1 for $p \in \Cc'$ and to 0 otherwise. Integration over $q$ is performed along the contour $\Cc_0$ starting at $q=0$ and ending at ${q_{\rm max}}$. The above integral equation differs from the original one, Eq.~\eqref{eq:d_Sproj_kern}, by the discontinuity added in the inhomogeneous part. However, since $\Cc'$ shrinks to zero when $\im s \to 0^+$, we see that in this limit, inhomogeneous parts of both equations become identical. Thus $d'_S(q,k)$ in the homogeneous term does not have discontinuities along $q \in \Cc_0$, and one can safely use this integration contour. This procedure is schematically illustrated in Fig.~\ref{fig:two-contours}. It can be represented as ``diving" with the contour $\Cc_0$ into the second sheet of OPE and emerging outside the area enclosed by the ``circle."

{\bf Extrapolation to the bound-state pole:}
As usual, in the solution attempt, one has to make sure that the contour used in Eq.~\eqref{eq:d_S_II} does not cross the corresponding domain of non-analyticity, $\bar{\Rc}_{\Cc_0}$. Finally, after computing $d_S(p,k)$ for $p \in \Cc_0$ one must determine whether the extrapolation momentum $p=q_b \in \bar{\Rc}_{\Cc_0}$ or not. In fact, from symmetry of the OPE, $G_S(p,k) = G_S(k,p)$, we see that the point $p=q_b$ is crossed by the cuts of $K(p,q)$ for $q = p_{i1}$ and $q = p_{i2}$, and thus belongs to $\bar{\Rc}_{\Cc_0}$. Therefore, to continue solution $d'_S(p,k)$ from $p \in \Cc_0$ to $p=q_b$, we have to employ the prescription of Brayshaw, Eq.~\eqref{eq:d_S_continuation}. We write,
    \begin{align}
    \label{eq:d_S_III}
    d'_{S}(q_b, q_b)
    =
    - G_S(q_b,q_b) 
    - \int_{\Cc_0} dq \, \bar{K}(q_b,q) \, d'_{S}(q, q_b) \, .
    \end{align}
The $\Delta(p,k)$ piece in the inhomogeneous term disappeared since $q_b \notin \Cc'$ and the $\theta$ function becomes zero. 

We observe that for $q \in \Cc_0$ the constant-momentum cuts of $K(q_b,q)$ have a more complicated shape. Moreover, with $q$ moving along $\Cc_0$ they evolve very differently than in the simple $q \in [0, q_{\rm max}]$ case presented in Fig.~\ref{fig:ancon1}. In general, it is difficult to follow their evolution analytically and find a solution to a condition equivalent to Eq.~\eqref{eq:cont-condition}. Fortunately, from the symmetry of $G_S$, we know they cross the $p=q_b$ point exactly twice, at $q = p_{i1,2}$, and thus the $\theta(q-q_{\rm cut})$ in $\bar{K}$ term has to be replaced by $\theta(q \in \Cc')$. The modified integration kernel becomes,
    \begin{align}
    \bar{K}(p=q_b,q) 
    = \frac{q^2}{ (2\pi)^2 \, \omega_{q}} \, \left[G_{S}(p=q_b, q) + \Delta(p,q) \theta(q \in \Cc') \right] \, \Mc_2( q) \, .
    \end{align}
This way discontinuity $\Delta(p,q)$ is added to the integration kernel for those values of $q$ for which $p$ is found on its second sheet. Together, Eqs.~\eqref{eq:d_S_II} and~\eqref{eq:d_S_III} allow one to continue the solution $d_S(p,q_b)$ from $\im s \leq 0$ to $\im s >0$ and extrapolate it to $p = q_b$. It concludes our discussion of the analytic continuation of $\Mc_{\varphi b}(s)$ below the $\varphi b$ threshold.

\subsection{Continuation above the two- and three-particle thresholds}
\label{sec:above-thresholds}

\begin{figure}[b]
    \centering
    \includegraphics[width=0.45\textwidth]{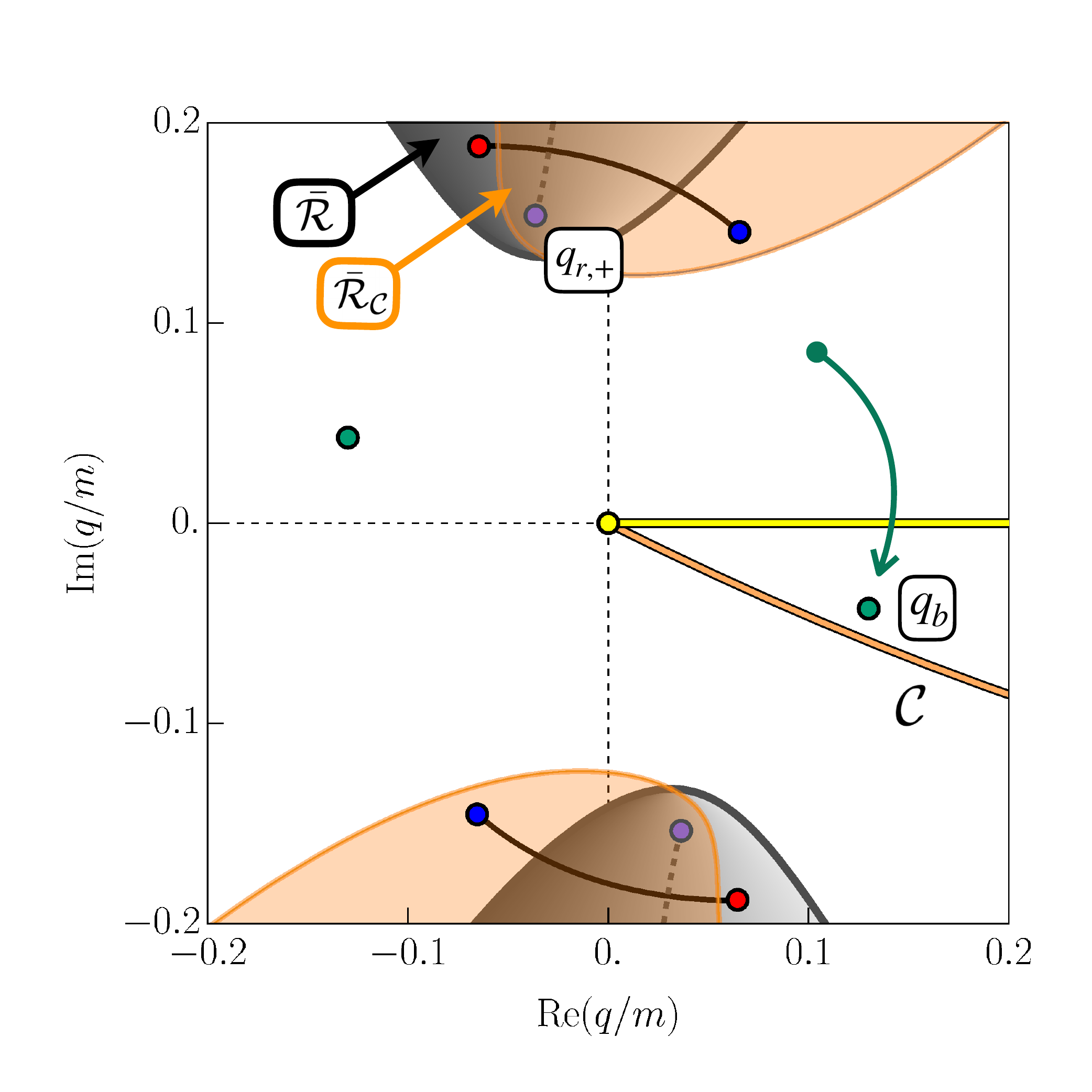}
    \caption{Singularities of both terms of the ladder equation for $s/m^2 = 8.9 - 0.05 i$. The external momentum is set to $k=q_b$ for $ma=6$. The threshold energy is $s/m^2 =s_{\varphi b} \approx 8.833$. Black lines represent cuts of the inhomogeneous term, $G_S(p=q,k)$. Pole is represented as a green dot. Singularities of the homogeneous term are shown as shaded areas. The $\Mc_2(q)$ branch cuts are highlighted with purple points. Associated cuts are shown as gray dashed lines. Contour $\Cc$ is given by parametrization $\gamma(t) = q_{\rm max} (t + 0.5 \, i t \, (t-1))$, $t \in [0,1]$.
    }
    \label{fig:above_phi_b}
\end{figure}

Above discussion can be applied to the ladder equation evaluated above the $\varphi b$ and $3\varphi$ thresholds. One has to study singularities of the inhomogeneous and homogeneous terms and decide whether contour deformation is required to continue the amplitude to the complex energy plane. The main difference to our previous considerations is that one can no longer ignore the singularities of $\Mc_2(q)$ in the kernel. In particular, the two-body right-hand cut in the $s$ variable has its source in the collision of the $q_b$ bound-state with the origin of the integration contour. Similarly, the three-body right-hand cut emerges when the $q_{\pm,r}$ branch cuts coincide with the $q=0$ point. We note that points $q_b, q_{\pm,r}$ do not depend on the $p$ variable but still depend on $s$. When considering $s > s_{3\varphi}$, they are found on the real axis: one copy on the positive and the other on the negative half. In the case of $q_{\pm,r}$ points, we orient the cut to run between them when $\im s = 0$. When $\im s \neq 0$, two cuts are given by the condition $\sigma_k > 4 m^2$.

When $s$ becomes complex, both $q_b$ and $q_{\pm,r}$ cross the real $q$ axis. Contour deformation that follows the pole allows one to probe the second sheet of the solution in the $s$ variable. When the contour is deformed between the $q_b$ and $q_{+,r}$ points above the $3 \varphi$ threshold, one may probe the second sheet associated with the three-body open channel. We note that in the $s > s_{3\varphi}$ case, we must not use the unmodified smooth cut-off prescription due to the essential singularities that coincide with $q_{\pm,r}$, as discussed in Sec.~\ref{sec:analytic-structure}.

From symmetry under exchange $p \leftrightarrow k$ in $G_S(p,k)$, we can rewrite the homogeneous term of the ladder equation, Eq.~\eqref{eq:d_S_explicit}, as,
    \begin{align}
     - \int\limits_0^{q_{\rm max}} \frac{d q \, q^2}{ (2\pi)^2 \, \omega_{q}} \int\limits_{-1}^1 dx \, \frac{H(p, q) \Mc_2(q) }{4 \beta_x(s,p) [q - p_{\text{cut},+}(s,p,x)] [q - p_{\text{cut},-}(s,p,x)]}  \, d_{S}(q, k) \, ,
    \end{align}
i.e., in a form where one ought to look for singularities of the integrand in the complex $q$ plane for $p \in \Cc$. Note that since both $q,p \in \Cc$ in the Nystr\"om method, this is equivalent to our previous analysis in terms of $p$. In the analytically continued solution, the integration contour should avoid the poles in $q$ and singularities of $H(p,q) \Mc_2(q)$. 

We show an example behavior and integration contour for the case $\re s > s_{\varphi b}$ and non-zero $\im s$ in Fig.~\ref{fig:above_phi_b}. We consider the case with the $\Mc_2(q)$ containing the two-body bound state pole and explore the $k=q_b$ case. The cuts are relatively far from the real axis; however, for $s > s_{\varphi b}$ and $\im s$ decreasing from positive to negative values, the bound-state pole crosses the integration path (green arrow). The contour is deformed to $\Cc$ to access the second sheet of the solution in the complex $s$ plane. Increasing $s$ above $s_{\varphi}$ point, $q_{+,r}$ crosses the real integration path as well. We note that the pole is positioned far from the domain of non-analyticity, and extrapolation of external momenta to this value does not pose any problem. In this work, we are mostly interested in bound-state physics and do not consider this case further. One finds application of similar ideas to the three-body physics in Refs.~\cite{Sadasivan:2020syi, Sadasivan:2021emk}. Finally, we note that the analytic continuation through the right-hand cut of $\Mc_{\varphi b}(s)$ can be achieved either via the contour deformation or by using the explicit expression, Eq.~\eqref{eq:phi-b-second-sheet}, derived from the $S$-matrix unitarity. 

\section{Integral equation solution}
\label{sec:results}

In this section, we present numerical solutions of the inhomogeneous and homogeneous ladder equation, Eqs.~\eqref{eq:d_Sproj_kern} and~\eqref{eq:homo-eq}. The results for $s \leq s_{\varphi b}$ and for complex $s$ are obtained using methods described in the preceding section. Before discussing the outcome of our calculation, it is worthwhile to summarize the major steps of the solution method:
\begin{enumerate}[(a)]
    \item \textbf{Definition of the kinematics:} One specifies the total invariant mass $s$ and external momenta $p$, $k$ for which one wants to compute the ladder amplitude $d_S(p,k)$ or the vertex function $\Gamma(p)$.
    \item \textbf{Complex analysis of the equation:} One performs the analysis of the structure of singularities of the inhomogeneous and homogeneous terms of the integral equation. Both are considered functions of $p$ for fixed $s$ and $k$. Their singularities in the complex $p$ plane are inherited by $d_S(p,k)$.
    \item \textbf{Definition of the integration contour:} If, for a fixed $p$, the singularities of the integrand cross the real $q$ axis one continuously deforms the integration contour to $\Cc$. The contour must start at $q=0$ and ends at $q=q_{\rm max}$. Moreover, $\Cc$ must avoid all of the singularities identified in the previous step: both the cuts of the inhomogeneous term and the domain of non-analyticity $\bar{\Rc}_{\Cc}$.
    \item \textbf{Numerical implementation:} One evaluates momenta $p,q$ on the integration path $\Cc$. The Nystr\"om method is applied by discretizing them along the contour and solving the resulting algebraic equation numerically.
    \item \textbf{Analytic continuation in $p$:} Once the solution of the algebraic problem is known, one extrapolates it from $p \in \Cc$ to the desired $p$ point, as chosen in step \textbf{(a)}. If $p \in \bar{\Rc}_{\Cc}$ one continues the solution by applying Brayshaw's method.
\end{enumerate}

The first step is self-explanatory. One needs to specify what set of kinematical variables is relevant/ interesting for the physical system under study. Here, our main interest lies in the amplitude $\Mc_{\varphi b}(s)$. We wish to identify the presence of the three-body bound states and test the amplitudes obtained using the relativistic FV formalism, Ref.~\cite{Romero-Lopez:2019qrt}. Therefore, we fix $p=k=q_b$ (equivalently $\sigma_p =\sigma_k = \sigma_b$) and study $s < s_{\varphi b}$. Following Ref.~\cite{Romero-Lopez:2019qrt}, we consider cases $ma = 2,6,16$, which describe two-body bound states of decreasing binding energy. We note that the same poles appear in coupled amplitudes, e.g., in $\Mc_{3\varphi}(s)$. We verify that by computing $d_S(p,k)$ for momenta corresponding to fixed $\sigma_p = \sigma_k = 2 m^2 \neq \sigma_b$. In this case, the singularity structure of the amplitude simplifies as the left-hand cuts are pushed deeper below the $\varphi b$ threshold. 

The second step of the procedure outlined above is essentially equivalent to the discussion of Sec.~\ref{sec:analytic-structure}. It is required to understand whether and how to avoid the singularities of the OPE and integral equation kernel and properly define analytically continued solutions.

In the third step, after identifying all relevant singularities in the $p$ variable, one needs to define a self-consistent integration contour. This was discussed in Sec.~\ref{sec:analytic-continuation}. The numerical procedure is based on the discretization of the integral equation and the solution of the resulting algebraic equation. The row and column indices of the kernel matrix correspond to $p$ and $q$. Thus, after discretization, both of them are evaluated on the same integration contour $\Cc$. The corresponding domain of non-analyticity, $\bar{R}_\Cc$ changes as $\Cc$ is deformed, and the two can not cross each other. This would invalidate the application of the Cauchy theorem and the analytic continuation of the solution.

The fourth step of the procedure requires a numerical implementation of the contour-deformed integral equation. If, for a given choice of $(p,k)$ and $s$ the cuts of the OPE are absent from the real integration axis, and $p$ belongs to the domain of analyticity of $d_S(p,k)$, one can adapt the numerical solution method from Ref.~\cite{Jackura:2020bsk}. Namely, one discretizes the real momentum interval in the integral equation as no contour deformation is required. Extension of the numerical methods to the contour-deformed integral equation is described in detail in App.~\ref{app:C}. In particular, we present an effective discretization method and define example self-consistent contours appropriate for the analytic continuation along the cut for $s < s_{L1}$.

Finally, similarly to the procedure presented in Ref.~\cite{Jackura:2020bsk}, the final solution $d_S(p,k)$, obtained for $p \in \Cc$ has to be extrapolated to the kinematic point of interest, e.g., $p=q_b$ in the case of the $\varphi b$ amplitude. This point does not have to belong to $\Cc$. However, if it belongs to $\bar{R}_\Cc$ one must analytically continue the solution to the domain of non-analyticity. Once the integration contour $\Cc$ is established, the simplest approach is to apply Brayshaw's procedure, potentially with modifications described in Sec.~\ref{sec:domain-of-non}.

\subsection{Results}

We now turn to the presentation of the solutions of the ladder equation. We first discuss the $\Mc_{\varphi b}(s)$ amplitude for real $s < s_{\varphi b}$ and the analytic continuation of the bound-state--spectator $\Kc_{\varphi b}$ matrix below the threshold. We compare our findings with the FV calculation of Ref.~\cite{Romero-Lopez:2019qrt} and identify the positions of the trimers. Second, we present the solution for the $d_S(p,k)$ amplitude at $\sigma_p = \sigma_k = 2 m^2$ and verify the unaltered presence of the three-body bound-state pole. Next, we present the $\Mc_{\varphi b}$ amplitude in the complex $s$ plane on the first and second Riemann sheets. We discuss the rotation of the $s < s_{L1}$ cut and identify the positions of the virtual-state poles. Next, we solve the homogeneous ladder equation and discuss the resulting three-body vertex functions $\Gamma(p)$. We comment on the cut-off dependence of our results by presenting plots for various regularization choices when appropriate.

\subsubsection{Amplitudes on the real axis}

\begin{figure}[t!]
    \centering
    \includegraphics[width=0.95\textwidth]{ 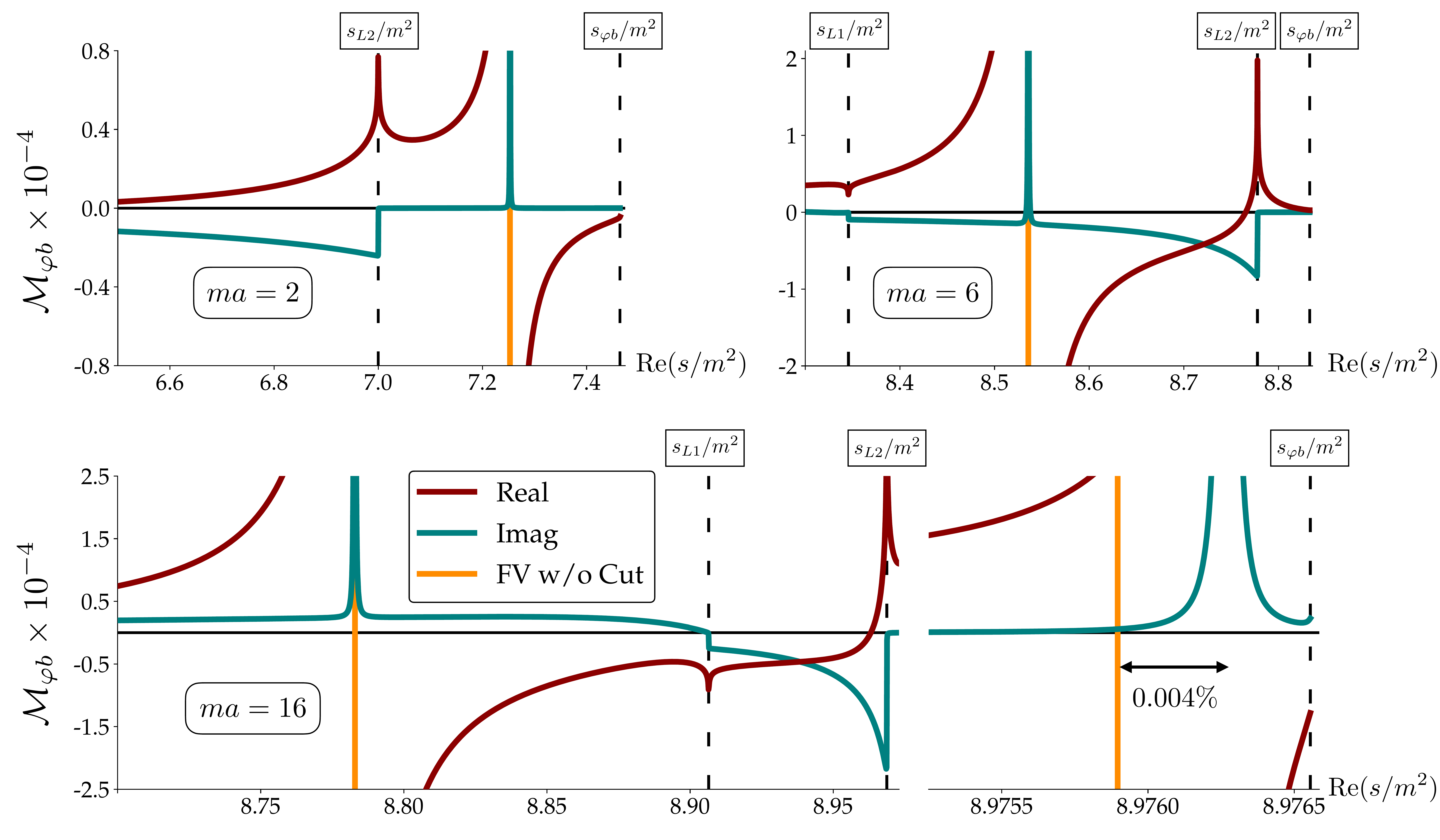}
    \caption{Solution for the $\varphi b$ scattering amplitude evaluated as function of $\re(s) < s_{\varphi b}$ slightly above the real axis, $s = \re(s) + i \delta$, where $\delta = 10^{-5}$. We set the scattering length to $ma = 2,6,16$. The figure shows the real (red) and imaginary (blue) parts of the scattering amplitude. Dashed lines indicate the positions of the branch points, while orange vertical lines highlight the position of the trimer poles found in the FV calculation~\cite{Romero-Lopez:2019qrt}. The amplitude is obtained using the $N=500$ Gauss-Legendre quadrature method described in App.~\ref{app:C}.}
    \label{fig:5-mphib_avg}
\end{figure}

We compute $\Mc_{\varphi b}(s)$ for $s \leq s_{\varphi b}$ and the scattering length $ma= 2, 6, 16$, similarly to Refs.~\cite{Jackura:2020bsk, Romero-Lopez:2019qrt}. To evaluate the amplitude slightly above the real axis, we apply the continuation through the left-hand cut starting at $s_{L1}$, as explained in Sec.~\ref{sec:an-con-circ}. We perform it to uncover possible bound-state poles on the real axis, below this point. We employ the smooth cut-off scheme defined in Eq.~\eqref{eq:cut-off} and the GL method described in App.~\ref{app:C}. 

Typically, we use momentum meshes of size $N=500$. We carefully study the convergence of the result in the mesh size and find our solutions are stable under large variations of $N$. A more detailed analysis of the systematic effects of the integral equation solutions is presented in App.~\ref{app:C}. We note that below the bound-state--spectator threshold, the system is no longer constrained by the usual $S$ matrix unitarity. Therefore, we do not offer a unitarity-based test of the quality of our solution, used in Ref.~\cite{Jackura:2020bsk}.

We show the results for the bound-state--spectator amplitude in Fig.~\ref{fig:5-mphib_avg}. The top-left panel corresponds to the $ma = 2$ case, which describes a deep two-body bound state of mass $m_b = \sqrt{3} \, m$. Corresponding threshold is placed at $s_{\varphi b} /m^2 \approx 7.4641 $, while the ``short" OPE cut branch points take integer values, $s_{L1}/m^2 = 4$, and $s_{L2}/m^2 = 7$. We find a three-body bound-state pole at $s_b / m^2 \approx 7.253$. In Ref.~\cite{Romero-Lopez:2019qrt}, the authors found the trimer energy to be $E/m = 2.693$, which corresponds to the same value of $s$. Their result was obtained using the finite-volume formalism, i.e., by application of the three- and two-body FV quantization conditions to the identical model: three scalar particles with the two-body amplitude $\Mc_2$ controlled by $a$ and $\Kc_{\text{df},3}$ set to zero. We present their result as the orange vertical line. Therefore, we find an excellent agreement with that independent study.

Moreover, in Ref.~\cite{Jackura:2020bsk}, the $\Mc_{\varphi b}(s)$ amplitude was computed as a solution of the same integral equation but for energies above $s_{\varphi b}$, for which no contour deformation was needed. In that work, the NLO effective-range expansion,
  \begin{equation}
  \label{eq:ERE}
  q_b \cot\delta_{\varphi b} = -\frac{1}{b_0} + \frac{1}{2} r_0 q^2_b \, ,
  \end{equation}
was fitted to the outcome leading to parameters $mb_0 \approx 6.4$ and $mr_0 \approx 2.3$. It implies the prediction for the trimer energy $s/m^2 \approx 7.284$, which is just $0.4\%$ from the correct result, and provides a numerical justification for the ERE approximation. It is an expected agreement since the trimer appears above the nearest left-hand branch point, $s_{L2}$, thus within the ERE radius of convergence.

\begin{figure}[t!]
    \centering
    \includegraphics[width=0.95\textwidth]{ 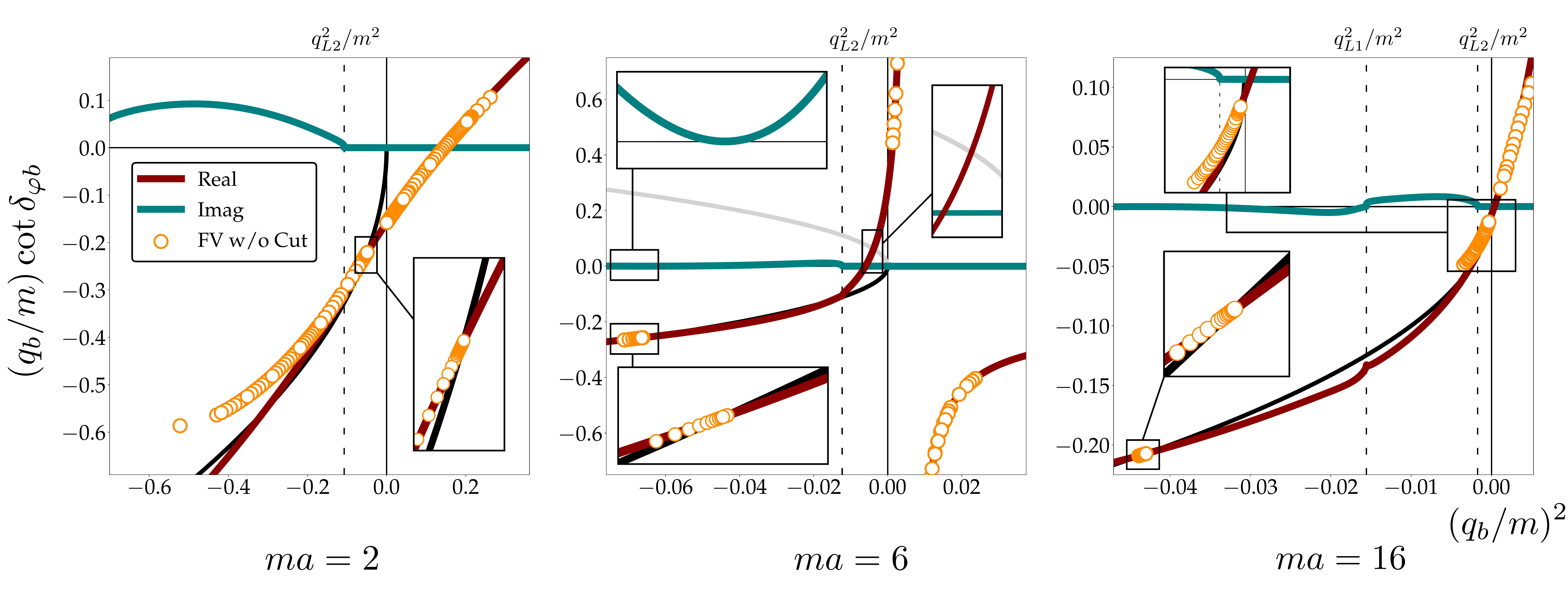}
    \caption{Function $q_b \cot \delta_{\varphi b}$ as a function of $q_b^2$ for fixed values of $ma$. The black line corresponds to the function $-|q_b/m|$. The grey line in the middle panel corresponds to the $+|q_b/m|$ line. Orange points were obtained in the FV computation. We highlight the OPE amplitude branch points with vertical, dashed lines.}
    \label{fig:5-qcotdel}
\end{figure}

Analytic continuation of the $q_b \cot \delta_{\varphi b}$ below the threshold is shown in Fig.~\ref{fig:5-qcotdel}. We present it as a function of $q_b^2$ rather than the total invariant mass $s$ so it can be easily compared to Fig.~7 of Ref.~\cite{Romero-Lopez:2019qrt}. For the $ma=2$ (left panel), the $3\varphi$ threshold corresponds to $(q_{3\varphi}/m)^2 \approx 0.361$, which is the highest value included in the plot, while the bound-state--spectator threshold is placed at $q_b=0$. Momenta associated with the branch points of the OPE amplitude are $(q_{L1}/m)^2 \approx -0.75$ and $(q_{L2}/m)^2 \approx -0.107$. 
 
The condition $q\cot\delta_{\varphi b} = -|q_b|/m$ corresponds to the trimer's pole position in the $q_b$ variable. We see that the real part of the $q\cot\delta_{\varphi b}$ crosses the $-|q_b|/m$ line in two places: first at $(q_1/m)^2 \approx -0.049$ and then at $(q_2/m)^2 \approx -0.311$. The first point corresponds to the already described three-body bound state at $s/m^2 \approx 7.4641$. For the second point, however, $\im(q_b \cot\delta_{\varphi b}) \neq 0$, due to the presence of ``short" OPE cut below the threshold. There is no trimer corresponding to this point.

Furthermore, let us observe that the finite-volume result starts diverging from our analytic solution for points below $(q_{L2}/m)^2$. It is expected since the occurrence of the OPE left-hand cut was not included in the two-body quantization condition used to analyze the FV energies. It leads to a power-law volume dependence effects unaccounted for by the formalism employed in Ref.~\cite{Romero-Lopez:2019qrt}. Our result is an explicit numerical confirmation that the presence of the left-hand cuts invalidates the standard two-body quantization condition, a problem recently pointed out in Ref.~\cite{Green:2021qol} in the context of actual lattice QCD results for the H-dibaryon channel. In Ref.~\cite{Raposo:2023nex}, the authors presented the first attempt to address it theoretically.

We proceed with a discussion of the $ma = 6$ case. The two-body bound state becomes considerably more shallow, with a mass of $m_b \approx 1.972 \, m $. The OPE cut runs from $s_{L1}/m^2 = 8.346$ to $s_{L2}/m^2 = 8.778$. Interestingly, the pole of the amplitude overlaps with the cut of the inhomogeneous term in the ladder equation. We find it at $s_b/m^2 \approx 8.5357$. Fitting the ERE to the physical amplitude, Ref.~\cite{Jackura:2020bsk} found a scattering length $m b_0 \approx -3.6$ (with $r_0$ set to zero). It corresponds to the bound-state pole at $s/m^2 \approx 8.486$, which is $0.6\%$ away from our result obtained by calculating the amplitude below the $s_{\varphi b}$ threshold. We observe that the ERE expansion yields a result deviating from the correct result by a value an order of magnitude worse than in the $ma=2$ case.

In the central panel of Fig.~\ref{fig:5-qcotdel}, we present the $q_b \cot \delta_{\varphi b}$ for the $ma=6$ case. The $3\varphi$ threshold and the OPE branch points are shown respectively at $(q_{3\varphi}/m)^2 \approx 0.037$, $(q_{L1}/m)^2 \approx -0.108$, and $(q_{L2}/m)^2 \approx -0.012$. The real part of the $q_b \cot\delta_{\varphi b}$ crosses the $-|q_b|/m$ line in two places, $(q_1/m)^2 \approx -0.015$ and $(q_2/m)^2 \approx -0.066$. Again we see that for $(q_1/m)^2$, the imaginary part has a finite value whereas, for $(q_2/m)^2$, it is zero (see the insets in the central panel of Fig.~\ref{fig:5-qcotdel}), thus leading to a trimer state. Zero in $\im (q_b \cot_{\varphi b})$ can be understood by inspecting Eq.~\eqref{eq:cot_delta}. Whenever, $\Mc_{\varphi b}$ has a pole, the imaginary part of $\Kc^{-1}$ disappears, since $i q_b$ is real. This behavior is not affected by the presence of the cut. It is interesting to find that the finite volume calculation correctly predicted this pole despite neglecting the cut structure of the OPE. We believe this is caused by the enhancement of the amplitude $\Mc_{\varphi b}$ in the vicinity of the trimer pole which makes the cut presence a negligible effect. It would be interesting to see how well would the FV quantization condition perform in the region $(q_b/m)^2 \in [-0.6,-0.1]$ for which there is no data available.

\begin{figure}[t!]
    \centering
    \includegraphics[width=0.95\textwidth]{ 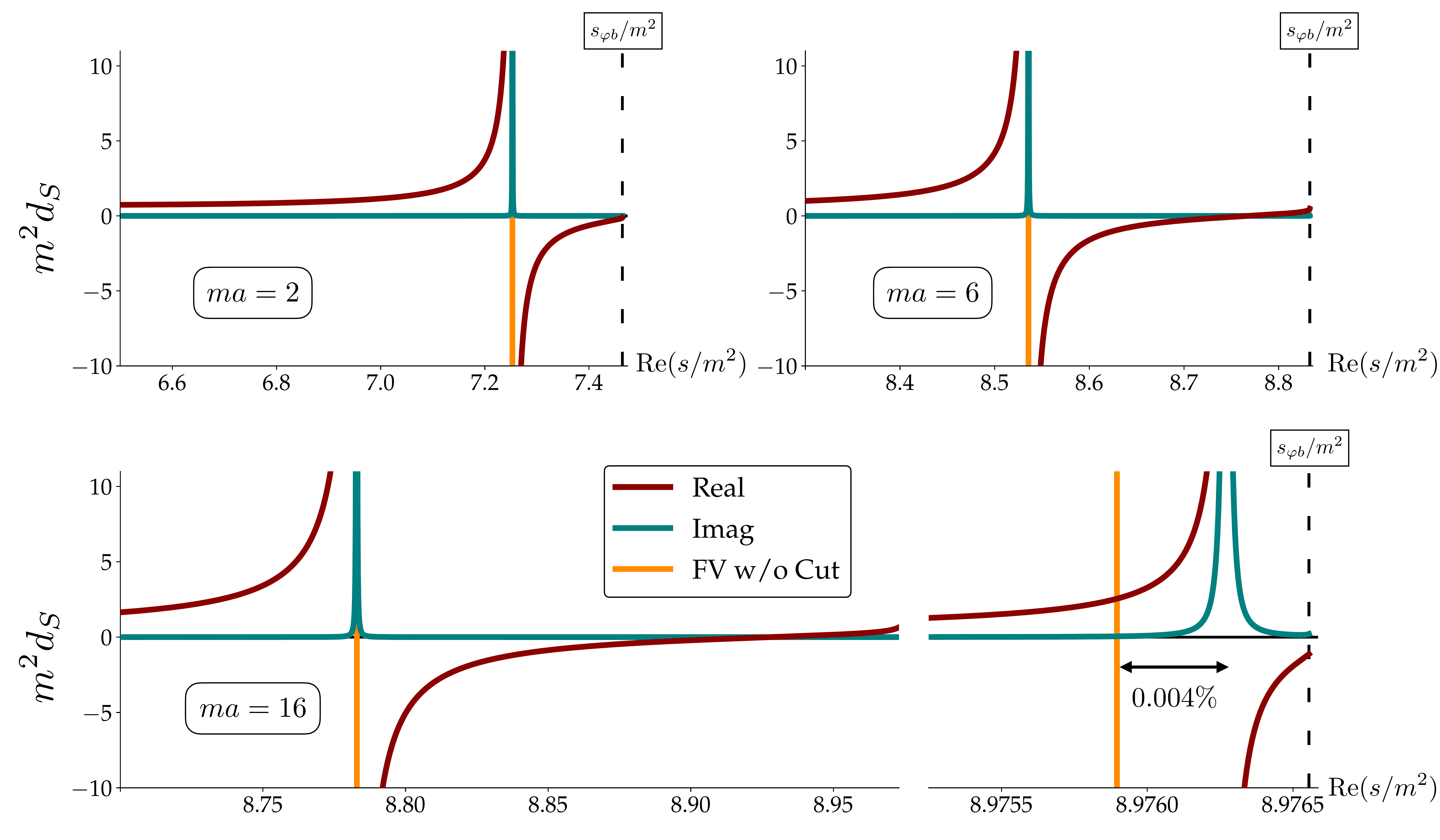}
    \caption{Amputated ladder amplitude, $d_S(s)$ as a function of $s = \re s + i\delta$, for the spectator momenta $(p,k)$ set by fixing the invariant $\sigma_{p,k} = 2m^2$. Three values of two-body scattering length $ma = 2,6,16$ are displayed. The extent of the $\re(s)$ axis is kept the same as of Fig.~\ref{fig:5-mphib_avg} for comparison and $\delta = 10^{-5}$. The amplitude was obtained for the smooth cut-off function and the GL method with $N=500$ using the real integration path.}
  \label{fig:5-dS_plots}
\end{figure}

Finally, we discuss the $ma = 16$ case, for which the two-body bound state becomes very shallow, with a mass $m_b \approx 1.9961 m$. We show the obtained amplitude on the bottom panel of Fig.~\ref{fig:5-mphib_avg}. The short cut goes between $s_{L1}/m^2 \approx 8.9065$ and $s_{L2}/m^2 \approx 8.9686$. The threshold is located $s_{\varphi b}/m^2 = 8.9766$. We find a clear indication of the three-body pole at $s_b/m^2 \approx 8.7829$. This result agrees well with the value $s/m^2 \approx 8.7829$ found for the same scattering length in Ref.~\cite{Romero-Lopez:2019qrt}.

\begin{table}[b]
    \centering
    \begin{ruledtabular}
    \begin{tabular}{cccccc}
       \multirow{2}{*}{$ma$} & \multirow{2}{*}{$s_{\varphi b}/m^2$} & \multicolumn{2}{c}{$s_b/m^2$} & \multicolumn{2}{c}{$s_v/m^2$} \\
           &   & Smooth & Hard & Smooth & Hard \\
       \hline
       $2$ & 7.4641 & 7.2530 & 6.8497 & — & 7.0007 \\
       $6$ & 8.8329 & 8.5357 & 8.3860 & 8.8158 & 8.8257 \\
       $16$ & 8.9766 & 8.7828, 8.9763 & 8.6900, 8.9755 & — & — \\
    \end{tabular}
    \end{ruledtabular}
    \caption{Positions of the bound and virtual states for different values of $ma$ and two choices of the UV regularization. Values of the $\varphi b$ threshold are listed for comparison.} 
    \label{tab:poles}
\end{table}

Reference~\cite{Romero-Lopez:2019qrt} also found a second, shallow trimer at position $s/m^2 \approx 8.9759$. We, too, observe this pole, at $s_b/m^2=8.9763$, which is only a $4.5 \cdot 10^{-3} \, \%$ deviation from the finite-volume result. This sub-percent agreement is emphasized in the bottom panel of Fig.~\ref{fig:5-mphib_avg}.

Near the $\varphi b$ threshold, we fit the amplitude using the  ERE expansion and obtain $mb_0 = 149.17$, $r_0 = 38.73$. This leads to an approximate prediction of the shallow bound state's location of $s_b/m^2 \approx 8.9763$. It is within $10^{-4} \, \%$ the value we obtained in our calculation. As one would expect, the  ERE can not predict the first (deeper) trimer since it breaks down before reaching this pole due to the presence of the OPE short branch cut. 

For a more direct comparison with the finite-volume results, we point the reader to the $q_b \cot \delta_{\varphi b}$ plot for $am=16$ shown on the right panel of Fig.~\ref{fig:5-qcotdel}. The short OPE branch points and the $3\varphi$ threshold are at $q_{L1}^2/m^2 \approx -0.016$, $q_{L2}^2/m^2 \approx -0.002$ and $q_{3\varphi}^2/m^2 \approx 0.005$ respectively. The two trimers correspond to $q^2/m^2 \approx -6.31 \times 10^{-5}$ and $q^2/m^2 = -0.043$. Again, we observe that the FV results do not reproduce the amplitude for momenta between the short OPE cut branch points. As in the case for $am=2$ and $am=6$, the two methods agree in the vicinity of the trimer poles.

Furthermore, to verify our determination of the three-body bound-state poles, we performed an additional computation, in which $d_S(p,k)$ was obtained for external momenta corresponding to fixed $\sigma_p = \sigma_k =2 m^2$. We remind the reader that for this value the ladder equation is solved without contour deformations, i.e., using a straight line in the $q$ variable as an integration path. In this case, the left-hand cuts of $d_S$ in the $s$ variable move far below the near vicinity of the $\varphi b$ threshold. However, the poles corresponding to physical states should still be visible at the same positions, potentially with different corresponding residues. We present the result of this test in Fig.~\ref{fig:5-dS_plots}. One can see excellent agreement both with the FV study and the values obtained from $\Mc_{\varphi b}$. We note that for $ma=6$, the short OPE cut no longer overlaps with the bound-state position, and it is possible to observe the pole presence clearly.

Finally, we repeat these calculations using a hard-cut-off prescription. We do not present the plots for this case since they do not offer any new insight into the behavior of the amplitudes. However, we provide positions of the bound-state poles for both the smooth and the hard cut-off regularizations in Tab.~\ref{tab:poles}.

\subsubsection{The complex plane amplitudes}

\begin{figure}[t!]
    \centering
    \includegraphics[width=0.98\textwidth, trim = {0 0 0 0}, clip]{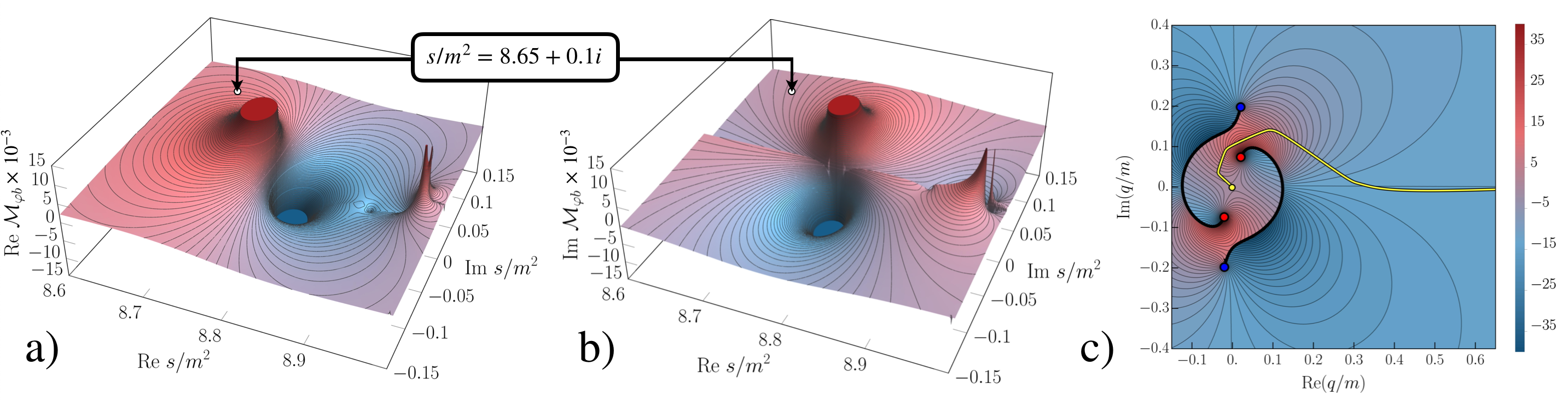}
    \caption{Amplitude $\Mc_{\varphi b}(s)$ as a function of the complex total invariant mass squared for $ma=16$. We observe the emergence of two trimer poles and the cut structure announced in Sec.~\ref{sec:analytic-continuation-A}. Panels \textbf{(a)} and \textbf{(b)} present real and imaginary parts, respectively. The left-hand cut, starting at $s_\circ$, is aligned with the real axis. We present the integration contour corresponding to such a choice in panel \textbf{(c)}. This panel presents the imaginary part of the OPE amplitude, $m^2 \im G(q,q_b)$, with its cuts highlighted by the black lines. It corresponds to the kinematic point $s/m^2 = 8.65 + 0.1i$ highlighted on panels (a) and (b). At this point, the amplitude has value $\Mc_{\varphi b} = 3021 + 746.6 i$.} 
    \label{fig:5-3D-amplitude-2}
\end{figure}

\begin{figure}[b]
    \centering
    \includegraphics[width=0.98\textwidth, trim = {0 5 0 1}, clip]{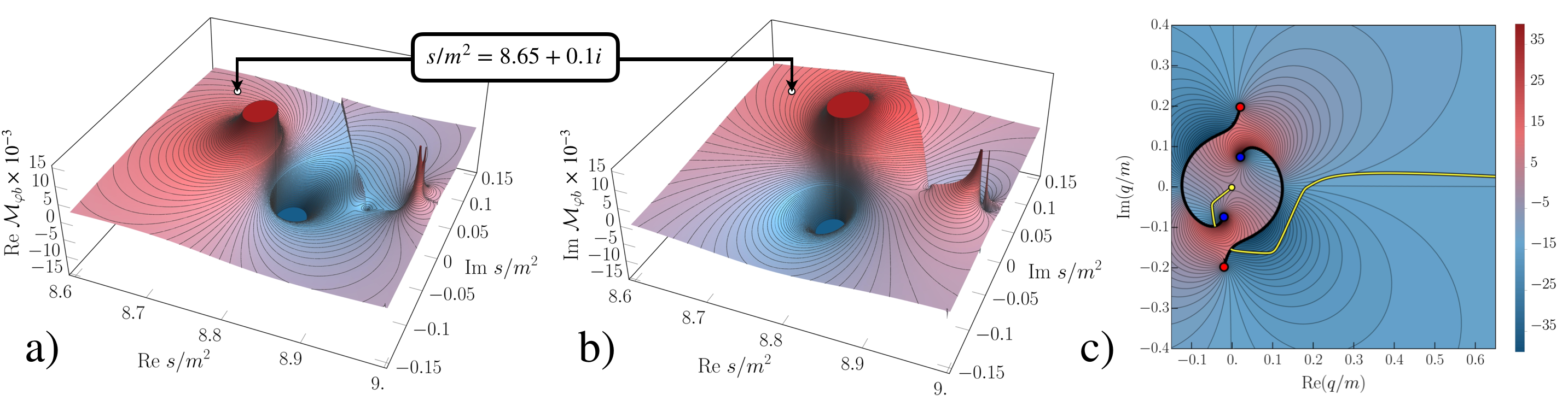}
    \caption{Same as Fig.~\ref{fig:5-3D-amplitude-2} but with the left-hand cut starting at $s_\circ$ rotated away from the real axis. We achieve it by using the integration contour presented in the panel \textbf{(c)}. At the point $s/m^2 = 8.65 + 0.1i$ the amplitude has value $\Mc_{\varphi b} = 2284 + 4106i$.}
    \label{fig:5-3D-amplitude}
\end{figure}

Now we discuss an extension of our result to the complex $s$ plane. In principle, the solution method is the same as for the amplitude evaluated slightly above the real axis. For the increasing imaginary value of $s$, one needs to deform the integration contour according to the motion of the singularities of the integration kernel. As explained in Sec.~\ref{sec:analytic-continuation}, continuous change from $\im s \leq 0$ to $\im s > 0$ may result in a discontinuity in the integration, e.g., related to the reflection of the cuts of the OPE amplitude. It manifests as a left-hand cut of the amplitude $\Mc_{\varphi b}$ starting at $s_\circ$. Following the prescription described in Sec.~\ref{sec:an-con-circ}, this cut can be rotated into the complex plane.

The amplitude $\Mc_{\varphi b}$ in the complex $s$ plane for $ma=16$ is presented in Fig.~\ref{fig:5-3D-amplitude-2}. It was obtained using the GL method with a mesh of $N=250$ nodes and the smooth cut-off choice in the OPE amplitude definition. In addition to the two poles already identified in the previous paragraphs, we observe additional singularities. The fixed, ``short" OPE cut from the inhomogeneous part of the equation is aligned with the real axis, running between $s_{L1}$ and $s_{L2}$ branch points. If needed, one can continue the amplitude through that cut by the deformation of the integration contour in the $x$ variable in Eq.~\eqref{eq:Gs_proj}, as described at the end of Sec.~\ref{sec:analytic-structure}.

\begin{figure}[t]
    \centering
    \includegraphics[width=0.99\textwidth, trim = {1 3 0 1}, clip]{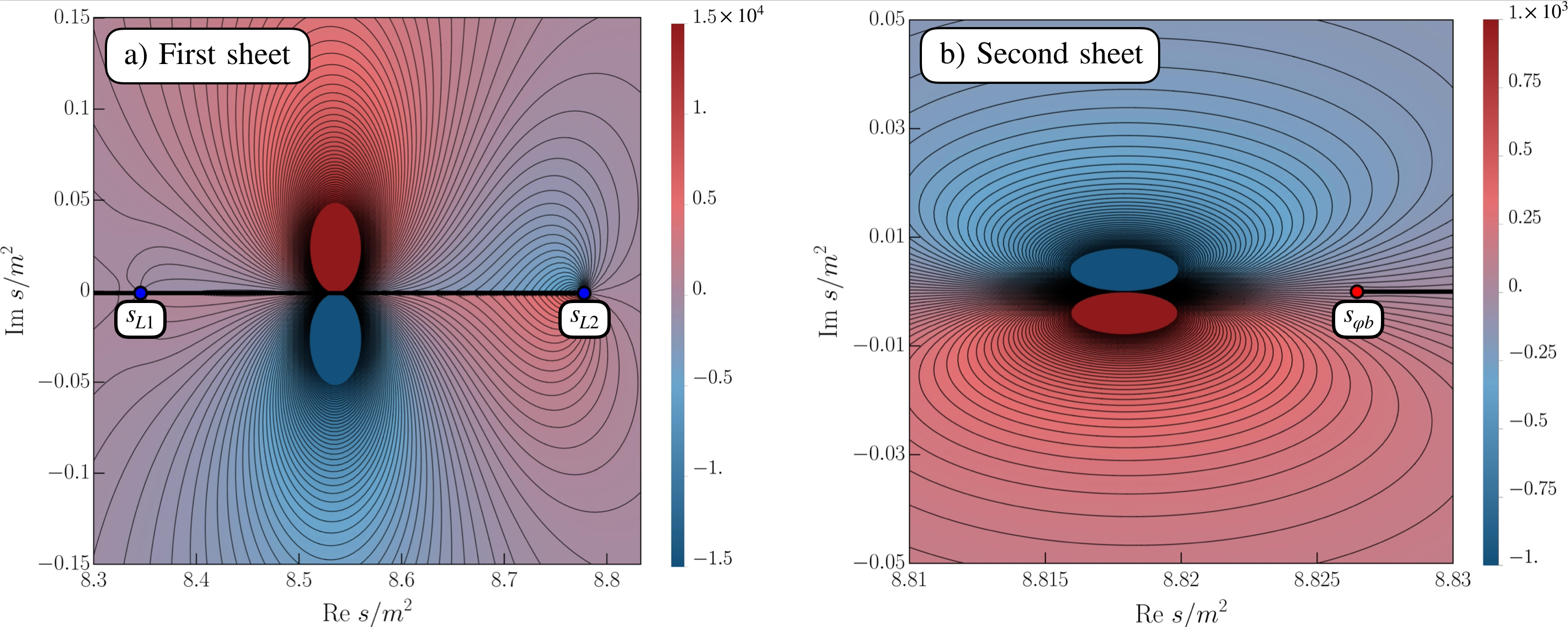}
    \caption{Imaginary part of the ${\varphi b}(s)$ amplitude for $ma=6$ and the smooth cut-off choice. The left panel shows the amplitude on the first Riemann sheet, $\Mc_{\varphi b}$, in the vicinity of the ``short" OPE cut and the trimer pole. The right panel presents the second-sheet amplitude, $\Mc_{\varphi b}^{\rm II}$, in the vicinity of the $\varphi b$ threshold and the virtual-state pole. We highlight the cuts with black lines.}
    \label{fig:5-3D-amplitude-3}
\end{figure}

Moreover, in panels (a) and (b) of Fig.~\ref{fig:5-3D-amplitude-2}, we see a 
left-hand cut starting at $s_\circ$, placed on the real axis. Panel (c) presents the corresponding integration contour circumventing the OPE cut from the top. The resulting cut structure agrees with the description of Sec.~\ref{sec:analytic-continuation-A}. We note that the parabola-like cuts can not be seen on the presented plots, as these are the cuts through which we continued the amplitude down to the smaller values of $\re s$, i.e., from Region 1 to Regions 2 and 3 of the complex $s$ plane, as shown in Fig.~\ref{fig:mphib-strucutre}.

For comparison, in Fig.~\ref{fig:5-3D-amplitude}, we present a result of continuing the lower-half amplitude through the left-hand cut. The rotation of the cut is performed according to the prescription of Sec.~\ref{sec:an-con-circ}. It ensures the unphysical singularity does not coincide with the trimer pole. In panel (c), we present a corresponding integration contour that allows for the analytic continuation through that cut for an example value of $s/m^2 = 8.65 + 0.1i$.

\begin{table}[b]
    \centering
    \begin{ruledtabular}
    \begin{tabular}{ccccc}
       $ma$ & \multicolumn{2}{c}{$|\zeta(q_b)|^2$} & \multicolumn{2}{c}{$|\Gamma_{\varphi b}|^2/m^2$} \\
       & Smooth & Hard & Smooth & Hard \\
       \hline
       $2$ & 0.923 & 2.289 & 321.4 & 797.1 \\
       $6$ & 6.257 & 7.945 & 826.9 & 1050 \\
       $16$ & 12.60, \, 0.1532 & 14.16, \, 0.4000 & 632.1, \, 7.686 & 710.2, \, 20.07 \\
    \end{tabular}
    \end{ruledtabular}
    \caption{Three-body vertex factors for the $\varphi b$-to-trimer state for two cut-off choices and scattering lengths $ma=2,6$, and $16$.}
    \label{tab:residue_mphib}
\end{table}

Having determined the amplitude in the complex $s$ plane, one may analytically continue it 
to the unphysical Riemann sheet of the right-hand cut starting at the $\varphi b$ threshold. It can be done straightforwardly by using unitarity and Eq.~\eqref{eq:phi-b-second-sheet} or by appropriately deforming the integration contour to avoid the $\Mc_2$ dimer pole, as explained in Sec.~\ref{sec:above-thresholds}. Having computed the amplitude on the second sheet, we now seek the virtual states.

An example plot of the second-sheet amplitude $\Mc_{\varphi b}^{\rm II}$ can be seen in Fig.~\ref{fig:5-3D-amplitude-3}. There, we plot the amplitude for $ma=6$ near the bound-state pole (left panel) and virtual-state pole (right panel). The virtual state can be also identified on the central panel of Fig.~\ref{fig:5-qcotdel} as the point where the amplitude crosses $+|q_b/m|$ line. This happens at $(q_v/m)^2 = -0.0016$. Considering pole trajectories as functions of $a$, it is possible to identify every bound-state pole on the physical Riemann sheet as a virtual state that crossed the threshold and ``escaped" the unphysical Riemann sheet through the unitarity cut. As we increase the two-body scattering length, $a$, we find the virtual state moves to the right, closer to the threshold $s_{\varphi b}$ and the virtual state of the $ma=6$ system becomes the second, shallow bound state found in the $ma=16$ case.

The positions of the identified virtual states are provided in Tab.~\ref{tab:poles} for both the smooth and hard cut-off functions. We look for those poles in the region $s_{L2} \leq s \leq s_{\varphi b}$ by solving Eq.~\eqref{eq:5-second-sheet-resonance}. We do not see any virtual states below the branch point $s_{L2}$—an indication that they escape the second Riemann sheet through the ``short" OPE cut to further sheets of the scattering amplitude. Using the smooth regularization prescription, we find only one virtual state—in the $ma=6$ case. For the hard cut-off, there is an additional state in the $ma=2$ case, right above the $s_{L2}/m^2 =7$ point. In both cases, we do not find virtual states for $ma=16$. Moreover, we do not see evidence of nearby resonances for these values of the scattering lengths.

\subsubsection{Three-body bound-state vertex functions}

Here we discuss the solutions of the homogeneous ladder equation that we use to compute residues of the amplitude at the three-body bound-state poles. In Eq.~\eqref{eq:dS_pole}, the residue is given by $\zeta(p) \zeta^*(k)$, and in Eq.~\eqref{eq:Gamma_vertex} it is related to the residue of $D_S(p,k)$, $\Gamma(p)\Gamma^*(k)$. The vertex factor of the $\Mc_{\varphi b}$ amplitude is defined in Eq.~\eqref{eq:phi_b_vertex}.

We calculate the vertex factors corresponding to the $\varphi b$-to-trimer state for three different two-body scattering lengths, $ma=2,6,16$. We use Eq.~\eqref{eq:homo-eq}. We set the external spectator momenta at the two-body bound state pole, $p=q_b$, and look for solutions of the eigenvalue equation at the trimer pole $s=s_b$, tabulated in Tab.~\ref{tab:poles}. Since setting $p=q_b$ makes singularities of the kernel cross the integration path, as discussed in Sec.~\ref{sec:analytic-continuation}, it is necessary to use a self-consistent, deformed contour $\Cc$. As a result, we obtain vertex factor $\zeta(p)$ for complex momenta $p \in \Cc$. Knowledge of this function along the contour allows for extrapolation to $p=q_b$. Note that values of $\zeta$ obtained this way are determined up to a multiplicative constant. Before the extrapolation, we fix the normalization of the vertex function by computing the value of the residue of the ladder amplitude $d_S(p',p')$ at $s=s_b$ and some $p' \in \Cc$. It is done by performing a simple linear fit to the $1/\re(d_S)$ function at this kinematic point. Resulting values of $|\zeta(q_b)|^2$ and $|\Gamma_{\varphi b}|^2$ are provided in Tab.~\ref{tab:residue_mphib}.

\begin{figure}[t]
    \centering
    \includegraphics[width=0.95\textwidth]{ 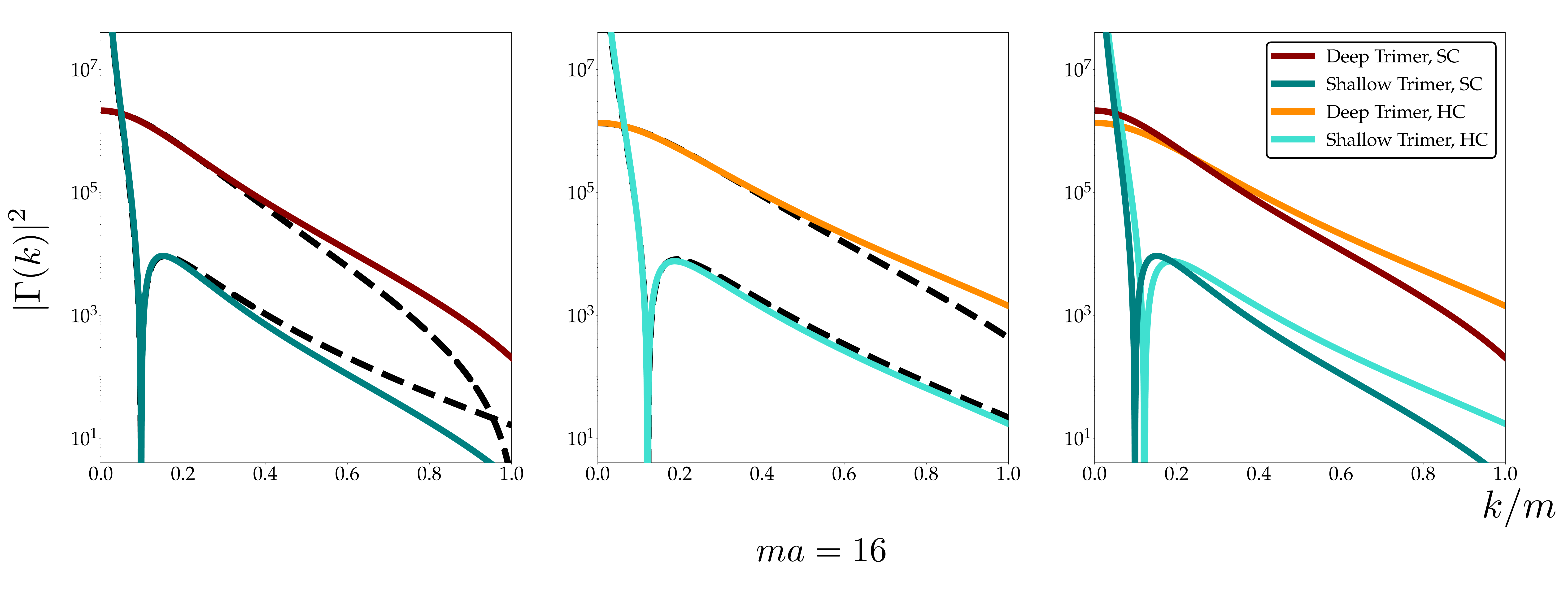}
    \caption{Vertex functions modulus squared $|\Gamma(k)|^2$ is plotted as a function of spectator momentum $k$ for two body scattering length $ma=16$. Two different UV regularization schemes are used. SC and HC denote smooth cut-off and hard cut-off, respectively. The black dashed line corresponds to the fit performed using $|\Gamma_{\text{NR}}(k)|^2$ predictions.}
    \label{fig:5-vertexfunctions}
\end{figure}

We also solve the homogeneous equation for the vertex function $\Gamma(k)$ considered as a function of arbitrary spectator momentum $k$. Inspecting the kernel of the homogeneous equation, we find that the OPE cut does not intersect the integration interval $q=[0,q_{\rm max}]$ if the desired external spectator momentum is real, $k \in [0,q_{\rm max}]$, and we set the total invariant mass to $s_b$ for the three considered values of $ma$. Thus, in this case, no contour deformation is needed to solve the homogeneous equation. The solutions are shown in Fig.~\ref{fig:5-vertexfunctions} for the two-body scattering length, $ma=16$, along with two choices of the UV regularization scheme. These vertex factors describe the coupling between the trimer and the three-particle state. The coupling becomes maximum when the spectator momentum $k\approx 0$. It decreases exponentially as the spectator momentum increases.

\begin{table}[b]
    \centering
    \begin{ruledtabular}
    \begin{tabular}{lccc}
       UV regularization & Trimer position $s_b/m^2$ & $|A|^2$ & $s_0$  \\
       \hline
       Smooth Cut-off (SC) & $8.7828$ (deep) & 2.68 & 1.37 \\
        &  $8.9763$ (shallow) & 6.75 & 0.98 \\
       \hline 
        Hard Cut-off (HC) & $8.6900$ (deep) & 2.32 & 1.32 \\
        & $8.9756$ (shallow) & 4.93 & 1.14 \\
    \end{tabular}
    \end{ruledtabular}
    \caption{Fit parameters for the NR vertex factors, Eq.~\eqref{eq:NR-vertex}, for $ma=16$.} 
    \label{tab:vertex_functions_mphib}
\end{table}

It is consistent with the expectation from the non-relativistic (NR) result in the unitary limit ($a\to \infty$), which was derived analytically in Ref.~\cite{Hansen:2016ync} and reproduced numerically in Ref.~\cite{Briceno:2018mlh}. In Fig.~\ref{fig:5-vertexfunctions}, we present a fit of our numerical result to the analytic form, 
    \begin{align}
    \label{eq:NR-vertex}
    |\Gamma_{\rm NR}(k)|^2
    =|c||A|^2 \frac{256\pi^{5/2}}{3^{1/4}} \frac{m^2\kappa_{\rm NR}^2}{k^2(\kappa_{\rm NR}^2+3k^2/4)} 
    \frac{\sin^2 \Big(s_0 \sinh^{-1} \big(\sqrt{3}k/2\kappa_{\rm NR} \big) \Big) }{\sinh^2(\pi s_0/2) } \, .
    \end{align}
Here, $\kappa_{\rm NR}$ is fixed by the energy of the system, $\sqrt{s} = 3m - \kappa_{\rm NR}^2$. In the unitary limit, two of the other parameters are fixed to be $s_0 = 1.00624$  and $|c| = 96.351$, while $A$ is expected to be close to $1$ in the unitary limit. 

Given that the results presented here lie sufficiently far from the unitary limit, we leave $s_0$ as a free parameter. We observe that modifying the definition of $\kappa_{\rm NR}$ to be $\sqrt{s} = \sqrt{s_{\varphi b} } - \kappa_{\rm NR}^2$ leads to a better description of $|\Gamma(k)|^2$ for these scattering lengths. Although this modification is no more than an empirical observation, it is reasonable given that for a finite scattering length, there are two thresholds, $s_{\varphi b}$ and $(3m)^2$. The closest one to the trimer is $s_{\varphi b}$, which could explain why the $\Gamma$ should be more sensitive to this threshold. In the unitary limit these two thresholds, of course, collapse onto each other.

By fitting $s_0$ and $A$ in the small $k/m \in [0,0.2]$ region, one can find qualitative similarities between the numerical results presented in Fig.~\ref{fig:5-vertexfunctions} and this functional form. The fit parameters are listed in Tab.~\ref{tab:vertex_functions_mphib}. The similarities are more striking for small values of $k$. As expected, this functional form fails to describe the whole range of momenta.

\section{Conclusions}
\label{sec:conclusions}

In this work, we discussed an analytic continuation of the bound-state--spectator amplitude, $\Mc_{\varphi b}$, below the threshold and to the complex energies, generalizing the study of Ref.~\cite{Jackura:2020bsk}. The amplitude is obtained from the relativistic three-body on-shell integral equation, considered in the ladder approximation and the $S$ partial wave only. The solution of the equation is reduced to the dimer-particle amplitude via the LSZ formula for the bound-state systems and studied as the function of a single complex variable, the total invariant mass $s$.

The three-body reaction amplitudes exhibit a more complicated analytic structure than their two-body equivalents. The additional complications are related to the contribution of the long-range, physical one-particle exchanges to the overall interaction. To understand this aspect of the model, we analyzed the analytical structure of the ladder equation in the kinematical region relevant to the study of bound-state physics. We found that the three-body equations are characterized by singularities that cross the integration interval forcing the deformation of the integration path into the complex plane. In particular, the logarithmic discontinuities of the OPE amplitude can form into a circular cut for a range of energies below the $\varphi b$ threshold. 

We explained how to analytically continue the integral equation via the combination of the contour deformation and explicit inclusion of the kernel discontinuities. As we explained, one can not use arbitrary integration paths and has to ensure a self-consistent choice, which defines the smooth continuation of the ladder amplitude to the domain of analyticity. To that end, we defined suitable integration contours that circumvent the relevant cuts and proposed a general scheme of the solution procedure. We presented a method to rotate unphysical left-hand cuts that allows one to extract the trimer pole positions and their residues. The discussion of analytic properties was supplemented by a description of numerical methods for solving the problem of interest. They rely on the replacement of the integral equation of interest with an algebraic system of $N$ equations. In addition to providing particular numerical routines, we discuss systematic effects and potential improvements of our techniques. We find that the computational procedures we use yield stable and reliable results for relatively small values of $N$.

Finally, we presented solutions for the ladder amplitude, $d_S$, and the dimer-particle amplitude $\Mc_{\varphi b}$ for three cases, $ma=2,6,16$ and found agreement with the finite-volume results of Ref.~\cite{Romero-Lopez:2019qrt} and the LO effective-range expansion of Ref.~\cite{Jackura:2020bsk}. We identified the three-body bound state poles at energies predicted by the finite volume formalism, together with associated trimer-to-$\varphi b$ couplings. We discussed the continuation of the $\varphi b$ amplitude to the complex energy plane and the unphysical sheet through the two-body unitarity cut to investigate the presence of the virtual-state poles. 

Nevertheless, our formal and numerical framework allows for a relatively simple application in future lattice QCD computations that will involve genuine resonances. Presented methods can be implemented in the procedure of analytic continuation through the three-body threshold cut to the Riemann sheets where the three-body resonances reside. It is possible to extend our analysis to systems where the two-body bound-state sub-channel is resonant instead and to higher partial waves. Although technically more complex, these cases are characterized by the same logarithmic cuts of the OPE amplitude and the analysis of Secs.~\ref{sec:analytic-structure} and~\ref{sec:analytic-continuation} remains unaltered. Continued studies in this direction will enable the extraction of the three-body resonances from the Lattice QCD.


\section{Acknowledgements}

The authors would like to thank J. Baeza-Ballesteros and F. Romero-L\'opez for pointing out the issue of the complex cut-off extensions, and A. Jackura, S. Sharpe, and A. Szczepaniak for many useful discussions. SMD is supported by U.S. Department of Energy Contract no.~DE-SC0011637. RAB and MHI acknowledge the support of the USDOE Early Career award, contract DE-SC0019229. MHI acknowledges the support from Jefferson Science Associates/Jefferson Lab graduate fellowship program.  


\appendix

\section{Ladder equation in terms of Lorentz invariants}
\label{app:B}

In this work, we presented the ladder equation using the momentum representation, i.e., considering the spectators' momenta, $(p,k)$ as kinematic arguments describing the $S$-wave scattering process. Equivalently, one may analyze it using the final and initial invariant mass squared of pairs, $\sigma_{p}, \sigma_{k}$. In practical applications, we find that the momentum representation proves more useful in the study of analytic continuation. It is because the OPE cuts take a simpler shape in this form. They wrap around the origin of the complex $q$ plane and have associated parity copies allowing for less problematic choices of the deformed integration contours. On the other hand, the invariants-space OPE cuts follow the movable upper integration limit and have a more complicated, fishing-hook-like shape. 

However, in some cases, the invariant-space equations are simpler to manipulate. One such case is a derivation of the positions of the OPE branch points. Ultimately, it is desirable to have two representations since one can prove more useful than the other in analyses concerned with different physical systems and the LQCD data. In particular, bound-state and resonance poles occur in $\Mc_2$ at fixed values of the two-body invariant mass, making it an intuitively better variable to consider. Moreover, the variables $\sigma_p, \sigma_k$ are Lorentz invariants and do not change for different values of the total invariant mass $s$ in contrary to momenta $(p,k)$. In this appendix, we concisely present the invariants representation of the ladder equation focusing on the analytical structure of the building blocks of the equation.

The $S$-wave projected ladder equation, Eq.~\eqref{eq:d_Sproj}, is written in terms of Lorentz invariants $\sigma_{p}, \sigma_k$, as,
    \beq
    \label{eq:App-d_Sproj}
    d_S(\sigma_{p}, \sigma_k) = - G_S(\sigma_{p}, \sigma_k) -
    \int\limits_{0}^{\sigma_{\rm max}} 
    d\sigma_q \, K(\sigma_{p}, \sigma_q) \, d_S(\sigma_q, \sigma_k) \, .
    \eeq
Variable $\sigma_q$ is the invariant mass squared of the intermediate pair in the OPE process. The integration kernel is,
    \beq
    K(\sigma_{p}, \sigma_q) = \frac{1}{2 \pi} G_S(\sigma_{p}, \sigma_q) \tau(s,\sigma_q)  \Mc_2(\sigma_q) \, ,
    \eeq
where the implicit $s$ dependence is assumed. The integration is performed in the interval $[0 , \sigma_{\rm max}]$, where $\sigma_{\rm max} = (\sqrt{s} -m)^2$. It corresponds to $q = 0$, while $\sigma_q = 0$ corresponds to $q = q_{\rm max}$ in the integral of Eq.~\eqref{eq:d_Sproj}.

The integration kernel contains three objects. The three-body phase space is,
    \beq
    \label{eq:three-body-phase-space}
    \tau(\sigma_q) = \frac{\lambda^{1/2}(s,\sigma_q, m^2)}{ 8\pi s} \, .
    \eeq
It has an explicit pole at $s = 0$ and the branch points at $\sigma_{\tau 1} = (\sqrt{s}-m)^2$ and $\sigma_{\tau2} = (\sqrt{s}+m)^2$. We orient both associated cuts to the right. In particular, for real $s$, this results in a single branch cut running between the two branch points. We note that the upper integration limit coincides with the former branch point. The two-body amplitude is given in Eq.~\eqref{eq:M2_general} as a function of $\sigma_q$. As can be seen, it has a left-hand cut at $\sigma_q = 0$ and a right-hand cut at $\sigma_q = 4 m^2$ required by the unitarity. It also develops a pole on the first complex sheet at $\sigma_b$. The $\Mc_2$ amplitude can be rewritten in a ``propagator" form that makes the presence of the pole explicit,
    \beq
    \label{eq:bound-state-amplitude}
     \Mc_2(\sigma_q) = \frac{R(\sigma_q)}{\sigma_q - \sigma_{b} - i \epsilon_b } \, ,
    \eeq
where the residue,
    \beq
    R(\sigma_q) = -(32 \pi)^2 \, \sigma_q \left(\Kc_2^{-1} + i \rho \right) \, .
    \eeq
In Eq.~\eqref{eq:bound-state-amplitude}, we included infinitesimal $i\epsilon_b$ in the denominator (different than the $i \epsilon$ in the OPE amplitude) to shift the pole position above the real $\sigma_q$ axis. It is necessary when solving for the physical amplitude, as discussed in Ref.~\cite{Jackura:2020bsk}, and is equivalent to the integration contour deformation. For $s = s_{\varphi b}$, the upper limit of the integration coincides with the pole, leading to the unitarity branch point in the ladder solution $d_S(\sigma_p,\sigma_k)$. For $\re s < s_{\varphi b}$, the integration interval does not coincide with the singularities of $\Mc_2$.

\begin{figure}[t]
    \centering
    \includegraphics[width=0.9\textwidth]{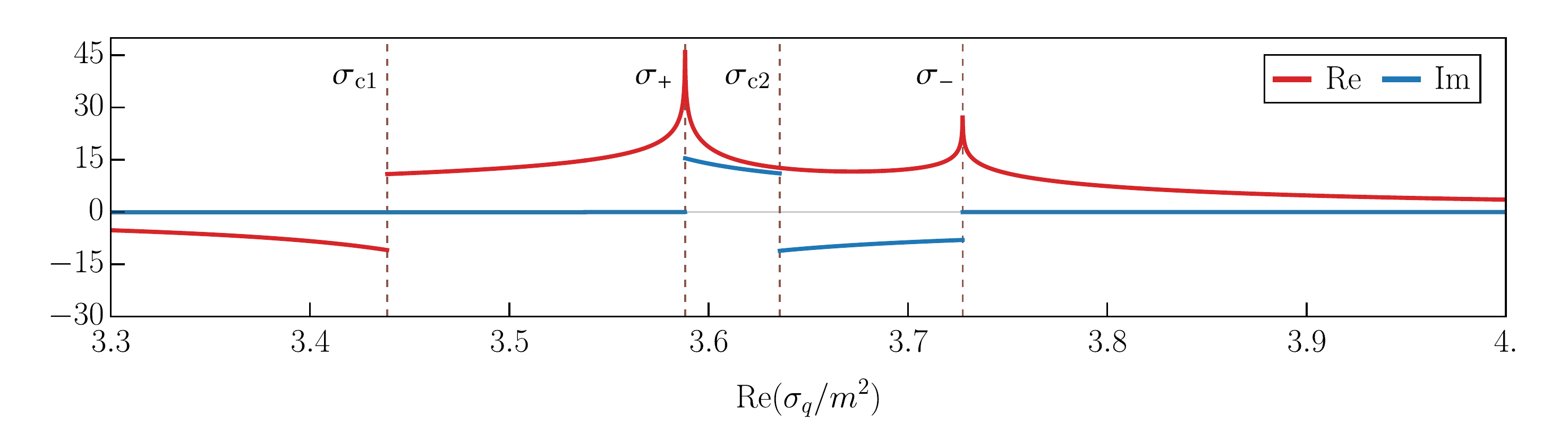}
    \caption{The OPE amplitude $G_S(\sigma_p, \sigma_q + i \delta)$ in units of $1/m^2$, evaluated slightly above the real $\sigma_q$ axis. The kinematic parameters are $s/m^2=8.3$ and $\sigma_{p} = \sigma_b$ for $ma = 16$. Infinitesimal $\delta = 10^{-4}$. Singularities described in the text and defined in Eqs.~\eqref{eq:sigma-plus-minus},~\eqref{eq:sigma_c1},~\eqref{eq:sigma_c1} are highlighted with dashed lines.}
    \label{fig:appA-OPE}
\end{figure}

\begin{figure}[t]
    \begin{center}
    \includegraphics[width=0.95\textwidth]{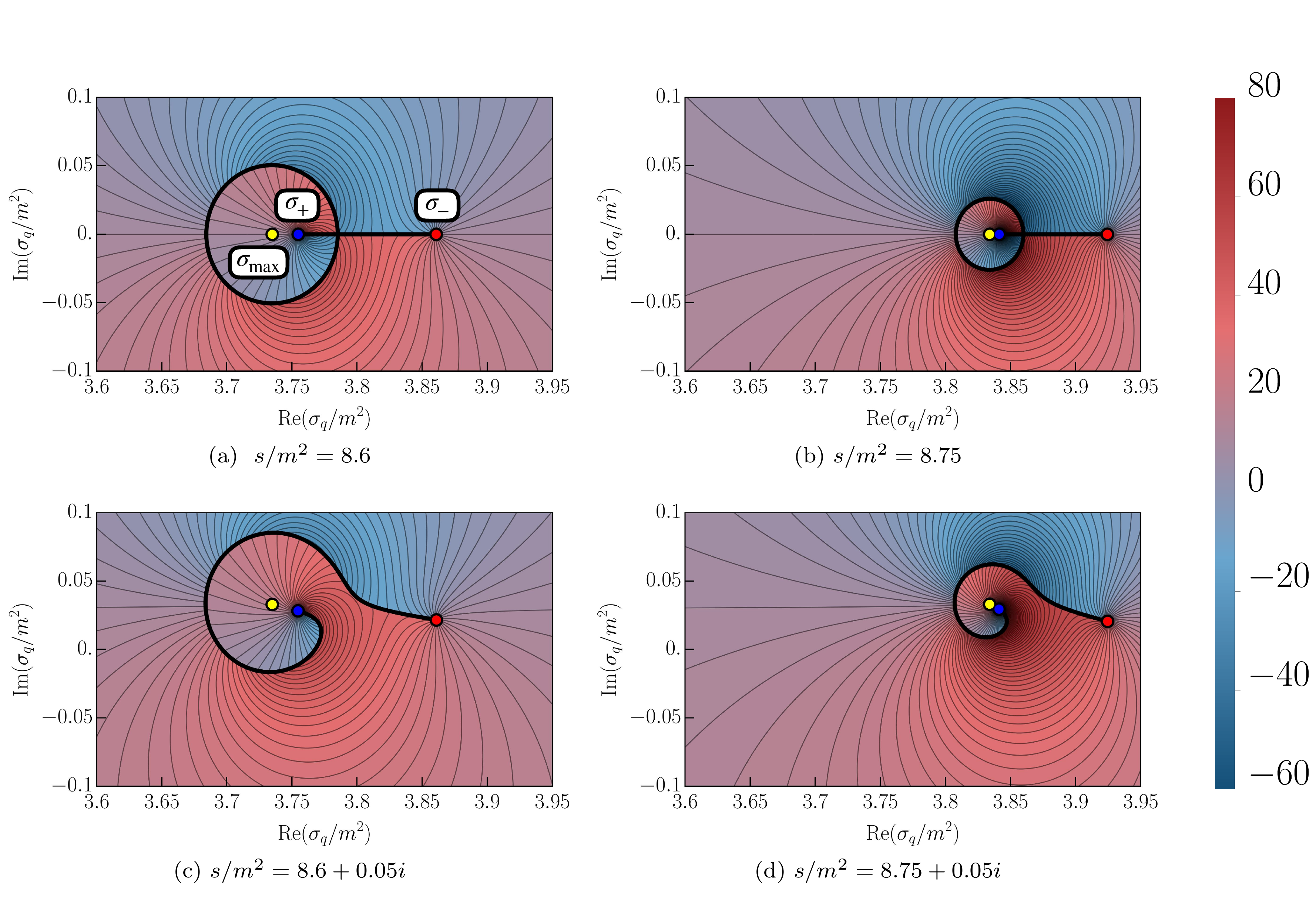}
    \caption{Cut structure of the imaginary part of the OPE, $\im G_S(\sigma_{q}, \sigma_k)$, in the complex $\sigma_q$ plane for typical value of kinematical variables. We use hard cut-off and plot the amplitude in units of $1/m^2$. We set $ma=16$ and consider $\re s < s_{\varphi b}$. Two-body invariant mass, $\sigma_k = \sigma_b \approx 3.984 m^2$. The red circle corresponds to $\sigma_-$ and the blue to the $\sigma_+$ branch point. The OPE cut (black curve) runs between these two points. We indicate the upper integration limit $\sigma_{\rm max} = (\sqrt{s}-m)^2$ by a yellow point. It is enclosed by the cut and contour deformation is required. We present four cases: \textbf{(a)} Purely real kinematical parameters, for which the cut takes a shape of a circle crossed by a line; \textbf{(b)} Purely real kinematical parameters, with $s$ closer to $s_{\varphi b} \approx 8.976 m^2$, for which the cut moves the right and shrinks; \textbf{(c)} The non-zero, positive imaginary part of $s$, for which the cut opens, moves upward and resembles a fishing hook; \textbf{(d)} Same as previously, but for larger $\re s$, the structure shrinks and $\sigma_+$ approaches $\sigma_{\rm max}$. In particular, when $\sigma_{k} = m( m + \sqrt{s})$, the OPE branch point coincides with the upper integration limit.}
    \label{fig:circular-cut-invariants}
    \end{center}
\end{figure}

The $S$-wave projection of the OPE amplitude is,
    \beq
    \label{eq:OPE-invariants}
    G_S(\sigma_{p}, \sigma_k) = - \frac{s \, H(\sigma_p,\sigma_k)}{ \lambda^{1/2}(s, \sigma_p, m^2) \lambda^{1/2}(s, \sigma_k, m^2) } 
    \log \left(\frac{ z(\sigma_p,\sigma_k) - \lambda^{1/2}(s, \sigma_p, m^2) \lambda^{1/2}(s, \sigma_k, m^2) }{ z(\sigma_p,\sigma_k) + \lambda^{1/2}(s, \sigma_p, m^2) \lambda^{1/2}(s, \sigma_k, m^2) } \right) \, ,
    \eeq
where the $z(\sigma_p,\sigma_k)$ function is defined as,
    \beq
    \label{eq:z-function}
    z(\sigma_{p}, \sigma_k) = 2 \, s\, (\sigma_{k} + i \epsilon) - (s + \sigma_{k} - m^2)(s + m^2 - \sigma_{p} )  \, .
    \eeq
This representation follows from Eq.~\eqref{eq:Gs_proj}, where one expresses the external spectator's momenta through the relation given in Eq.~\eqref{eq:momentum_invs}. The function $H(\sigma_{p'}, \sigma_{p})$ is the smooth/ hard regularization scheme, as described in the paragraph containing Eq.~\eqref{eq:cut-off}. In Fig.~\ref{fig:appA-OPE}, we present $G_S(\sigma_p, \sigma_q)$ as a function of real-valued $\sigma_q$, choosing the smooth cut-off, defined in Eq.~\eqref{eq:cut-off}.

The OPE amplitude, Eq.~\eqref{eq:OPE-invariants}, considered as a function of $\sigma_p$ for fixed $s$ and $\sigma_k$, has two logarithmic branch points connected with a cut. Its parametrization is obtained from the condition,
    \beq
    \label{eq:invariants-pole-condition}
    z(\sigma_p,\sigma_k) + x \, \lambda^{1/2}(s, \sigma_p, m^2) \lambda^{1/2}(s, \sigma_k, m^2)  = 0 \, ,
    \eeq
which is an equation satisfied by the pole positions of the integrand in the right-hand side of Eq.~\eqref{eq:Gs_proj}. Solving for $\sigma_p$ yields,
    \beq
    \label{eq:sigma-space-OPE-cut}
    \sigma_{\text{cut},\pm}(s,\sigma_k,x) = (s+m^2) + \frac{ 2 s \sigma_k (s+\sigma_k-m^2) \pm \sqrt{s x^2 \lambda(s,\sigma_k, m^2)} \sqrt{4 m^2 B_x - \sigma_k B_1} }{B_x} \, ,
    \eeq
where function
    \beq
    B_x \equiv B_x(s,\sigma_k) = (x^2-1) \lambda(s,\sigma_k, m^2) - 4 s \sigma_k \, .
    \eeq
We have set $\epsilon=0$. The above formula is analogous to the momentum-space parametrization of Eq.~\eqref{eq:OPE-cut-parametrization}. Equation~\eqref{eq:sigma-space-OPE-cut} is symmetric with respect to $x \to -x$ change; thus, we can take $x$ in the $[0,1]$ interval. Two solutions labeled ``$\pm$" do not describe two ``parity copies" of the cut, but two smoothly connected halves of the same cut attached to a different branch point. We call the branch points $\sigma_{\pm}$, and obtain them from the above parametrization by setting $x=1$,
    \beq
    \label{eq:sigma-plus-minus}
    \sigma_{\pm} = \sigma_{\text{cut},\pm}(s,\sigma_k,1) = \frac{1}{2} (s - \sigma_k + 3 m^2) \pm  16 \pi \, \rho(\sigma_k) \, \lambda^{1/2}(s, \sigma_k, m^2) \, .
    \eeq
The expression for $\sigma_{\pm}$ has singularities in $\sigma_k$ and $s$ since the second term of Eq.~\eqref{eq:sigma-plus-minus} contains both the triangle function and two-body phase space. They have practical consequences for the implementation of the integral equation solution. For example, the ordering between $\re \sigma_{\pm}$ (i.e., which point is on the left and which on the right in the complex plane) depends on the relative value of $\im s$ and $\im \sigma_k$. It affects the choice of the integration contour; considering only real, positive values of $\sigma_k<4m^2$ and $\re s<s_{\varphi b}$, the $\lambda^{1/2}$ has a cut in $s$ below $(\sqrt{\sigma_k}+m)^2$. For $s \to s^*$, the real parts of branch points transform into each other, $\re \sigma_- \leftrightarrow \re \sigma_+$.

Similarly to the momentum-representation OPE amplitude, for $\im s = 0$, the branch cut wraps around the real $\sigma_{p}$ axis, resulting in the circular cut, as seen in Fig.~\ref{fig:circular-cut-invariants}. It occurs when $s$ is decreased below the value of $s$ given in Eq.~\eqref{eq:s-cut-condition-2}, at which the branch point $\sigma_+$ collides with $\sigma_{\rm max} = (\sqrt{s}-m)^2$. The cut encloses the upper integration limit $\sigma_{\rm max}$. For a non-zero imaginary part of $s$ or $\sigma_k$ (or non-zero $\epsilon$), the circle ``opens".

One finds the point where the cut passes the real axis by looking for the solution of condition \eqref{eq:invariants-pole-condition} with a vanishing imaginary part. We can rewrite it as,
   \beq
    \label{eq:invariants-pole-condition_v2}
    \frac{ z(\sigma_p,\sigma_k) }{\lambda^{1/2}(s, \sigma_p, m^2) \lambda^{1/2}(s, \sigma_k, m^2)} + x   = 0 \, .
    \eeq
Using the fact that $x$ is purely real, the crossing in the real axis satisfies, 
    \beq
    \label{eq:invariants-condition-im-z}
    \im \left[z(\sigma_p,\sigma_k) \left( \lambda^{1/2}(s,\sigma_p, m^2) \lambda^{1/2}(s,\sigma_k, m^2) \right)^* \right] = 0 \, .
    \eeq
Assuming real $s < (\sqrt{\sigma_k}+m)^2$, $\sigma_p < \sigma_{\rm max}$ and $\sigma_k$, we simplify it to,
    \beq
    \re z(\sigma_p, \sigma_k) = 0 \, ,
    \eeq
by noticing that the $\lambda^{1/2}(s,\sigma_p, m^2) \lambda^{1/2}(s,\sigma_k, m^2)$ factor is purely imaginary. Thus, we find that the circular cut crosses the real $\sigma_p$ axis at,
    \beq
    \label{eq:sigma_c1}
    \sigma_{c1} = \frac{(s - m^2) (s + m^2 - \sigma_{k})}{(s - m^2 + \sigma_{k})} \, .
    \eeq
For complex $s$, as $\im s \to 0$, the cut approaches the real $\sigma_p$ axis at another point, which we call $\sigma_{c2}$. Referring to panel (a) of Fig.~\ref{fig:circular-cut-invariants} for illustration, it is the point where the line and the circle cross each other. To express it in terms of $s$ and $\sigma_k$, we write $\sigma_p = \sigma_{c2} + i \delta$, where $i \delta$ is a positive, infinitesimal imaginary part. It constitutes a parametrization of the line tangent to the cut near the real $\sigma_p$ axis. Again, we start from the condition \eqref{eq:invariants-condition-im-z}. For real $s < (\sqrt{\sigma_k}+m)^2$, this
becomes, 
    \beq
    \delta \, (s + \sigma_k - m^2 ) \, \im \lambda^{1/2}(s,\sigma_p, m^2) + (2 s \sigma_k - (s+ \sigma_k - m^2) (s+m^2 - \sigma_{c2})) \re \lambda^{1/2}(s,\sigma_{p},m^2) = 0 \, ,
    \eeq
where this time $\lambda^{1/2}(s,\sigma_p, m^2)$ has in general non-zero real and imaginary parts. We expand the triangle function around $\delta = 0 $,
    \beq
    \lambda^{1/2}(s, \sigma_p, m^2) = \lambda^{1/2}(s, \sigma_{c2}, m^2) - \frac{i \delta \left(m^2+s-\sigma_{c2}\right)}{\lambda^{1/2}(s, \sigma_{c2}, m^2)} + \Oc(\delta^2) \, ,
    \eeq
which, assuming $\lambda^{1/2}(s, \sigma_{c2}, m^2)$ is purely imaginary, leads to,
    \beq
    -2 s \left(-2 m^4+m^2 (2 s+\sigma_k)+\sigma_k (\sigma_{c2}-s)\right) + \Oc(\delta) = 0  \, .
    \eeq
Neglecting terms of order $\delta$, the solution of the equation becomes,
    \beq
    \label{eq:sigma_c2}
    \sigma_{c2} = \frac{\left(m^2-s\right) \left(2 m^2-\sigma_{p} \right)}{\sigma_{p}} \, .
    \eeq
We note that $\sigma_{c1}, \sigma_{c2}$ correspond to $q_{c1}$ and $q_{c2}$ given in Eqs.~\eqref{eq:mom_c1}, \eqref{eq:mom_c2} The generalized values, $q'_{c1,2}$, provided in App.~\ref{app:C}, correspond to the points where the OPE branch cut, considered in the complex $\sigma_q$ plane, crosses a line $\im(\sqrt{s} - m)^2 =$ const. Control over the functional form of those points is essential when preparing the deformed integration contour, which enters the closed circle for $\im s=0$ through the $\sigma_{c2}$ or $q_{c2}$.

\section{Short introduction to analytic continuation}
\label{app:A}

This appendix should serve as a pedagogical summary of concepts used in Sec.~\ref{sec:analytic-continuation}, where we discuss an analytic continuation of the ladder equation. It is based on Refs.~\cite{cohen2007complex, Eden:1966dnq, burkhardt1969dispersion, lang1985complex} which may be consulted for more details. 

To understand our treatment of the integral equation, it is beneficial to consider a simpler case of an analytic continuation of a complex integral. We define a generic,
    \beq
    I(z) = \!\!\!\int\limits_{\Cc(w_1,w_2)} \!\!\!f(w, z) \, dw \, ,
    \eeq
where the integrand $f(w, z)$ is a complex function of argument $w$ and depends on a complex parameter $z$. Integration is performed over a path $\Cc(w_1,w_2)$ which starts at $w_1 = w_1(z)$ and ends at $w_2 = w_2(z)$. As indicated, these two points can also depend on $z$. The homogeneous term of the ladder equation, Eq.~\eqref{eq:d_Sproj_kern}, has an analogous form; however, we do not know the equivalent of $f(w,z)$ beforehand, since $d_S$ is an unknown of the integral equation.

If we know the analytic structure of $f(w, z)$, we can infer the analytic structure of $I(z)$. In general, singularities of $I(z)$ appear for those values of $z$ for which: a) $f(w,z)$ has explicit, $w$-independent singularities in $z$; b) $z$-dependent singularity in $w$ coincides with the lower limit of integration, $w_1$; c) $z$-dependent singularity in $w$ coincides with the upper limit of integration, $w_2$; d) two movable singularities of $f(w,z)$, pinch the integration contour; e) movable singularities of $f(w,z)$ require contour deformation to complex infinity. We note it is sufficient to know singularities of $f(w,z)$ to establish singularities of $I(z)$ and not the value of $f(w,z)$ at every point of the complex plane.

We illustrate this with a typical example of a real integral,
    \beq
    \label{eq:integral-example}
    I(x) = \int_{-1}^{1} \frac{dw}{w - x} \, ,
    \eeq
where the integration variable lies on the real axis between $w_1 = -1$ and $w_2=+1$. For $x$ outside of the integration range, we can easily evaluate the integral and obtain,
    \beq
    \label{eq:integral-example-1}
    I(x) = \log \left( \frac{x-1}{x+1} \right) \, , ~~~~ x \in (-\infty, -1) \cup (1,\infty)  \, .
    \eeq
The integral is not defined for $x \in [w_1,w_2]$ due to the pole singularity at $w = x$. However, having the explicit functional form, given in Eq.~\eqref{eq:integral-example-1}, it is possible to assign a meaning to this function in this range. Namely, we promote the real $I(x)$ to a function $I(z)$ of a complex variable $z$ which is equal to $x$ on the real axis. Due to the multi-valued nature of the complex logarithm, $I(z)$ has two branch points, at $w_1$ and $w_2$, and two associated cuts. These can be chosen arbitrarily, corresponding to different definitions of the function on the first Riemann sheet. For example, we can align both cuts with the real axis and orient them to the right, which results in a single branch cut in the interval $[w_1,w_2]$. For this choice, the function is undefined on this short segment of the real axis, which is clear since the original integral in \eqref{eq:integral-example} was ill-defined there. 

\begin{figure}[t]
    \centering
    \includegraphics[width=0.8\textwidth]{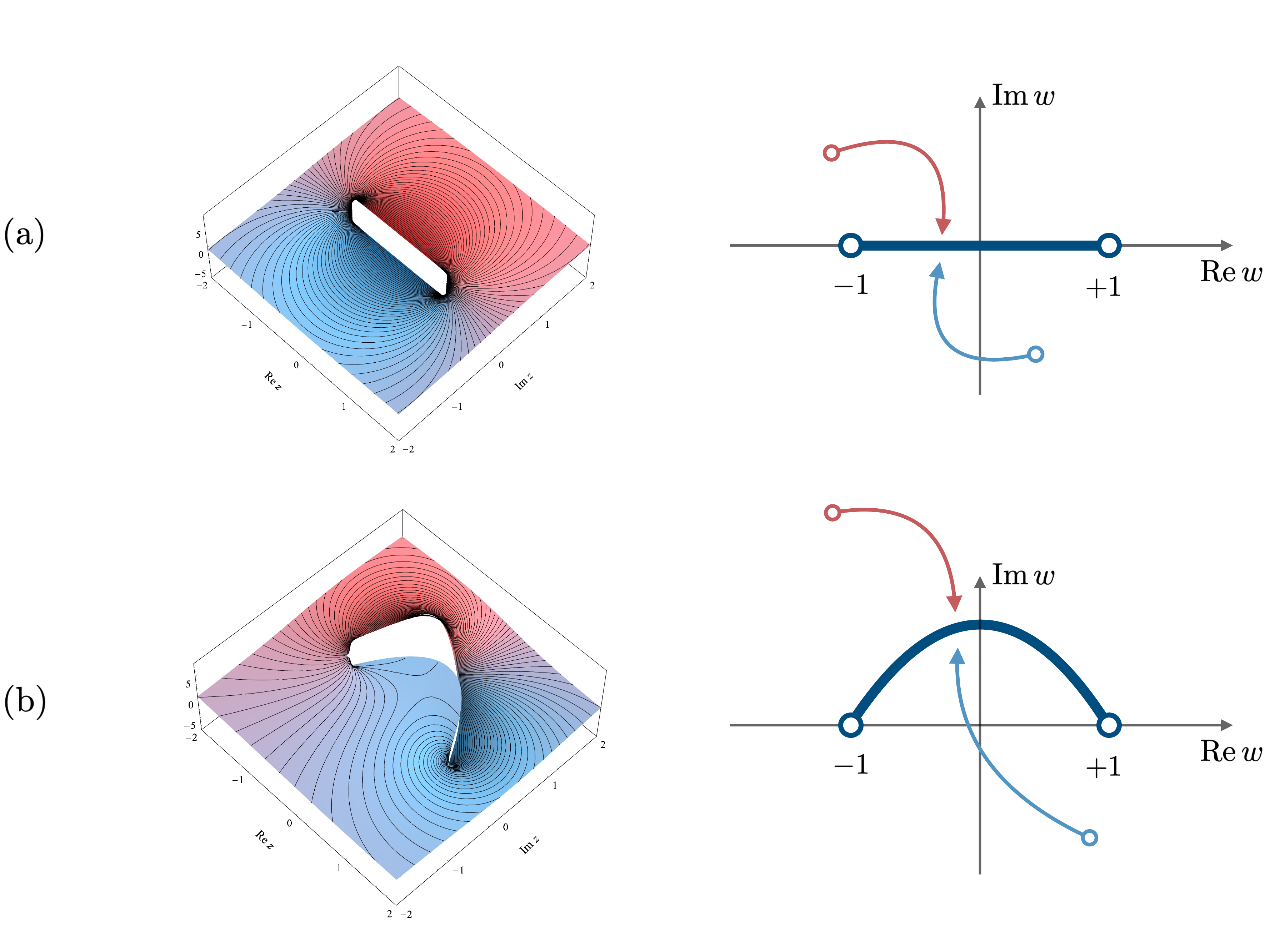}
    \caption{\textbf{Left:} Imaginary part of the integral $I(x)$ in Eq.~\eqref{eq:integral-example}, defined as a complex function. \textbf{Right:} Corresponding integration path $\Cc(-1,+1)$. Continuation of the function to the real interval $[-1,+1]$ requires contour deformation. Function (b) is given there by the function (a) evaluated on the Riemann sheet connected to the bottom half-plane.}
    \label{fig:basic_complex_example}
\end{figure}

However, the cuts can be oriented in other directions, e.g., to cover $(-\infty, -1) \cup (1,\infty)$, such that the function has a well-defined value $I(x)$ for $-1 < x < 1$. To establish a relation between the complex function $I(z)$ with its cuts moved away from $[-1,+1]$ and the original defining integral, we can promote it to a complex integral along a general complex contour $\Cc(-1,+1)$. The chosen integration path determines the cut structure of the resulting $I(z)$. The pole of the integrand leads to singularity only if it coincides with the integration path; thus, if the integration contour avoids the interval $[-1,+1]$, the integral is well-defined there. The function varies continuously as we cross the $[-1,+1]$ interval vertically and becomes equal to its value on the nearest Riemann sheet of the ``principal" definition, according to the Cauchy theorem, see Fig.~\ref{fig:basic_complex_example}. 

It is easy to see that in the example of Eq.~\eqref{eq:integral-example}, we encounter cases (b) and (c). The $x$-dependent pole of $f(w,x)=1/(w-x)$ coincides with $w_1$ and $w_2$, resulting in the two branch points of the logarithm at these points. These branch points have a fixed position that cannot be altered by contour deformation, as every deformed path must begin and end at the same locations. Consequently, the shape of the new branch cut is determined by the deformed integration path.

In addition to illustrating branch cuts emergence, the above example suggests how a contour deformation allows one to extend the definition of $I(z)$ to a point $z_0$ where it was originally ill-defined. In other words, contour deformation defines the analytic continuation of $I(z)$ to a new region of the complex plane. This fact is well-known in the general S-matrix theory \cite{Eden:1966dnq} and has been widely used in the phenomenological studies of the scattering processes, e.g., see Ref.~\cite{Mai:2020ltx, ALBALADEJO2022103981, Doring:2009yv}.

In Secs.~\ref{sec:analytic-structure} and \ref{sec:analytic-continuation}, we explain how singularities of the solution $d_S(p,k)$, considered as a function of $s$ emerge from conditions (a)-(e) applied to the integration kernel $K(p,q)$ and the unknown function $d_S(q,k)$. For example, $K$ contains the two-body amplitude $\Mc_2$ that exhibits an $s$-dependent, complex pole, $q=q_b$. Collision of $q_b$ with the lower integration endpoint, $q=0$, leads to the unitarity branch point at $s = s_{\varphi b}$ [condition (b)]. Collision of the OPE branch point $p_+$ with $q=0$ leads to a branch point at $s = (m^2 - \sigma_k)^2/m^2$ [condition (b)]. Moreover, the final amplitude inherits an explicit ``short" cut from the OPE amplitude in the inhomogeneous term, considered as a function of $s$ for fixed $(p,k)$ [condition (a)]. 

To summarize, although we do not know the solution $d_S(p,k)$ the defining integral equation can be used to infer singularities of the amplitude even without solving it. Moreover, the ladder amplitude can be analytically continued to the kinematic regions of interest via the contour deformation.

\section{Numerical methods}
\label{app:C}

In this section, we describe numerical methods that were used to obtain analytically continued solutions presented in Sec.~\ref{sec:results}. Similarly to Ref.~\cite{Jackura:2020bsk} we employ the Nystr\"om method \cite{10.1007/BF02547521, delves1988computational}, i.e., we discretize momentum variables and rewrite the problem as an algebraic equation. The methods presented below are applicable in more general studies of the three-body integral equations. They are relatively well-known but we describe them here to increase the reader's ease in reproducing the results presented in this study.

\subsection{Definition of the deformed contour}

For reader's convenience, below we reproduce the partial-wave projected, amputated ladder equation, Eq.~\eqref{eq:d_Sproj},
    \beq
    \label{eq:C-ladder-equation}
    d_S(p, k) = - G_S(p, k) -
    \int\limits_{0}^{q_{\text{max}}} 
    \hspace{-2pt} dq
    \, K(p, q) \, d_S(q, k) \, .
    \eeq
where the integration kernel, $K(p, q)$ is defined in Eq.~\eqref{eq:K_Sproj}. We indicated the finite range of the integration with the upper limit $q_{\text{max}}$, which is defined by the cut-off function, Eq.~\eqref{eq:cut-off}. The motion of the OPE cuts in the complex $p$ and $q$ planes necessitates contour deformation in Eq.~\eqref{eq:C-ladder-equation} for a large range of values of $s$ and $p$. In the following, we consider values of kinematic variables $(p, k; s)$ for which the real-$q$ axis is crossed by a cut. To compute the solution, we deform the integration path,
    \beq
    \label{eq:contour-deformation}
    \int\limits_{0}^{q_{\text{max}}} dq \longrightarrow 
    \int_{\Cc} = \int_0^{1} dt \, \gamma'(t) \, .
    \eeq
The complex contour $\Cc$ is defined by a parametrization $q = \gamma(t)$ where real parameter $t \in [0,1]$. For a given set of kinematic variables, the curve has fixed endpoints, $\gamma(0) = 0$, and $\gamma(1) = q_{\text{max}}$. The integral equation becomes,
    \beq
    \label{eq:C-ladder-equation-2}
    d_S(p, k) = - G_S(p, k) -
    \int_{0}^{1} 
    \hspace{-2pt} dt \, \gamma'(t)
    \, K(p, \gamma(t)) \, d_S(\gamma(t), k) \, .
    \eeq
The Nystr\"om method is applied to the ladder equation in the above form.

\begin{table}[t!]
    \centering
    \begin{ruledtabular}
    \begin{tabular}{ccc}
       Node  & Im$(s)\leq$0 & Im$(s)>$0 \\ \hline
       $q_0 $  & 0 & 0 \\
       $q_1 $  & $- \frac{2}{3\sqrt{2}} (1 + 
        i) |q_{c2}|$ & $- \frac{2}{3\sqrt{2}} (1 + 
        i) |q_{c2}|$  \\
       $q_2 $  & $ \frac{1}{2} \re\left( p_+ - p_- \right) - |q_{c2}| i$ & $-p_{\text{cut},+}(s, q_b, -x_0)$  \\
       $q_3 $  & $\frac{2}{3}(1 - 2 i) |q_{c2}|$ & $p_{\text{cut},+}(s, q_b, x_0)$ \\
       $q_4 $  & $\frac{3}{2} |q_{c2}|$ & $\frac{2}{3}(1 - 2 i) |q_{c2}|$ \\
       $q_5 $  & $q_{\text{max}}$ & $\frac{3}{2} |q_{c2}|$ \\
       $q_6 $  & $-$ & $q_{\text{max}}$ \\
    \end{tabular}
    \end{ruledtabular}
    \caption{Example nodes defining piece-wise linear contour $\Cc$ for positive and negative values of Im$(s)$. Momentum $q_{c2}$ is defined in Eq.~\eqref{eq:mom_c2} and momentum $p_\pm$ in Eq.~\eqref{eq:mom-plus-minus}. Both are evaluated at $\sigma_{k} = \sigma_b$. Moreover, $ x_0 = |\re \left( z(q_{c2}, q_b) / 2 q_{c2} q_b \right) |$.} 
    \label{tab:nodes}
\end{table}

We note that every self-consistent contour that avoids singularities of the OPE and the integration kernel is a legitimate choice. In practice, the contour used in the solution routine must evolve with values of $s$, $k$, and the scattering length $a$, since the position of the OPE cuts depends on these parameters. Due to the complicated shapes of the cuts we employ contours defined in a piece-wise linear manner, which allows for more control than explicitly given, fixed functions. A contour is defined by a set of $n+1$ nodes $\{q_i\}_{i \in [0,n]}$, which connect lines constituting the integration path. The $i$-th line is defined as,
    \beq
    \gamma_i(t) = \left(\frac{q_{i+1}-q_i}{t_{i+1} - t_i} \right) \, t + \frac{t_{i+1} q_i - t_i q_{i+1}}{ t_{i+1} -t_i } \, , ~~ t \in [t_i,t_{i+1}] \, , ~~i=0,1,\dots,n-1 \, ,
    \eeq
where $t_0 = 0 < t_1 < \dots < t_{n-1} < t_n = 1$.
Two example sets of nodes for two different cases of Im$(s)$ are given in Tab.~\ref{tab:nodes}. They are suitable for $k = q_b$, and a relatively large range of complex $s$ and positive $a$. Example contours created using these nodes are shown in Fig.~\ref{fig:two-contours}. Note that for $\im s>0$ the $G_S$ amplitude is evaluated on the second sheet between points $q_2$ and $q_3$. One can use contours that have a different number of nodes depending on the shape of the cut and other practical considerations.

We note that points $q_{c1}$ and $q_{c2}$, derived in Eqs.~\eqref{eq:mom_c1},~\eqref{eq:mom_c2} are used in the definition of both contours. They roughly describe the size of the ``circle" and thus are useful in devising an integration path that avoids the OPE amplitude cuts. Although we are satisfied with this prescription, one can also generalize those points to a case when $\sigma_k$ and $s$ are complex. This describes the ``open" circle scenario. The generalized points are called $q'_{c1}$ and $q'_{c2}$. They are derived from the condition $\im [z(p,k)/2 pk] = 0$. Below, we show an example derivation of $q'_{c1}$; the other point is obtained analogously. First, we observe that the above condition implies,
    \beq
    \im [z(p,k) p^* k^*] = 0 ~~~\Rightarrow~~~ \im z(p,k) \re k^* + \re z(p,k) \im[ k^*] = 0 \, .
    \eeq
since $p = q'_{c1}$ is real. (For $q_{c2}$ we assume purely imaginary $p=q'_{c2}$.) We observe that,
    \beq
    z(p,k) = \sigma_k - 2 (\sqrt{s} - \omega_k) \omega_p \, .
    \eeq
Thus,
    \beq
    \im z(p,k) = \im \sigma_k - 2 \omega_p \im[\sqrt{s} - \omega_k] \, , ~~~ \re z(p,k) = \re \sigma_k - 2 \omega_p \re[\sqrt{s} - \omega_k] \, .
    \eeq
This leads to a linear equation for $\omega_p$, which can be solved,
    \beq
    \omega_p = \frac{1}{2} \frac{\im[\sigma_k k^*]}{\im[(\sqrt{s}-\omega_k)k^*]} \, .
    \eeq
Thus position where the OPE cut crosses the real axis is,
    \beq
    q'_{c1} = \sqrt{ \frac{1}{4} \left(\frac{\im[\sigma_k \, k^* ]}{ \im [ (\sqrt{s} - \omega_k) \, k^*]} \right)^2 - m^2 } \, .
    \eeq
Similarly, we can obtain a point where it crosses the imaginary axis,
    \beq
    q'_{c2} = \sqrt{ \frac{1}{4} \left(\frac{\re[\sigma_k \, k^* ]}{ \re [ (\sqrt{s} - \omega_k) \, k^*]}\right)^2 - m^2 } \, .
    \eeq
Reflection of these points with respect to the origin of the complex momentum plane gives the remaining crossover points of the OPE. We note that for real $s$, $q_{c1}=q'_{c1}$, but $q_{c2} \neq q'_{c2}$, since $q_{c2}$ is not a point of crossover.

\begin{figure}[t]
\centering
\includegraphics[width=0.9\textwidth]{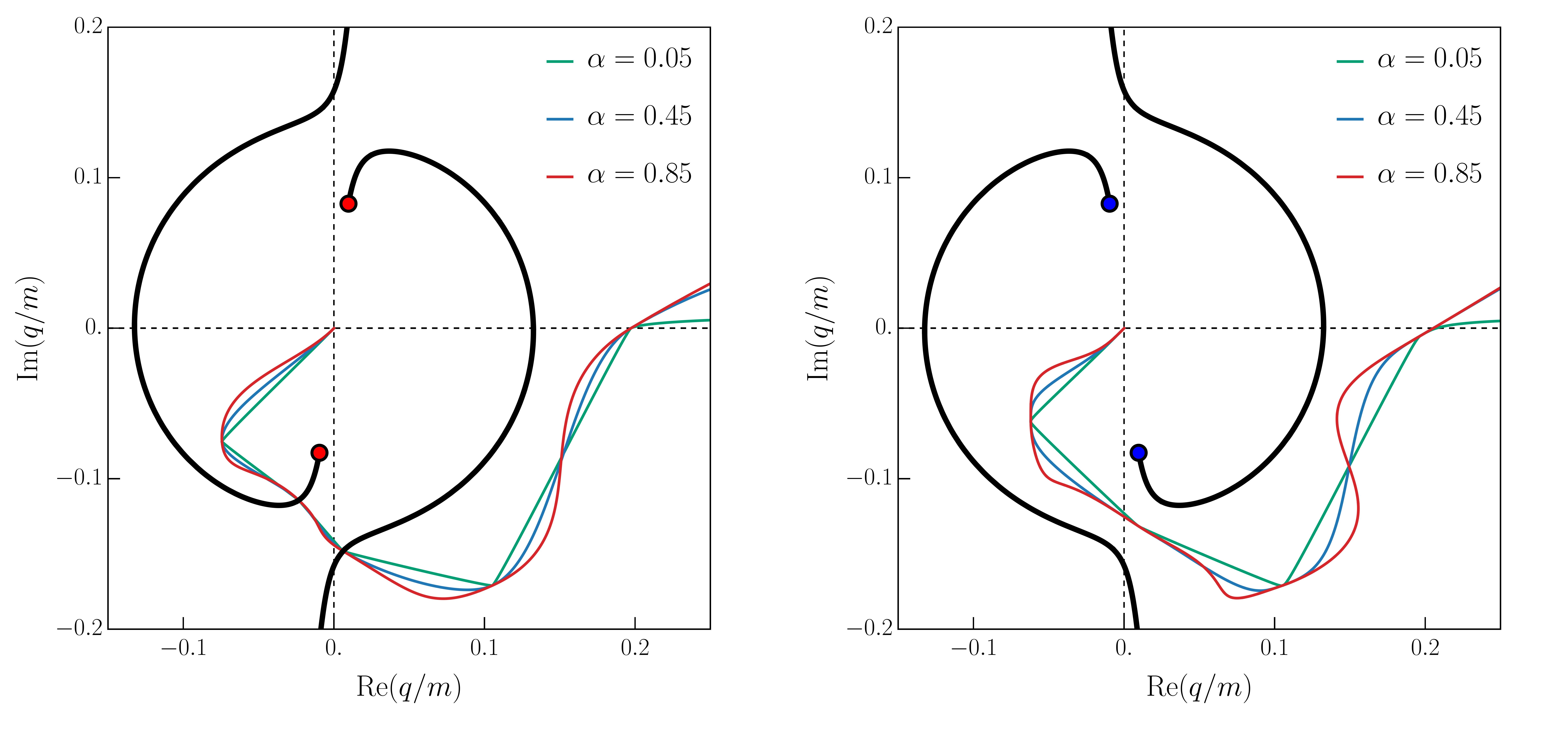}
\caption{ Example interpolation of the piece-wise linear contour with a smooth one for various values of parameter $\alpha$. Black lines represent cuts of $G_S(q_b,q)$ for $ma=16$. Left panel: $s/m^2 = 8.6 + 0.05i$. Right panel: $s/m^2 = 8.6 - 0.05i$.}
\label{fig:smooth_contour}
\end{figure}

In certain cases, we find that ``smoothing" the integration contour leads to a better numerical convergence of the amplitudes. Derivative $\gamma'(t)$ in Eq.~\eqref{eq:C-ladder-equation} is discontinuous for the piece-wise linear path, which might prevent one from using certain types of quadratures when discretizing the integral equation. To smooth out the $\gamma(t)$ function around points $q_i = \gamma(t_i)$ one may, for example, perform an interpolation of the contour using cardinal Hermite splines \citep{Schoenberg:1973}. To achieve continuity of $\gamma''(t)$ we use the 5th order polynomials, defined as,
    \beq
    \left [ \begin{array}{c}
    p_1(t)  \\
    p_2(t)  \\
    p_3(t)  \\
    p_4(t)  \\
    p_5(t)  \\
    p_6(t)
    \end{array} \right] = \left [ \begin{array}{cccccc}
     1 & 0 & 0 & -10 & 15 & -6 \\
     0 & 0 & 0 & 10 & -15 & 6 \\
     0 & 1 & 0 & -6 & 8 & - 3 \\
     0 & 0 & 0 & -4 & 7 & -3 \\
     0 & 0 & 1/2 & -3/2 & +3/2 & -1 \\
     0 & 0 & 0 & 1/2 & -1 & 1/2
    \end{array} \right] 
    \left [ \begin{array}{c}
    1  \\
    t  \\
    t^2  \\
    t^3  \\
    t^4 \\
    t^5   
    \end{array} \right] \, .
    \eeq
The integration contour is given by,
    \beq
    \gamma_i(t) = p_1(y_i) \, q_i + p_2(y_i) \, q_{i+1} + p_3(y_i) \, q'_{i} + p_4(y_i) \, q'_{i+1} + p_5(y_i) \, q''_{i} + p_6(y_i) \, q''_{i+1} \, ,
    \eeq
for $t \in [t_i,t_{i+1}]$ and $i=0,1,\dots,n-1$. Here $y_i = (t - t_i)/(t_{i+1}-t_i)$. The first and second tangents are defined as,
    \beq
    \label{eq:tangents}
    q'_i &=& \alpha \, \left(\frac{q_{i+1} - q_i}{t_{i+1} - t_i} + \frac{q_{i} - q_{i-1}}{t_{i} - t_{i-1}} \right) \, , \\
    \label{eq:2nd_tangents}
    q''_i  &=& \alpha \, \left(\frac{q'_{i+1} - q'_i}{t_{i+1} - t_i} + \frac{q'_{i} - q'_{i-1}}{t_{i} - t_{i-1}} \right) \, \, .
    \eeq
For $i=0$ we use only the first and for $i=n$ second term in \eqref{eq:tangents}, and define $q''_0 = q''_n = 0$. Real parameter $\alpha$ defines the ``tension" of the interpolating curve. For $\alpha=0$ one obtains the piece-wise linear contour. An example of smooth contours is given in Fig.~\ref{fig:smooth_contour}.

\subsection{Discretization of spectator momenta}

To apply Nystr\"om method to the Eq.~\eqref{eq:C-ladder-equation-2}, at fixed $s$ and $k$, one has to fix the integration contour $\Cc$, and evaluate the $p$ variable on $\Cc$, i.e., rewrite $p = \gamma(u)$, $u \in [0,1]$. One then discretizes both real variables, $t$ and $u$, to rewrite the integral equation as a matrix one. In the simplest numerical approach, we use a uniform mesh of $N+1$ points: $p_i = \gamma(u_i)$ and $q_i= \gamma(t_i)$, where $i \in \{0,..., N\}$ and $t_i = u_i = i/N$. Thus, each linear path of a contour contains the number of discrete points proportional to its length. This represents a simple extension of the ``brute force" method from Ref.~\cite{Jackura:2020bsk}. The integral is replaced with a sum,
    \beq
    \label{eq:discretized_ladder}
    \bm{d}_i = - \bm{G}_i -
    \sum_{j=0}^{N-1}
    \bm{K}_{ij} \, \bm{d}_{j} \, ,
    \eeq
where, 
    \beq
    \label{eq:integral_discretization1}
    \bm{d}_i &=& d_S(\gamma(u_i), p) \, , \\
    \label{eq:integral_discretization2}
    \bm{G}_i &=& G_S(\gamma(u_i), p) \, , \\
    \label{eq:integral_discretization3}
    \bm{K}_{ij}(s) &=& \left[ \gamma(t_{j+1}) - \gamma(t_j) \right] \, K(\gamma(u_i), \gamma(t_j)) \, .
    \eeq
We used bold font to indicate that $d_S$, $G_S$, and kernel $K$ became vectors and a matrix in the discrete $(u_i,t_j)$ space. In Eq.~\eqref{eq:integral_discretization3}, we employed the simplest rectangular rule with a forward derivative. One can also apply other methods (e.g. trapezoidal, Simpson, etc.) and use the exact value of $\gamma'(t)$ at a discrete point $t_j$. The solution of the algebraic equation is,
    \beq
    \label{eq:algebraic-solution}
    \bm{d}_i^{(\text{sol})} = - \sum_{j=0}^{N-1} \, [ \one +  \bm{K}]^{-1}_{ij} \, \bm{G}_j \, .
    \eeq
Assuming we know $\bm{d}^{(\text{sol})}$, the final amplitude is obtained by extrapolating the solution to the momentum of interest, e.g., $p=q_b$,
    \beq
    \label{eq:discretized_interpolation}
    \Mc_{\varphi b}(s) = - g^2 \, G_S(q_b, q_b) - g^2 \,
    \sum_{j=0}^{N-1} \, 
    \left[ \gamma(t_{j+1}) - \gamma(t_j) \right] \, 
    K(q_b, \gamma(t_j)) \, \bm{d}^{(\text{sol})}_j \, .
    \eeq
The conceptually simple rectangular rule is an elementary numerical technique that yields improving results with larger $N$. However, its convergence with the matrix sizes is relatively slow and can be accelerated with alternative, more sophisticated discretization techniques. For instance, one can use Gaussian quadratures \cite{golberg, atkinson_1997}, or spline-based method \cite{Glockle:1982, Horacek:1977pv, Jackura:2020bsk}. We find that Gauus-Chebyshev (GC) and Gauss-Legendre (GL) quadratures can be easily employed and offer a great improvement in the convergence of the solutions. Conceptually, implementation of a Gauss quadrature is achieved by replacing,
    \beq
    \label{eq:quadratures-basics}
    \int_{0}^{1} \, dt \, g(t) = \sum_{n=0}^{N-1} w_i \, g(t_i) \, ,
    \eeq
for a function $g(t)$. Here, $t_i$ and $w_i$ are pre-computed mesh points and corresponding weights, respectively. In practice, this amounts to the replacement of Eq.~\eqref{eq:integral_discretization3} with,
    \beq
    \label{eq:gl_discretization}
    \bm{K}_{ij}(s) &=& \frac{1}{2} \, w_j \, \gamma'(t'_j) \, K(\gamma(u'_i), \gamma(t'_j)) \, .
    \eeq
Since the GC and GL quadratures are defined for the integration interval $t \in [-1,1]$, we map linearly $[-1,1] \to [0,1]$, hence the $1/2$ factor in the equation above. Primed variables are obtained from the Gauss points $t_j$ as $t'_j = (1 + t_j)/2$. In our C++ implementation of the ladder equation, we use available GL quadratures (weights and points) from Ref.~\cite{Bogaert:2014}. The solution in the GL method is still given by Eq.~\eqref{eq:algebraic-solution} while extrapolation $p \to q_b$ and $\Mc_{\varphi b}$ is achieved through,
    \beq
    \label{eq:gl_interpolation}
    \Mc_{\varphi b}(s) = - g^2 \, G_S(q_b, q_b) - \frac{1}{2} \, g^2
    \sum_{j=0}^{N-1} \, 
    w_i \, \gamma'(t'_j) \, 
    K(q_b, \gamma(t'_j)) \, \bm{d}^{(\text{sol})}_j \, .
    \eeq

Discretization procedures described in this subsection are also applied to the homogeneous version of the ladder equation, Eq.~\eqref{eq:homo-eq}. Position of the three-body bound state pole in $s$ is obtained either from the determinant condition, Eq.~\eqref{eq:determinant-condition} or identification of zeros of $1/\re( d_S(p,p))$ for some choice of external momenta. In both cases, we accomplish it numerically by using the secant method with precision $\Delta s = 10^{-12}$.

\subsection{Analysis of the systematic effects}

In Ref.~\cite{Jackura:2020bsk}, the authors studied systematic effects of the numerical approaches by considering two limits: matrix size $N$ going to infinity, and the two-body pole position shift, $i\epsilon$, going to zero. Here, we do not deal with poles coinciding with the integration contour, which usually cause numerical instabilities. Thus, the analysis of systematic effects is greatly simplified and the precision of the solutions is improved compared to the previously studied, more demanding case.

\begin{figure}[b!]
    \centering
    \includegraphics[width=0.99\textwidth]{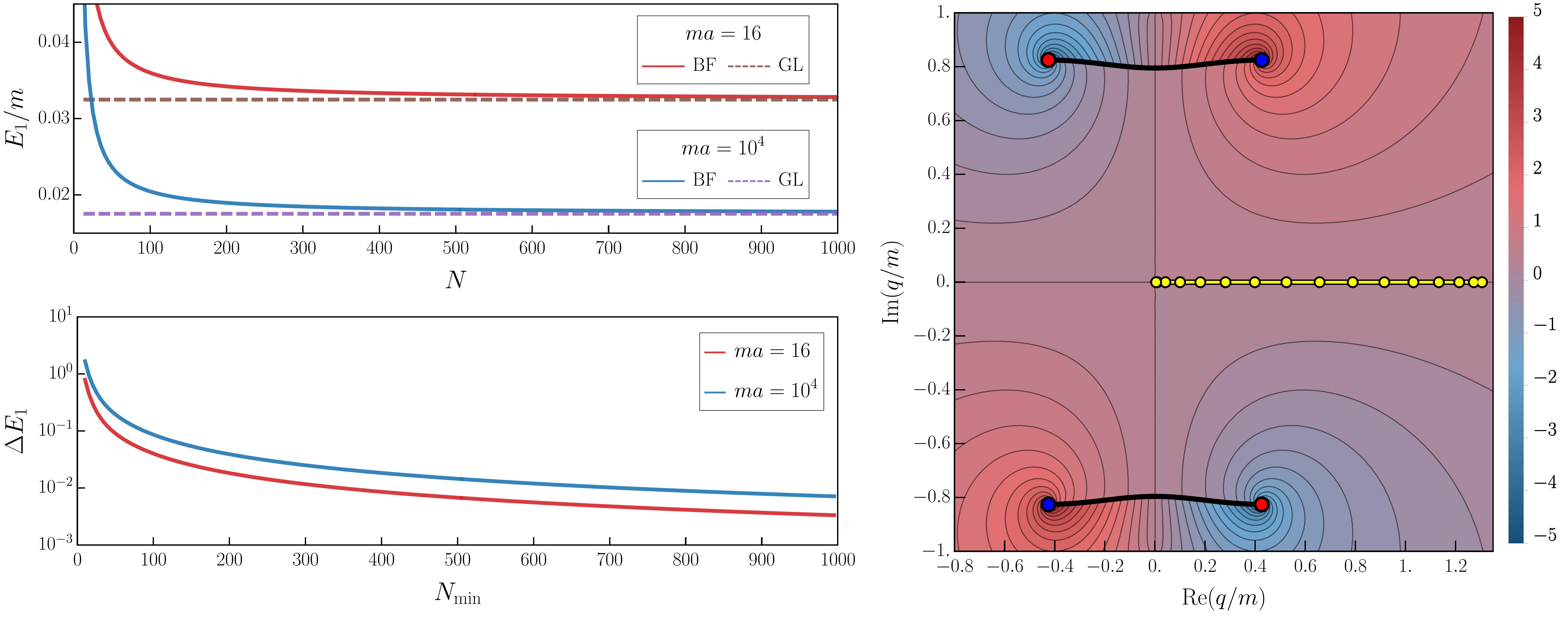}
    \caption{Mesh size dependence of the binding energy of the deepest three-body bound state, $E_1$, obtained from the BF and GL methods. On the top-left panel, we present convergence for two cases, $ma=16$, and $ma=10^4$. On the bottom-left panel, we show the corresponding value of the relative error, $\Delta E_1$ as a function of the $N_{\text{min}}$, as described in the text. On the right panel, we present a contour used to obtain these results, with an example of Gaussian nodes for $N=15$. The depicted function is $\im G(p,q)$ for $\sigma_{p} = 2m^2$, with cuts represented by black lines. It is given in units of $1/m^2$. The smooth cut-off was used.}
    \label{fig:convergence}
\end{figure}

\begin{figure}[t!]
    \centering
    \includegraphics[width=0.99\textwidth]{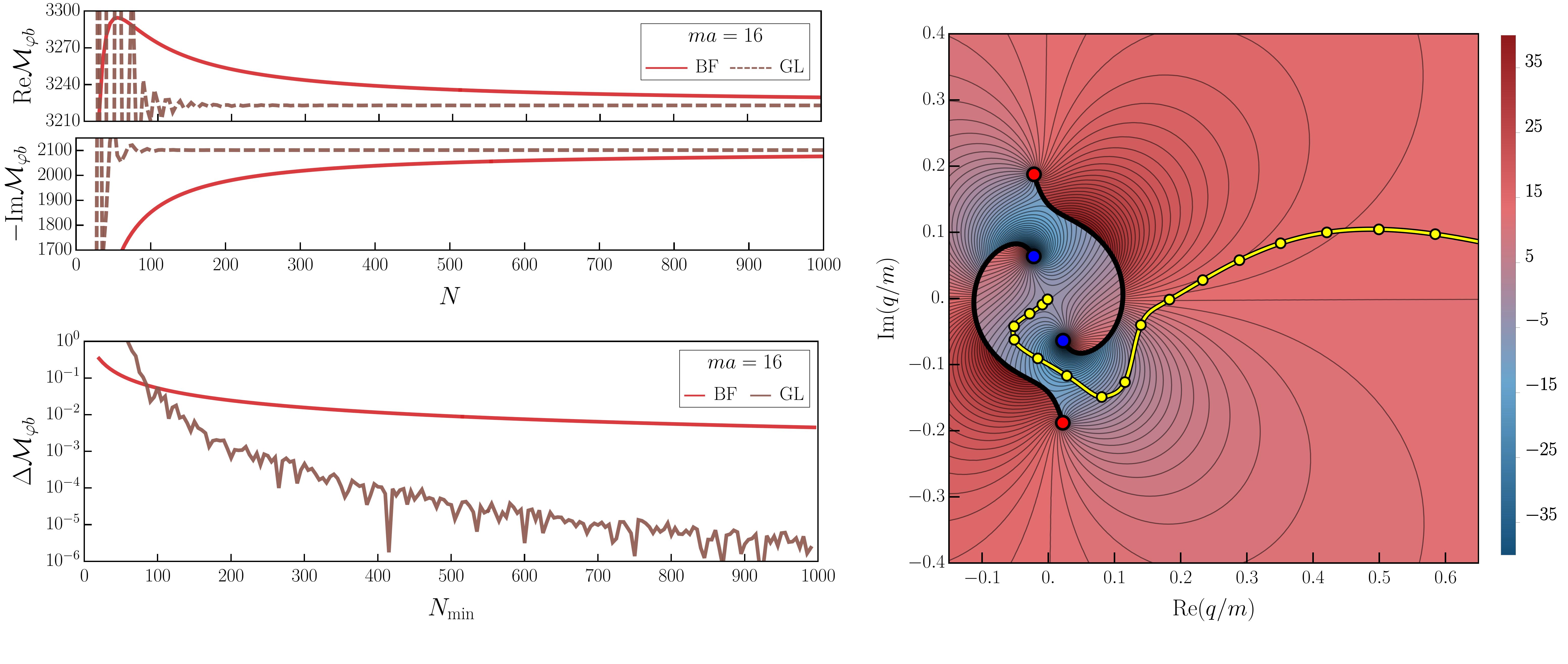}
    \caption{Mesh size dependence of the $\Mc_{\varphi b}(s)$ for $s/m^2 = 8.7 - 0.1 i$, obtained from the BF and GL methods. On the top-left panel, we present the convergence of the real and imaginary parts of the amplitude. On the bottom-left panel, we show the corresponding value of $\Delta \Mc_{\varphi b}$ as a function of the $N_{\text{min}}$. On the right panel, we present a smooth contour ($\alpha=0.45$) used to obtain these results, highlighting example Gaussian nodes for $N=30$. The depicted function is $\im G(q_b,q)$, with cuts represented by black lines. It is given in units of $1/m^2$. The smooth cut-off was used. }
    \label{fig:convergence1}
\end{figure}

We find that the GL method leads to a fast convergence of the results when smooth integration contours are used. Typically, a mesh of $N \approx 100$ points is sufficient to obtain results that cease to depend on the matrix size within desired precision. Piece-wise linear contours may cause unwanted oscillations of $\bm{d}^{(\text{sol})}$ considered as a function of $N$. In this case, the GL method amplitudes still converge faster than the one obtained from the BF results, but it is harder to analyze them systematically. We find, the BF method always leads to a smooth controllable dependence of the $N$-dependent amplitudes, regardless of the type of contour. However, it requires the implementation of extrapolation to continuum, $N \to \infty$. Practically, this means one has to calculate the amplitude $\bm{d}^{(\text{sol})}$ using a set of few matrix sizes, usually of the order $N \approx 10^3$, and then fit the result with the polynomial formula,
    \beq
    \label{eq:extrapolation-fit}
    \bm{d}^{(\text{sol})}(N) = A + \frac{B}{N} \, ,
    \eeq
where the asymptote $A = \bm{d}^{(\text{sol})}(\infty)$ is taken as the continuum result \cite{Jackura:2020bsk}. Higher orders of $1/N$ can be included to improve convergence. From this perspective, the GL method is much more effective, since it allows one to use a single, relatively small value of $N$ to obtain the desired outcome with high confidence. We note, that convergence of the BF method can be improved via different means, e.g., Richardson extrapolation \cite{delves1988computational}; however, we do not implement any acceleration techniques in this work.

The analysis of the numerical uncertainty of $\bm{d}^{(\text{sol})}$ and its extrapolations can be performed as described in Chap.~4 of Ref.~\cite{delves1988computational}. Since the estimated error of our results is satisfactorily small, we use simpler, rough estimates. We note that the convergence of the results typically depends on the distance between the singularities of the kernel and the integration path. For instance, three-body pole positions are obtained from the ladder equation at external momenta $p',p$ for which the OPE cuts are far from the integration range. In the GL method, this leads to a relative difference of the order $10^{-4}$\% between the $N=15$ and $N=1000$ results, and virtually no difference between $N=50$ and $N=1000$ values. Thus, for the bound-state pole positions, we take the finite-$N$ GL result with the error given by the precision of the root-finding algorithm, which we set to $\Delta s = 10^{-12}$. We find that the extrapolated BF result converges to the GL one when large matrices are used for the fit. 

In Fig.~\ref{fig:convergence}, we show convergence of the binding energy of the ground-state timer, $E_1 = \sqrt{s_{\varphi b}} - \sqrt{s_b}$, with matrix size $N$. The bottom left panel shows the relative difference $\Delta E_1 = 100 \times \left| E_{1,\text{BF}}(N_{\text{min}}) - E_{1,\text{GL}} \right|/E_{1,\text{GL}}$, where $E_{1,\text{BF}}(N_{\text{min}})$ is the extrapolated BF result obtained from fitting the Eq.~\eqref{eq:extrapolation-fit} in the interval $[N_{\text{min}},1000]$. We see that, as the larger matrices are used in the fit, the extrapolated BF result converges to the GL one, reaching an acceptable relative difference of $10^{-2}$\% at $N_{\text{min}} \approx 500$. Since the BF method requires computation at several values of $N$ to achieve this level of agreement, we point to a significant advantage of the GL over the BF method.

When the OPE cuts approach the origin of the complex $q$ plane and enclose the lower limit of the integration, the convergence of the results becomes slower. The GL method amplitudes exhibit oscillatory behavior with $N$ and do not stabilize entirely at any finite value of the matrix size. However, despite this behavior, they still converge very quickly with oscillations damped by orders of magnitude within a relatively small range of $N$. Due to the oscillations, one can not easily extrapolate the GL values, e.g., by using a version of Eq.~\eqref{eq:extrapolation-fit}. Instead, for a given value of $s,k$, one computes $\bm{d}^{(\text{sol})}$ at a few close values of $N$, and takes their average as the final result, with the largest difference between the two of used values as an error estimate. Applying this procedure to different values of complex $s$ reveals that for sufficiently large $N$, the error estimate is much smaller than $10^{-2}$\%, allowing one to use a finite-$N$ result as a sufficient approximation of the continuum one. 

For illustration, in Fig.~\ref{fig:convergence1} we present example results for the amplitude $\Mc_{\varphi b}$ computed at $s/m^2 = 8.7 - 0.1 i$ and $ma=16$. In the top left panel, for $N<100$ we see large oscillations of the GL amplitude, that are quickly damped and hardly noticeable for larger values of $N$. The bottom panel shows the ``quality measure" of the solution, $\Delta \Mc_{\varphi b} = 100 \times \left| (\Mc_{\varphi b}(N_{\text{min}}) - \Mc_{\varphi b} ) /\Mc_{\varphi b} \right|$, for both methods. We assume that the ``correct solution", $\Mc_{\varphi b}$, is well approximated by an average of GL results obtained for $N=950,955,...,1000$. For the BF method, the $\Mc_{\varphi b}(N_{\text{min}})$ is the extrapolated result obtained from fitting Eq.~\eqref{eq:extrapolation-fit} in the interval $[N_{\text{min}},1000]$. For the GL method $\Mc_{\varphi b}(N_{\text{min}})$ is an average of three values of the amplitude computed at matrix sizes $N=N_{\text{min}}, N_{\text{min}}+5,$ and $N_{\text{min}}+10$. We see that the GL method offers a reduction of such defined error by several orders of magnitude compared to the BF approach at a given matrix size. The actual improvement depends on the value of $s$ and the contour smoothness parameter $\alpha$.

In this study, we consider a sub-percent precision of our results as entirely satisfactory. Such uncertainty is much smaller than anticipated errors from the lattice data that would enter our integral equations through the inclusion of non-zero $\Kc_{\text{df},3}$. In most applications, we choose to use the GL method with matrix size $N=500$, which should result in a relative error of at most $10^{-2}$\%. As discussed above, when the OPE cuts are far from the integration contour, like in the case of $\sigma_p = \sigma_k = 2m^2$ which we used to extract bound-state pole positions, the error is expected to be many orders of magnitude smaller.

\bibliographystyle{apsrev4-1}
\bibliography{main}

\end{document}